\documentstyle[11pt]{article}

\makeatletter
\@addtoreset{equation}{section}
\makeatother

\makeatletter
\def\citen#1{%
\edef\@tempa{\@ignspaftercomma,#1, \@end, }
\edef\@tempa{\expandafter\@ignendcommas\@tempa\@end}%
\if@filesw \immediate \write \@auxout {\string \citation {\@tempa}}\fi
\@tempcntb\m@ne \let\@h@ld\relax \def\@citea{}%
\@for \@citeb:=\@tempa\do {\@cmpresscites}%
\@h@ld}
\def\@ignspaftercomma#1, {\ifx\@end#1\@empty\else
   #1,\expandafter\@ignspaftercomma\fi}
\def\@ignendcommas,#1,\@end{#1}
\def\@cmpresscites{%
 \expandafter\let \expandafter\@B@citeB \csname b@\@citeb \endcsname
 \ifx\@B@citeB\relax 
    \@h@ld\@citea\@tempcntb\m@ne{\bf ?}%
    \@warning {Citation `\@citeb ' on page \thepage \space undefined}%
 \else
    \@tempcnta\@tempcntb \advance\@tempcnta\@ne
    \setbox\z@\hbox\bgroup 
    \ifnum0<0\@B@citeB \relax
       \egroup \@tempcntb\@B@citeB \relax
       \else \egroup \@tempcntb\m@ne \fi
    \ifnum\@tempcnta=\@tempcntb 
       \ifx\@h@ld\relax 
          \edef \@h@ld{\@citea\@B@citeB }%
       \else 
          \edef\@h@ld{\hbox{--}\penalty\@highpenalty
            \@B@citeB }%
       \fi
    \else   
       \@h@ld\@citea\@B@citeB
       \let\@h@ld\relax
 \fi\fi%
 \def\@citea{,\penalty\@highpenalty\hskip.13em plus.1em minus.1em}%
}
\def\@citex[#1]#2{\@cite{\citen{#2}}{#1}}%
\def\@cite#1#2{\leavevmode\unskip
  \ifnum\lastpenalty=\z@\penalty\@highpenalty\fi
  \ [{\multiply\@highpenalty 3 #1
      \if@tempswa,\penalty\@highpenalty\ #2\fi 
    }]\spacefactor\@m}
\makeatother

\def\dalemb#1#2{{\vbox{\hrule height .#2pt
        \hbox{\vrule width.#2pt height#1pt \kern#1pt
                \vrule width.#2pt}
        \hrule height.#2pt}}}
\def\square{\mathord{\dalemb{6.8}{7}\hbox{\hskip1pt}}}

\let\a=\alpha \let\b=\beta   \let\e=\epsilon
    
 \let\m=\mu

\def\nn{\nonumber} \def\bd{\begin{document}} \def\ed{\end{document}}
\def\ds{\documentstyle} \let\fr=\frac \let\bl=\bigl \let\br=\bigr
\let\Br=\Bigr \let\Bl=\Bigl 
\let\bm=\bibitem
\let\na=\nabla
\def\tU{{\widetilde U}}
\let\pa=\partial \let\ov=\overline
\def\ie{{\it i.e.\ }} 
\newcommand{\be}{\begin{equation}} 
\newcommand{\ee}{\end{equation}} 
\def\ba{\begin{array}}
\def\ea{\end{array}}
\def\ft#1#2{{\textstyle{{\scriptstyle #1}\over {\scriptstyle #2}}}}
\def\fft#1#2{{#1 \over #2}}
\def\F#1#2{{ F_{#1}^{(#2)} }}
\def\cF#1#2{{ {\cal F}_{#1}^{(#2)} }}
\def\del{\partial}
\def\R{\rlap{\rm I}\mkern3mu{\rm R}}
\def\sst#1{{\scriptscriptstyle #1}}
\def\oneone{\rlap 1\mkern4mu{\rm l}}
\def\e7{E_{7(+7)}}
\def\td{\tilde}
\def\wtd{\widetilde}
\def\im{{\rm i}}
\def\bog{Bogomol'nyi\ }
\def\sw{{\sst W}}
\def\st{{\sst T}}
\newcommand{\ho}[1]{$\, ^{#1}$}
\newcommand{\hoch}[1]{$\, ^{#1}$}
\newcommand{\bea}{\begin{eqnarray}} 
\newcommand{\eea}{\end{eqnarray}} 
\newcommand{\ra}{\rightarrow}
\newcommand{\lra}{\longrightarrow}
\newcommand{\Lra}{\Leftrightarrow}
\newcommand{\ap}{\alpha^\prime}
\newcommand{\bp}{\tilde \beta^\prime}
\newcommand{\tr}{{\rm tr} }
\newcommand{\Tr}{{\rm Tr} } 
\newcommand{\NP}{Nucl. Phys. }
\newcommand{\tamphys}{\it Center for Theoretical Physics,
Texas A\&M University, College Station, Texas 77843}
\newcommand{\ens}{\it Laboratoire de Physique Th\'eorique de l'\'Ecole
Normale Sup\'erieure\hoch{3}\\
24 Rue Lhomond - 75231 Paris CEDEX 05}
\newcommand{\sissa}{\it SISSA, Via Beirut No. 2-4, 34013 Trieste,
Italy}

\newcommand{\nanchang}{\it 
Institute of Modern Physics, Nanchang University, Nanchang, China}

\newcommand{\auth}{H.  L\"u\hoch{\dagger},  
C.N. Pope\hoch{\ddagger\sharp1}, 
T.A. Tran\hoch{\ddagger} and K.-W. Xu\hoch{\star2}}

\begin{document}
\begin{flushright}
\hfill{CTP TAMU-32/97}\\
\hfill{LPTENS-97/37}\\
\hfill{SISSARef. 98/97/EP}\\
\hfill{hep-th/9708055}\\
\hfill{August, 1997}\\
\end{flushright}

\vspace{-10pt}

\begin{center}
{\large {\bf Classification of $p$-branes, NUTs, Waves and Intersections}}

\vspace{20pt}
\auth

\vspace{15pt}

{\hoch{\dagger}\ens}

\vspace{10pt}
{\hoch{\ddagger}\tamphys}

\vspace{10pt}
{\hoch{\sharp}\sissa}

\vspace{10pt}
{\hoch{\star}\nanchang}

\vspace{30pt}
\underline{ABSTRACT}
\end{center}

          We give a full classification of the multi-charge
supersymmetric $p$-brane solutions in the massless and massive maximal
supergravities in dimensions $D\ge2$ obtained from the toroidal
reduction of eleven-dimensional supergravity.  We derive simple
universal rules for determining the fractions of supersymmetry that
they preserve.  By reversing the steps of dimensional reduction, the
$p$-brane solutions become intersections of $p$-branes, NUTs and waves
in $D=10$ or $D=11$.  Having classified the lower-dimensional
$p$-branes, this provides a classification of all the intersections in
$D=10$ and $D=11$ where the harmonic functions depend on the space
transverse to all the individual objects.  We also discuss the
structure of U-duality multiplets of $p$-brane solutions, and show how
these translate into multiplets of harmonic and non-harmonic
intersections.

{\vfill\leftline{}\vfill
\vskip	10pt
\footnoterule
{\footnotesize	\hoch{1} Research supported in part by DOE 
Grant DE-FG03-95ER40917 and the \vskip	-12pt} \vskip 10pt
{\footnotesize \hoch{\phantom{1}} EC Human Capital and Mobility
Programme under contract ERBCHBGT920176. \vskip -12pt} \vskip 14pt
{\footnotesize \hoch{2} Research supported in part by the Chinese
National Science Foundation. \vskip -12pt} \vskip 14pt
{\footnotesize
        \hoch{3} Unit\'e Propre du Centre National de la Recherche
Scientifique, associ\'ee \`a l'\'Ecole Normale Sup\'erieure \vskip -12pt}
                       \vskip 10pt
{\footnotesize \hoch{\phantom{3}} et \`a l'Universit\'e de Paris-Sud 
\vskip -12pt}} 

\pagebreak
\setcounter{page}{1}

\section{Introduction}

    The BPS-saturated soutions in supergravity theories play an
important r\^ole in M-theory or string theory, since it is believed
that they will survive at the full quantum level, and they can therefore
provide information about the non-perturbative structure.  The
easiest way to find such BPS solutions is by looking for extremal
$p$-brane solitons, either in the $D=11$ or $D=10$ supergravities
themselves, or in their Kaluza-Klein reductions to lower dimensions.
Since the Kaluza-Klein reduction procedure itself preserves all of the
original supersymmetry, in the case of toroidal compactifications, it
follows that the lower-dimensional $p$-brane solitons can be
re-interpreted back in the higher dimension as supergravity solutions
that preserve the same fractions of supersymmetry as they do in the lower
dimension.  The simplest $p$-brane solitons preserve one half of the
original supersymmetry.  These may be characterised as solutions that
can be supported by a single charge, carried by a single field
strength in the supergravity theory.  The metric functions, and all
the other non-vanishing fields in the solution, are expressed in terms
of a single function that is harmonic in the space transverse to the
$p$-brane's world-volume.  There also exist families of related
solutions, obtained by acting with U-duality \cite{ht} transformations,
which are the discretised Cremmer-Julia (CJ) \cite{cj1,cj2}
global symmetry groups of the lower-dimensional supergravities.  These
families are in general more complicated, with more than one field
strength becoming active.  However, since U-duality commutes with
supersymmetry, they will still preserve the same one half fraction of
the supersymmetry.  At the classical level, the U-duality transformed
solutions are effectively just more complicated presentations of the
same single-field-strength solutions, and so it is useful to introduce
the notion of a ``simple'' single-charge $p$-brane as one where a
single field strength carries the charge.  In general, the
higher-dimensional solution that one arrives at by taking such a
lower-dimensional $p$-brane, and retracing the steps of Kaluza-Klein
reduction, or ``oxidising'' the solution, will not necessarily have the form
of a $p$-brane soliton.  It may instead be like a continuum of
$p$-branes distributed over a hypersurface in the compactifying space,
or else like a gravitational wave or a Taub-NUT-like monopole
solution.  Nonetheless, the oxidised solution will still preserve one
half of the supersymmetry, and so it is still a configuration that
should enjoy a preferred status in the full quantum theory.

     There also exist more complicated $p$-brane solitons in the
lower-dimensional theories, which carry more than one kind of charge.
In its simplest form, such an $N$-charge solution is characterised by
$N$ independent harmonic functions on the transverse space, one for
each of the charges.  The solution is ``diagonalised,'' in the sense
that each charge is associated with a particular field strength, and
each of these is expressed purely in terms of the harmonic function
for its charge.  Again, more complicated solutions can be found, by
acting with U-duality transformations.  It is therefore useful again
to introduce the notion of ``simple'' $N$-charge $p$-branes, meaning
the ones that are in the diagonal form described above.  The fractions
of supersymmetry that are preserved by the simple $N$-charge
$p$-branes are smaller than for the single-charge cases.  Those with
$N=2$ preserve $\ft14$, those with $N=3$ preserve $\ft18$, and the
story becomes more involved for those with $N\ge4$.  If multi-charge
$p$-branes are oxidised back to $D=10$ or $D=11$, they describe more
complicated kinds of configurations than do the oxidations of the
single-charge examples discussed above.  They can, however, in general
be interpreted as intersections \cite{pt,tsey1} of the basic
$p$-branes, waves and NUTs mentioned previously.  Again, of course,
the intersecting solutions preserve the same fractions of
supersymmetry as do their lower-dimensional $p$-brane reductions.

    Without the guidance of the lower-dimensional $p$-branes, it would
not be easy to obtain an orderly understanding and characterisation of
supersymmetric solutions in $D=10$ or $D=11$.  In particular, in the
lower dimensions the global $E_{11-D}$ symmetries of the maximal
massless supergravities can be used to generate families of
supersymmetric solutions from the simple multi-charge $p$-branes.
Such symmetries are non-manifest in $D=10$ or $D=11$, and it would,
for example, be no easy task to recognise families of
eleven-dimensional solutions that are related, from the
four-dimensional viewpoint, by $E_7$ U-duality transformations of
four-dimensional multi-charge $p$-brane solutions.  The best way to
benefit from the organising power of the U-duality symmetries is to
study the higher-dimensional supersymmetric solutions from the
lower-dimensional viewpoint.  Large classes of these lower-dimensional 
solutions are provided by the $p$-branes that we discussed above, and it
is these that will provide the focus of our attention in this paper.

    We shall present a full classification of all the $p$-brane
solutions in all the maximal (massless and massive) supergravities in 
dimensions $2\le D \le 11$ that are obtained by dimensional reduction of
eleven-dimensional supergravity.  We discuss their oxidations to $D=10$
and $D=11$, and present simple rules for obtaining the higher-dimensional
solutions.  We also give a detailed discussion of the supersymmetry of
all the solutions, and derive simple rules for determining the fractions
of preserved supersymmetry.  

    We begin in section 2 by giving the bosonic Lagrangians for the
dimensionally-reduced massless maximal supergravities in $D\ge3$, and
summarising the form of the simple multi-charge $p$-brane solutions.
In section 3, we show how these solutions can be oxidised back to
$D=10$ or $D=11$.  In fact the end-product of the oxidation of a
multi-charge solution is easily obtained by the mechanical application
of elementary rules.  In section 4, we give a complete classification
of all the simple multi-charge $p$-branes in all the massless and
massive supergravities in $D\ge3$.  This is extended in section 5 with
a derivation of the bosonic sectors of the maximal two-dimensional
massless and massive supergravities, and a classification of their
multi-charge $p$-brane solutions.  Simple $N$-charge solutions with
different values of $N$ belong to different U-duality multiplets.  The
classification of simple multi-charge solutions hence subsumes the
classification of different U-duality $p$-brane multiplets.  In
section 6 we give an analysis of the supersymmetry of all the
solutions in $2\le D\le11$.  Included in this discussion is a detailed
study of how the successive introduction of additional charges affects
the supersymmetry.  In general, when a new charge is added in an
existing $N$-charge solution to give a simple $(N+1)$-charge solution,
it can have the following effects on the preserved supersymmetry.  One
possibility is that the new charge does not further break the
supersymmetry.  In this case, the same charge but with the opposite
sign will break all the supersymmetry.  The one remaining possible
effect is that the introduction of the charge leads to a halving of
the supersymmetry of the $N$-charge solution. In this case, the sign
of the new charge is immaterial.  Note that in all cases the new
solution is still extremal, and there is no force between the charges.
In section 7, we summarise the results of the classification of
harmonic intersections in M-theory and string theory, corresponding to
the oxidations of the simple multi-charge $p$-branes obtained earlier
in the paper.  In section 8, we extend the discussion of $p$-brane
solutions to include the multiplets that are filled out by acting with
U-duality on the simple solutions.  In particular, we consider
examples in $D=9$ and $D=8$ that are related to simple solutions by
means of $SL(2,\R)$ transformations.  We obtain their oxidations in
$D=10$ and $D=11$, and show that they can be viewed as non-harmonic
intersections of the basic $p$-branes, waves and NUTs.  We also
discuss solutions for all possible pairs of field strengths, showing
that there are three categories.  We end our paper with a conclusion
in section 9.  In an appendix, we list all the field configurations
for all the simple $(N\ge3)$ charge solutions using 1-form field
strengths.

\section{Review of maximal supergravities and $p$-branes}

\subsection{Maximal supergravities in $D\le 11$}

     In this section, we present the toroidal dimensional reductions of
the bosonic sector of $D=11$ supergravity, whose Lagrangian takes the
form \cite{cjs}
\be
{\cal L} = \hat e \hat R -\ft1{48} \hat e\, \hat F_4^2 +\ft16 \ast(\hat 
F_4\wedge\hat F_4\wedge \hat A_3)
\ .\label{d11lag}
\ee
The subscripts on the potential $A_3$ and its
field strength $F_4=dA_3$ indicate the degrees of the differential forms.
We may reduce the theory to $D$ dimensions in a succession of 1-step
compactifications on circles.  At each stage in the reduction, say from
$(D+1)$ to $D$ dimensions, the metric is reduced according to the standard
Kaluza-Klein prescription
\be
ds_{\sst D+1}^2 = e^{2\a\varphi} \, ds_{\sst D}^2 + e^{-2(D-2)\a\varphi}\, 
(dz+{\cal A}_1)^2\ ,\label{metred}
\ee
where the $D$ dimensional metric, the Kaluza-Klein vector potential ${\cal
A}_1={\cal A}_{\sst M}\, dx^{\sst M}$ and the dilatonic scalar $\varphi$
are taken to be independent of the additional coordinate $z$ on the
compactifying circle.  The constant $\a$ is given by
$\a^{-2} = 2(D-1)(D-2)$, and the parameterisation of the metric is such
that a pure Einstein action is reduced again to a pure Einstein action.
(This choice is possible for the descent down as far as $D=3$, but when
$D=2$ it is no longer possible to choose an Einstein-frame parametrisation
of the metric.  We shall discuss this special case in section 5.)
Gauge potentials are reduced according to the prescription $A_n(x,z)=
A_n(x) + A_{n-1}(x)\wedge dz$, implying that
\be
F_n\longrightarrow dA_{n-1} + dA_{n-2}\wedge dz = dA_{n-1} -
dA_{n-2}\wedge {\cal A}_1 + dA_{n-2}\wedge (dz+{\cal A}_1)\ .\label{cs}
\ee
Thus while the dimensionally-reduced field strength $F_{n-1}$ is
defined by $F_{n-1}=dA_{n-2}$, the reduction of $F_n$ is defined by
$F_n=dA_{n-1} - dA_{n-2}\wedge {\cal A}_1$, and it is this
gauge-invariant field strength that appears in the lower-dimensional
gauge-field kinetic term.  These Kaluza-Klein modifications to the
lower-dimensional field strengths become progressively more
complicated as the descent through the dimensions continues.  Their
presence significantly restricts the possible solutions for $p$-branes
and intersecting $p$-branes, as we shall see later.

     It is easy to see that the
original eleven-dimensional fields $g_{\sst{MN}}$ and $A_{\sst{MNP}}$ in 
(\ref{d11lag}) will give rise to the following fields in $D$ dimensions:
\bea
g_{\sst{MN}} &\longrightarrow & g_{\sst{MN}}\ ,\qquad \vec\phi\ ,\qquad 
{\cal A}_1^{(i)}\ ,\qquad {\cal A}_0^{(ij)} \ ,\nn\\
A_3 &\longrightarrow & A_3\ ,\qquad A_2^{(i)}\ , \qquad A_1^{(ij)}\ ,
\qquad A_0^{(ijk)}\ ,\label{dfields}
\eea
where the indices $i, j, k=1,\ldots, 11-D$ run over the $11-D$
internal toroidally-compactified dimensions, starting from $i=1$ for the
step from $D=11$ to $D=10$.  The potentials $A_1^{(ij)}$ and 
$A_0^{(ijk)}$ are automatically antisymmetric in their internal indices,
whereas the 0-form potentials ${\cal A}_0^{(ij)}$ that come from the
subsequent dimensional reductions of the Kaluza-Klein vector potentials
${\cal A}_1^{(i)}$ are defined only for $j>i$.  The quantity $\vec\phi$
denotes the $(11-D)$-vector of dilatonic scalar fields coming from the
diagonal components of the internal metric.
The Lagrangian for the bosonic $D$-dimensional toroidal
compactification of eleven-dimensional supergravity then takes the form
\cite{lpsol}
\bea
{\cal L} &=& eR -\ft12 e\, (\del\vec\phi)^2 -\ft1{48}e\, e^{\vec a\cdot 
\vec\phi}\, F_4^2 -\ft{1}{12} e\sum_i 
e^{\vec a_i\cdot \vec\phi}\, (F_3^{(i)})^2
-\ft14 e\, \sum_{i<j} e^{\vec a_{ij}\cdot \vec\phi}\, (F_2^{(ij)})^2
\label{dgenlag}\\
&& -\ft14e\, \sum_i e^{\vec b_i\cdot \vec\phi}\, ({\cal F}_2^{(i)})^2
-\ft12 e\, \sum_{i<j<k} e^{\vec a_{ijk} \cdot\vec \phi}\,
(F_1^{(ijk)})^2 -\ft12e\, \sum_{i<j} e^{\vec b_{ij}\cdot \vec\phi}\,
({\cal F}_1^{(ij)})^2 + {\cal L}_{\sst{FFA}}\ ,\nn
\eea
where the ``dilaton vectors'' $\vec a$, $\vec a_i$, $\vec a_{ij}$, 
$\vec a_{ijk}$,
$\vec b_i$, $\vec b_{ij}$ are constants that characterise the couplings of
the dilatonic scalars $\vec \phi$ to the various gauge fields.  
They are given by \cite{lpsol}
\bea
&&F_{\sst{MNPQ}}\qquad\qquad\qquad\qquad\qquad\qquad\qquad\qquad
{\rm vielbein}\nonumber\\
{\rm 4-form:}&&\vec a = -\vec g\ ,\nonumber\\
{\rm 3-forms:}&&\vec a_i = \vec f_i -\vec g \ ,\nonumber\\
{\rm 2-forms:}&& \vec a_{ij} = \vec f_i + \vec f_j - \vec g\ ,
\qquad\qquad\qquad\qquad\qquad \,\,\, \,\vec b_i = -\vec f_i \ ,
\label{dilatonvec}\\
{\rm 1-forms:}&&\vec a_{ijk} = \vec f_i + \vec f_j + \vec f_k -\vec g
\ ,\qquad\qquad\qquad\qquad\vec b_{ij} = -\vec f_i + \vec f_j\ ,\nonumber \\
{\rm 0-forms:}&& \vec a_{ijk\ell} =\vec f_i +\vec f_j+\vec f_k +\vec f_\ell
-\vec g \ ,\qquad\qquad\quad\ \  \vec b_{ijk}=-\vec f_i +\vec f_j +\vec f_k\ .
\nonumber
\eea
The explicit expressions for the vectors $\vec g$ and $\vec f_i$, which 
have $(11-D)$ components in $D$ dimensions, are given in \cite{lpsol}.  For
our purposes, it is sufficient to note that they satisfy the relations
\be
\vec g \cdot \vec g = \ft{2(11-D)}{D-2}, \qquad
\vec g \cdot \vec f_i = \ft{6}{D-2}\ ,\qquad
\vec f_i \cdot \vec f_j = 2\delta_{ij} + \ft2{D-2}\ .\label{gfdot}
\ee
We have also included the dilaton vectors $\vec a_{ijk\ell}$ and $\vec
b_{ijk}$ for ``0-form field strengths'' in (\ref{dilatonvec}), although they
do not appear in (\ref{dgenlag}), because they fit into the same general
pattern and they will arise later when we consider $(D-2)$-brane solutions
(\ie domain walls) \cite{clpst} in section 4.2. 
The field strengths are associated with the gauge potentials in the
obvious way; for example $F_4$ is the field strength for $A_3$, $F_3^{(i)}$
is the field strength for $A_2^{(i)}$, {\it etc}.  The complete expressions
for the Kaluza-Klein modifications to the various field strengths are
given in \cite{lpsol}, as are the cubic Wess-Zumino terms ${\cal L}_{FFA}$ 
coming from the $F_4\wedge F_4\wedge A_3$ term in the eleven-dimensional 
Lagrangian (\ref{d11lag}).

     In the subsequent sections, we shall be making extensive use of the
results presented here, in order to discuss various aspects of $p$-brane
solitons in toroidally-compactified M-theory and type II strings.

\subsection{Review of supersymmetric $p$-branes}

        In this subsection, we review the form of $p$-brane soliton
solutions in maximal supergravities in all the dimensions $3\le D \le
11$; we shall derive the analogous results in $D=2$ later in section
5.  These various $p$-brane solutions preserve certain fractions of the
supersymmetry, owing to which it is believed that they also will also
be contained in the spectrum of type II strings or M-theory.  As is
well known, massless maximal supergravities in $D$ dimensions have
$E_{11-D}$ global symmetries \cite{cj1,cj2}, which provide a powerful
organising principle for the solitonic solutions.  However, not all
the solutions of a given dimension form one single multiplet under
$E_{11-D}$.  For example, solutions that preserve different fractions
of the supersymmetry must clearly belong to different multiplets,
since the global symmetry commutes with supersymmetry.

     It is non-trivial to solve the general equations of motion
following from the $D$-dimensional supergravity Lagrangians presented
in the previous subsection.  Moreover, one wishes to avoid laboriously
solving for solutions that are nothing but U-duality transformations
of already obtained solutions. Thus we shall simplify the problem by
starting with truncated Lagrangians that contain just the fields that
will play a r\^ole in the construction of the particular solitonic
$p$-branes under consideration.  The $p$-brane solutions of these
truncated Lagrangians that we shall construct will also be solutions
of the original theory.  (Of course not all solutions of the truncated
Lagrangian will be solutions of the original one, since the truncation
of the theory itself is not in general a consistent one, and so there
is a non-trivial check to verify that a $p$-brane solution of the
truncated system is indeed also a solution of the original one; we
shall discuss this later.)  The truncated Lagrangians are of the
following form, comprising dilatonic scalar fields and $n$-index
antisymmetric tensor field strengths \cite{lpmult}:
\be
{\cal L} = eR -\ft12 e(\del \vec \phi)^2 -\fft{1}{2n!}e
\sum_{\a=1}^{N} e^{\vec c_\a\cdot \vec\phi} (F_n^\a)^2\ ,\label{nlag}
\ee 
where $F_n^\a=dA_{n-1}^\a$.  In writing the truncated Lagrangian in
this form, it is understood that the field strengths $F_n^\a$ are
taken from some subset of the field strengths appearing in
(\ref{dgenlag}), possibly with dualisations.  When a particular field
strength $F_n$ in (\ref{nlag}) is exactly the same as a field strength
appearing in (\ref{dgenlag}), the corresponding dilaton vector is
given by (\ref{dilatonvec}). However, a field strength might be
related by dualisation to one of the original fields in
(\ref{dgenlag}).  In such a case, an original field $F_{D-n}$ with
kinetic term $e^{\vec c\cdot\vec\phi}\, F_{D-n}^2$ in (\ref{dgenlag})
would be represented by a new field $F_n = e^{\vec c\cdot\vec\phi}\,
*F_{D-n}$, where $*$ is the Hodge dual, with kinetic term $e^{-\vec
c\cdot\vec\phi}\, F_n^2$ in (\ref{nlag}).  Thus the dilaton vector for
this field $F_n$ in (\ref{nlag}) will be of the opposite sign to the
dilaton vector of the original field $F_{D-n}$, in (\ref{dgenlag}).
Later, in sections 4 and 5, we shall still use the original fields to label
the sets of field configurations for multi-charge $p$-brane solutions.
We shall use a ``$*$'' to indicate that dualisation of that particular
field is to be performed in obtaining the associated truncated
Lagrangian.  This implies that the dilaton vector associated with the
starred field strength has the opposite sign to the one given in
({\ref{dilatonvec}) for the original field.

     We shall show later that the $p$-brane solutions
for this truncated system, where in particular there are no Wess-Zumino terms
or Kaluza-Klein modifications to the field strengths, are also solutions of
the original system described by (\ref{dgenlag}) provided that the dilaton
vectors $\vec c_\a$ satisfy the dot product relations \cite{lpmult}
\be
M_{\a\beta} = \vec c_\a \cdot \vec c_\beta = 4 \delta_{\a\beta}
-\fft{2(n-1)(D-n-1)}{D-2}\ .\label{mdot}
\ee
We can then obtain
electric $(n-2)$-branes or magnetic $(D-n-2)$-branes, with metrics
\be
ds^2 = \Big(\prod_{\a=1}^{N} H_\a\Big)^{-\ft{\td d}{D-2}}\,   
dx^\mu dx^\nu \eta_{\mu\nu} + \Big(\prod_{\a=1}^N H_\a\Big)^{\ft{d}{D-2}}  
\, dy^m dy^m\ ,\label{dhsol}
\ee
where the $H_\a$ are harmonic functions depending on the coordinates
$y^m$ of the $(D-d)$-dimensional space transverse to the
$d$-dimensional world-volume, and $\td d=D-d-2$.  Note that one can
only find solutions for the truncated Lagrangian (\ref{nlag})
involving $N$ independent harmonic functions if the dilaton vectors
satisfy the conditions (\ref{mdot}) \cite{lpmult}.  The $p$-branes
are supported by field strengths $F_n^\a$ that all carry either electric or
magnetic charges:
\bea
{\rm electric}:&& F_n^\a = d^dx \wedge dH_\a^{-1}\ ,\nonumber\\
{\rm magnetic}:&& F_n^\a = e^{-\vec c_\a\cdot\vec\phi}*(d^dx\wedge
dH_\a^{-1})\ .\label{fans}
\eea
The dilatonic scalars $\vec\phi$ are given by
\be
\vec\phi = \ft12 \epsilon \, \sum_\a \vec c_\a\, \log H_\a\ ,\label{dilsol}
\ee
where $\epsilon=1$ for electric solutions and $\epsilon=-1$ for
magnetic solutions.  Although in terms of the field strengths $F^\a_n$
appearing in the truncated Lagrangian (\ref{nlag}) the solutions are
either purely electric or purely magnetic, in terms of the original
fields in (\ref{dgenlag}) the solutions will carry both electric and
magnetic charges if dualisations of the kind we discussed below
(\ref{nlag}) have been performed.  (This can be seen from the fact
that an electric $p$-brane solution supported by $F_n$ is identical to
the magnetic $p$-brane solution supported by its original field
$F_{D-n}$, and {\it vice versa} \cite{lpss1}.)  These may be called dyonic
solutions of the second kind \cite{lpsol}, describing the situation where
each individual field strength carries purely an electric or purely a
magnetic charge.  There are also dyonic solutions of the first kind,
where a given field strength carries both electric and magnetic
charges.  In fact, the solutions discussed above cover all possible
simple multi-charge $p$-brane solutions that can arise from
supergravities, with the exception of dyonic solutions of the first
kind.  However, only one such solution arises that is supersymmetric,
namely the dyonic string in $D=6$ \cite{dfkr}, whose structure is well
understood.
   
      The solutions given above range from ($-1$)-branes (instantons) to
$(D-2)$-branes (domain walls).  For an isotropic $p$-brane, the
harmonic functions are given by $H_\a=1 + |Q_\a|\, r^{-\td d}$ where
$r=\sqrt{y^my^m}$, and the ADM mass per unit $p$-volume is $M=\sum_\a
|Q_a|$, where $Q_\a$ are the charges carried by the field strengths
$F_n^\a$.  These formulae assume that the dilatonic scalars vanish
asymptotically at infinity.  If instead they approach the constant
values $\vec\phi_0$ asymptotically, then we will have
\be
H_\a=1+\fft{|Q_\a|\, e^{-\ft12 \epsilon\, \vec c_\a\cdot 
\vec \phi_0}}{r^{\td d}}
\ ,\qquad M=\sum_\a^N |Q_\a|\, e^{-\ft12 \epsilon \vec c_\a\cdot 
\vec\phi_0}\ .
\ee
Owing to the quadratic nature of field strength kinetic terms in the
Lagrangian, for each charge there is a sign choice to be made, which
determines whether it contributes positively or negatively to the
mass.  We always make the choice where it contributes positively, so
that the solution is free from naked singularities.   Note that a
$p$-brane with positive mass but the opposite charge can be viewed as
an anti-$p$-brane.

        When all the $N$ charges $Q_\a$ are equal (we consider the case
where $\vec\phi_0=0$ for simplicity), the harmonic functions $H_\a$ in
(\ref{dhsol}) become equal.  Under these circumstances, it is easy to
see from (\ref{dilsol}) that all except one combination of the dilatonic
scalars, namely $\vec\phi= N^{-1}\,\sum_\a \vec c_\a\, \phi$,
 will become zero, and at the same time all the participating field
strengths will become equal, $F^\a= F/\sqrt{N}$.  
The resulting single-scalar configuration is 
a solution of the truncated Lagrangian
\be
{\cal L} = e
R -\ft12 e\, (\del\phi)^2 -\fft1{2n!} e\, e^{a\phi} F^2 \ ,\label{slag}
\ee
and is given by \cite{lpss1}
\bea
ds^2 &=& H^{-\ft{4\td d}{\Delta(D-2)}}\, dx^\m\, dx^\nu\, \eta_{\mu\nu} +
H^{\ft{4 d}{\Delta(D-2)}}\,(dr^2 + r^2 \, d\Omega^2)\ ,\nn\\
e^{\ft{\epsilon\Delta}{2a}\phi}&=& H\ ,\label{ssol}
\eea
where $\Delta=4/N$,  and
\be
a^2=(\vec c_\a)^2 =\Delta -\fft{2d\tilde d} {D-2}\ .\label{avalue}
\ee
In maximal supergravities (massless or massive), single-charge
solutions all have $\Delta=4$.  In fact the $\Delta=4/N$ solutions
(\ref{ssol}) can be viewed as bound states of $\Delta=4$ solutions,
with zero-binding energy \cite{r,kklp,drbound}.  To construct
$N$-charge solutions using $\Delta=4$ building blocks, each associated
with a harmonic function, the dilaton vector dot products (\ref{mdot})
must be satisfied. Extremal $p$-brane solutions (\ref{ssol}) in
various supergravities in different dimensions were constructed in the
past \cite{dghr,str,dust,dl1,chs,guv,k1,cy,dklreview,lpss1,lpsol}.

     The various $p$-brane solutions obtained above have distinct
behaviours with respect to supersymmetry, which are best characterised
by looking at the eigenvalues of the \bog matrix, \ie the
anticommutator of the $D=11$ supercharges, which we shall discuss in
detail in section 6.  In particular, if there are $k$ zero
eigenvalues, then the solution preserves a fraction $k/32$ of the
original $D=11$ supersymmetry.  The non-zero eigenvalues provide
additional information that characterises the solutions.  This is
because the eigenvalues are invariant under U-duality, and so this
provides a way of recognising families of $p$-branes that lie in
different U-duality multiplets.  In particular, this leads to the
conclusion that solutions for different values of $N$ belong to
different U-duality multiplets.  Acting with U-duality on these simple
solutions, we can fill out complete U-duality multiplets.  The
U-duality transformations of the simple solutions will always give
solutions involving $N'$ field strengths with $N'\ge N$.  If $N'=N$,
then the new solution also satisfies the equations of motion following
from a truncated Lagrangian of the form (\ref{nlag}) (but with a
different set of field strengths retained in the truncation).  In fact
these sets of solutions with $N'=N$ form multiplets under the Weyl
subgroup of the U-duality group \cite{lpsweyl}. On the other hand if
$N'>N$, then the solutions will be of a more complicated form, where
the contributions from the Kaluza-Klein modifications and Wess-Zumino
terms cannot be ignored.  The cases where $N'=N$ are characterised by
the fact that the number of non-zero charges is equal to the number of
independent harmonic functions in the solution.  We refer to these as
the ``simple'' multi-charge solutions.  Thus the classification of
simple multi-charge solutions subsumes the classification of the
different U-duality $p$-brane multiplets.  Some related discussions of
solution multiplets and their supersymmetry have been given in
\cite{aafft}.  We shall for now concentrate on the Weyl-group
multiplets of simple multi-charge solutions; these are the ones that
are directly associated with the harmonic intersections of $p$-branes.
We return to the discussion of general U-duality multiplets in section
8.

\section{Oxidation rules for $p$-branes}

     The classification of supersymmetric $p$-branes in all 
dimensions $D\le11$ is an important problem in its own right, since these
BPS-saturated solutions are expected to describe the perturbative and
non-perturbative states of the fully quantised string theories.  In addition,
by reversing the Kaluza-Klein reduction procedure and ``oxidising'' them 
back to $D=10$ or $D=11$, they provide a convenient classification of 
BPS-saturated solutions in the original higher-dimensional theories.  In
special cases, the oxidation of an isotropic  single-charge $p$-brane in 
$D$ dimensions will again give rise to an isotropic single-charge solution
in $D=10$ or $D=11$.  In other cases, the end product of the oxidation 
can be a line, or more generally a hyperplane, of $p$-branes.  More 
complicated possibilities also arise when the charge of the single-charge
$p$-brane in $D$ dimensions is carried by a field strength derived from
the higher-dimensional metric in the Kaluza-Klein reduction process.  In
such cases, the end product of the oxidation will be a wave-like solution
or a Taub-NUT like solution, rather than what one would normally regard as
a $p$-brane.  Nevertheless, in all these cases the higher-dimensional end
product preserves the same fraction $\ft12$ of the supersymmetry as does
the $D$-dimensional $p$-brane from which it is derived.  This is because
the Kaluza-Klein reduction procedure itself breaks none of the supersymmetry.
Thus all these higher-dimensional end products deserve to be considered in
their own right, since they will describe quantum-protected states in the
$D=10$ string or $D=11$ M-theory.  

    More complicated situations arise when we begin with multi-charge
solutions in $D$ dimensions.  These will give rise to oxidation end
products that can be described as intersections \cite{pt,tsey1} of the
various $p$-branes, waves and NUTs mentioned above.  The particular
combinations of these basic ingredients that arise in any given
situation are governed by the details of the set of charges in the
lower-dimensional solution.  To be more precise, one can read off the
combination in the final end product by looking first at the
individual sets of end products associated with each individual charge
in the $D$-dimensional $p$-brane solution.  The oxidation end product
of the entire multi-charge solution will then be described in terms of
intersections of these sets of ingredients.  Again, the fraction of
supersymmetry that is preserved by the intersecting solution in $D=10$
or $D=11$ will be the same as the fraction that is preserved by the
multi-charge solution in $D$ dimensions.

\subsection{Intersections in M-theory}

    Let us first consider oxidations
to $D=11$.  All the lower-dimensional single-charge solutions give rise to
one of the following four kinds of solution:
\bea
{\rm Membrane}:\!\!\!&&\!\!\! 
ds_{11}^2 = H^{-\ft23}\, (-dt^2 + d\vec x^2 + 
d\vec z_\sw^2) + H^{\ft13}\, (d\vec y^2 +
d\vec z_\st^2)\ ,\nn\\
\hbox{5-brane}:\!\!\!&&\!\!\! 
ds_{11}^2 = H^{-\ft13}\, (-dt^2 + d\vec x^2 + 
d\vec z_\sw^2) + H^{\ft23}\, (d\vec y^2 +
d\vec z_\st^2)\ ,\label{11s}\\
{\rm Wave}:\!\!\!&&\!\!\! 
ds_{11}^2 = -H^{-1}\, dt^2 + H\, (dz_\sw + (H^{-1}-1)\, dt)^2
+ (d\vec y^2 +d\vec z_\st^2)\ ,\nn\\
{\rm NUT}:\!\!\!&&\!\!\! 
ds_{11}^2 =-dt^2 + d\vec x^2 + 
d\vec z_\sw^2 + H^{-1}\, \omega^2  
+ H\, d\vec y^2\ .\nn
\eea 
The notation here is as follows.  The $(11-D)$ internal coordinates
$z^i$ are divided into two categories, $z^i=(\vec z_\sw, \vec z_\st)$,
namely those that acquire the interpretation of world-volume
coordinates in $D=11$ and those that become transverse space
coordinates.  The harmonic functions depend on the $\vec y$ transverse
coordinates only.  (These were the transverse coordinates in the
original $D$-dimensional $p$-brane solution.) Note that here we have
generalised the concept of world-volume and transverse space
to include waves and NUTs.

     In the first two cases in (\ref{11s}), the world-volume
dimensions are 3 and 6 respectively, with $p$ of the spatial
coordinates being the original ones $\vec x$ of the $D$-dimensional
solution. The membrane and 5-brane solutions arise when the
$D$-dimensional $p$-brane is supported by an electric or magnetic
charge respectively for a field strength originating from $F_4$ in
$D=11$.  The wave solution arises when the $D$-dimensional $p$-brane
is a black hole or instanton carrying an electric charge for a 
Kaluza-Klein vector.
The NUT solution arises when a field strength coming from the metric
carries a magnetic charge.  There are in fact three distinct
subclasses to consider, depending on whether the $p$-brane in $D$
dimensions is a $(D-4)$-brane, a $(D-3)$-brane or a $(D-2)$-brane.
These will be supported by a field strength of degree 2, 1 or 0 coming
from the $D=11$ metric.  (The last case is associated with a more
general kind of Kaluza-Klein reduction which we shall discuss in
section 4.2.)  The 1-form $\omega$ in (\ref{11s}) is given by 
\bea
\hbox{$(D-4)$-brane}:&& \omega = dz_\st + Q\, \cos\theta\, d\varphi\ ,
\quad d\vec y^2= dr^2 + r^2 d\theta^2 + r^2\sin^2\theta\, d\varphi^2 
\ ,\label{d3b1}\\ 
\hbox{$(D-3)$-brane}:&& \omega = dz_\st^1 + Q\, y^1 \,
dz_\st^2 \ , \quad\quad d\vec y^2 = (dy^1)^2 + (dy^2)^2\ ,\label{d3b2}\\ 
\hbox{$(D-2)$-brane}:&& \omega = dz_\st^1 + Q
z_\st^3\, dz_\st^2\ , \quad\quad d\vec y^2 = (dy^1)^2 \ .\label{d3b3}
\eea 
Thus one, two or three internal coordinates respectively acquire
the interpretation of lying in the transverse space, with the
remainder lying in the world-volume.

     Having given the possible forms of end products of the oxidation
of single-charge $p$-branes to $D=11$, it remains to present the rules
that determine the precise end products for each lower-dimensional
single-charge $p$-brane.  We derive these by noting that in a single
step of double dimensional reduction, the degree of the field strength
reduces by 1 if it carries an electric charge, while remaining
unchanged if it instead carries a magnetic charge.  When a step of
double dimensional reduction is reversed, the compactification
coordinate joins the higher-dimensional world-volume.  Conversely, in
a step of vertical reduction, the degree of the field strength is
unchanged if it carries an electric charge, but is reduced by 1 if it
carries a magnetic charge.  Upon reversing the vertical reduction, the
compactification coordinate joins the transverse space.  The complete
results can now be presented in the form of two tables, Table (1a) for
the $p$-branes supported by field strengths derived from $F_4$ in
$D=11$, and Table (1b) for $p$-branes supported by field strengths
coming from the metric in $D=11$.\footnote{The electric solutions
supported by 1-form field strengths are instantons, or
``$(-1)$-branes.''  In these cases, there is no world-volume in the
$D$-dimensional configuration, and all $D$ dimensions are spatial
transverse coordinates.  In the present discussion, this situation
arises when one of the compactification coordinates $z^i$ is actually
the time coordinate of the eleven-dimensional theory, implying that
the lower-dimensional theory is then formulated in a space of
Euclidean signature.  The reduction on the time coordinate
automatically results in the 1-form field strengths appearing with the
opposite sign to the usual one in the $D$-dimensional Lagrangian, a
feature that is in fact necessary in order to describe the instanton
solutions.  Note that in this discussion, there is never any need to
perform any Euclideanisation by hand; the positive-definite metric
signature arises from compactification of the time coordinate.  It is
interesting that the only theory where an explicit Euclideanisation
would be needed in order to be able to construct single-charge
instanton solutions is the ten-dimensional type IIB supergravity.
However, the status of such a Euclideanisation is rendered
problematical by the fact that the self-duality condition on the
5-form field strength requires a Lorentzian signature for the
spacetime.}

\bigskip\bigskip

\centerline{
\begin{tabular}{|c|c|c|c|c|c|c|}\hline
& $F_4$ & $ F_3^i$ & $F_2^{ij}$ & $F_1^{ijk}$& $F_0^{ijk\ell}$ & Endpoint
\\ \hline
Electric $\vec z_\sw=$ & -- &$z^i$ &$z^i, z^j$ & $z^i, z^j, z^k$ &
N/A & Membrane \\ \hline
Magnetic $\vec z_\st=$ & -- &$z^i$ &$z^i, z^j$ & $z^i, z^j, z^k$ &
$z^i, z^j, z^k, z^\ell$& 5-brane \\ \hline
\end{tabular}}
\bigskip

\centerline{Table (1a): Oxidations to M-branes}
\bigskip

     The tables indicates how the internal compactification
coordinates divide between the world-volume and the transverse space
after the $D$-dimensional $p$-brane is oxidised to $D=11$.  Where
world-volume coordinates $\vec z_\sw$ are listed, the remaining
unlisted coordinates $\vec z_\st$ are associated with the transverse
space, and {\it vice versa}.  The indices $i,j,\ldots$ on the internal
coordinates run from 1 to $(11-D)$, starting with $i=1$ for the
reduction from $D=11$ to $D=10$.  The ``0-form field strengths''
$\F0{ijk\ell}$ in Table (1a) and $\cF0{ijk}$ in Table (1b) are like
cosmological terms in the $D$-dimensional Lagrangian, and arise from
generalised Kaluza-Klein reductions, as we shall discuss in section
4.2.

\bigskip\bigskip

\centerline{
\begin{tabular}{|c|c|c|c|c|}\hline
 & ${\cal F}_2^i$ & $\cF{1}{ij}$ & $\cF{0}{ijk}$ & Endpoint \\ \hline
Electric $\vec z_\sw=$ & $z^i$ & $z^i,z^j$ & N/A & Wave \\ \hline
Magnetic $\vec z_\st=$ & $z^i$ & $z^i, z^j$ & $z^i, z^j, z^k$ & 
NUT \\ \hline
\end{tabular}}
\bigskip

\centerline{Table (1b):  Oxidations to M-waves and M-NUTs}
\bigskip

   The asymmetry between the membranes and 5-branes in Table (1a), and
between the waves and NUTs in Table (1b), arises because the electric
solutions supported by $\F0{ijk\ell}$ or $\cF0{ijk}$ would be
$(-2)$-branes, which do not seem to have any meaning.  The metric and
the field strength $F_4$ in $D=11$ can be easily determined by
retracing the steps of the dimensional reduction given in
(\ref{metred}) and (\ref{cs}).

    Having discussed the oxidation of single-charge solutions to
$D=11$, we are now in a position to discuss the multi-charge
solutions.  It is manifest from the form (\ref{dhsol}) and
(\ref{dilsol}) of the $D$-dimensional $N$-charge solution, and the
structure (\ref{metred}) of the Kaluza-Klein reduction of the metric,
that the oxidation to $D=11$ will give a metric where each
non-vanishing metric component will be a product $\prod_\a
H_\a^{m_\a}$ of certain specific
powers $m_\a$ of the $N$ independent harmonic functions $H_\a$.  An easy way
to calculate these powers is by first
considering the oxidations of the $N$ individual single-charge
components, corresponding to all harmonic functions being 1 except for
the one associated with the single charge under consideration.  These 
single-charge oxidations immediately give the exponents $m_\a$ in the
products in the metric components. 

    We shall illustrate the above procedure with a few examples.   First,
consider a dyonic string solution in $D=6$ \cite{dfkr}.  It is easy to
verify that such
2-charge solutions can be constructed provided that the electric charge $Q_e$
and the magnetic charge $Q_m$ are carried by the same 3-form field strength
in $D=6$.  For definiteness, we shall consider the case where it is $\F31$
that carries these charges, \ie
\be
Q_e = \int e^{\vec a_1\cdot\vec\phi} \ast \F31\ , \qquad
Q_m = \int \F31\ ,
\ee
where the dilaton vector $\vec a_1$ is given by (\ref{dilatonvec}).
From (\ref{dhsol}) and (\ref{dilsol}), we see that the dyonic solution in
$D=6$ is given by
\bea
ds_6^2 &=& (H_e \, H_m)^{-\ft12}\, (-dt^2 + dx^2) +  (H_e \, H_m)^{\ft12}\,
d\vec y^2\ ,\nn\\
\vec \phi &=& \ft12 \vec a_1\, \log(H_e/H_m)\ ,\label{dyonic}\\
\F31 &=& d^2x\wedge dH_e^{-1} + e^{-\vec a_1\cdot \vec\phi}\,
\ast(d^2x\wedge dH_m^{-1}) \ ,\nn
\eea
where the harmonic functions are given by $H_e=1+Q_e/y^2$ and $H_m=1+Q_m/y^2$,
and $y^2=\vec y\cdot\vec y$.  

     If $H_m=1$ or $H_e=1$, the solution is purely 
electric or purely magnetic,
and from (\ref{11s}) and the oxidation rules in Table (1a) we see that the
corresponding end products in $D=11$ have metrics $ds_{11}^2(e)$ and
$ds_{11}^2(m)$ given by
\bea
ds_{11}^2(e) &=& H_e^{-\ft23}\, (-dt^2 + dx^2 + dz_1^2) + H_e^{\ft13}\,
(d\vec y^2 + dz_2^2 +\cdots +dz_5^2)\ ,\\
ds_{11}^2(m) &=& H_m^{-\ft13}\, (-dt^2 + dx^2 + dz_2^2+\cdots +dz_5^2) + 
H_m^{\ft23}\, (d\vec y^2 + dz_1^2)\ .
\eea
They respectively describe a membrane distributed uniformly over the 4-plane 
$(z_2,z_3,z_4,z_5)$ and a 5-brane distributed uniformly over the line $z_1$.  
Since there is a unique answer for the products of harmonic
functions for each coordinate direction in the oxidation of the dyon, it must
be that they are simply the products of the $H_e$ and $H_m$ factors for each
coordinate direction in the two limits above.  Thus the dyonic string
oxidises to give
\bea
ds_{11}^2 &=& H_e^{-\ft23}\, H_m^{-\ft13}\, (-dt^2+dx^2) +
 H_e^{\ft13}\, H_m^{\ft23}\, d\vec y^2 \nn\\
&& + H_e^{-\ft23}\, H_m^{\ft23}\, dz_1^2 + H_e^{\ft13}\, H_m^{-\ft13}\,
(dz_2^2 +\cdots + dz_5^2)\ ,\label{25int}
\eea 
which describes a membrane intersecting a 5-brane.

     For another example, consider a 3-charge extremal black hole in $D=5$.
As we shall show in section 4.1, there are two different kinds of 
configuration of charges that can support this solution.  One involves 
electric charges carried by $\{\F2{ij},\F2{k\ell},\F2{mn}\}$, where the
internal indices are all different (there are actually 15 different sub-cases
here, corresponding to all possible choices of index values).  The other
configuration, which we can denote by $\{\F2{ij}, *\F3i, \cF2j\}$, where
$i$ and $j$ are different, has electric charges carried by $\F2{ij}$ and
$\cF2j$, and a magnetic charge carried by $\F3i$ (here, there are 30 sub-cases
corresponding to different choices for the internal indices).  In both cases, 
if the three charges are set equal, the solution reduces to the 
Reissner-Nordstr{\o}m extremal black hole in $D=5$.  The oxidation of the 
first case to $D=11$ is straightforward, and gives three intersecting 
membranes, whose spatial world-volume coordinates are $(z^i,z^j)$, 
$(z^k,z^\ell)$ and $(z^m,z^n)$ respectively.  The second case oxidises to
give an intersection of a membrane, a 5-brane and a wave.  Consider the
example $\{\F2{12}, *\F31, \cF22\}$, with charges denoted by $Q_e$, $Q_m$ and 
$Q_w$ respectively; from the oxidations
of the individual components given in (\ref{11s}), and from Tables (1a) and
(1b), we immediately see that this 3-charge solution oxidises to give
\bea
ds_{11}^2 &=& -H_e^{-\ft23}\, H_m^{-\ft13}\, H_w^{-1}\, dt^2 +
H_e^{-\ft23}\, H_m^{\ft23}\, dz_1^2 + H_e^{\ft13}\, H_m^{\ft23}\,
d\vec y^2\nn\\
&& + H_e^{-\ft23}\, H_m^{-\ft13}\, H_w\, (dz_2 +(H_w^{-1}-1) dt)^2 +
H_e^{\ft13}\, H_m^{-\ft13}\, (dz_3^2+\cdots+dz_6^2)\ .
\eea

    As the above examples illustrate, the procedure of oxidising a
given multi-charge $p$-brane in $D$ dimensions back to intersections
in $D=11$ is a completely straightforward and mechanical one.  These
intersections are of a type where all the harmonic functions depend
on the $\vec y$ coordinates that are transverse to the world-volumes
of all the constituents.\footnote{There are also other kinds of
intersections, where each harmonic function only depends on the
``relative transverse space'' of coordinates transverse only to the
constituent associated with the harmonic
\cite{kr1,bbj,gkt,bdejs,roo,bdejs2}.  Thus in these intersections,
each harmonic function depends on totally non-overlapping subsets of
transverse-space coordinates, and so these solutions do not
dimensionally reduce to $p$-branes.  For related reasons, it is not
clear that any notion of mass or tension can be given for such
configurations.  Thus their relevance as quantum states in string
theory is unclear.    In this paper, unless indicated
otherwise, intersections will be assumed to be of the kind that do
dimensionally reduce to $p$-branes.}  Thus the task of classifying all
such intersections in $D=11$ is subsumed by the task of classifying
all $D$-dimensional multi-charge $p$-branes.  In turn, this latter
classification problem reduces simply to the task of finding all
possible sets of dilaton vectors, defined in (\ref{dilatonvec}), that
satisfy the dot-product relations (\ref{mdot}).  There is only one
remaining subtlety, namely that our discussion so far has been
restricted to reductions down to $D=3$.  Since some intersections in
$D=11$ can only be described in terms of oxidations of multi-charge
solutions in $D=2$, we will only have a complete classification scheme
after having obtained a construction for $p$-branes in $D=2$.  We
shall address this in section 5.  Note that although we have focussed
attention on the simple multi-charge $p$-branes where the number of
non-zero charges is equal to the number of independent harmonic
functions, it is straightforward also to oxidise the more complicated
solutions that are related to the simple ones by U-duality rotations
that lie outside the Weyl group.  We shall discuss this in more detail
in section 8.

\subsection{Intersections in type IIA string theory}
   
     So far, we have considered oxidations of lower-dimensional
$p$-brane solutions to eleven-dimensional M-theory.  It is also of
interest to view the lower-dimensional solutions instead from the
standpoint of ten-dimensional string theory.  For example, we can
categorise lower-dimensional $p$-brane solutions according to whether
they are supported by NS-NS, or R-R, or mixed sets of ten-dimensional
fields.  In particular, $p$-branes carrying R-R charges acquire the
interpretation of being D$p$-branes \cite{pol1}.   While M-theory is
intrinsically non-perturbative, the oxidation of $p$-branes to
ten-dimensional string theory allows us to distinguish between
perturbative and non-perturbative string states.

    Massless type IIA supergravity can be viewed as the first step in
the dimensional reduction of $D=11$ supergravity.  Thus it is
convenient to describe its fields, and their subsequent dimensional
reduction, in the same notation as we used for the reductions of
eleven-dimensional supergravity itself.  From the viewpoint of the
type IIA string, these divide into NS-NS fields $g_{\sst{MN}}$, $\phi$
and $A_2^{(1)}$, and R-R fields $A_3$ and ${\cal A}_1^{(1)}$.  This
separation into NS-NS and R-R fields is preserved under the subsequent
steps of dimensional reduction.  It follows that in $D$ dimensions,
the breakdown of fields into NS-NS and R-R is as follows \cite{lpsweyl}:
\bea
{\rm NS-NS}:&& A_2^{(1)} \quad A_1^{(1\a)} \quad A_0^{(1\a\beta)}
\quad {\cal A}_1^{(\a)} \quad {\cal A}_0^{(\a\beta)}\quad \vec \phi
\quad g_{\mu\nu}\ ,\label{nsns}\\
{\rm R-R}:&& A_3 \quad A_2^{(\a)} \quad A_1^{(\a\beta)} \quad
      A_0^{(\a\beta\gamma)} \quad {\cal A}_1^{(1)} \quad
     {\cal A}_0^{(1\a)}\ ,\label{rr}
\eea
where we have decomposed the internal index $i$ as $i=(1,\a)$.
All the lower-dimensional single-charge $p$-brane solutions, upon oxidation
back to $D=10$, will therefore give rise to one of the following eight
kinds of solution, which we subdivide into four NS-NS and 4 R-R:

\bigskip
\centerline{NS-NS single-charge endpoints:}
{\vskip -20pt}
\bea 
{\rm String}:\!\!\!&&\!\!\! 
ds_{10}^2 = H^{-\ft34}\, (-dt^2 + d\vec x^2 + 
d\vec z_\sw^2) + H^{\ft14}\, (d\vec y^2 +
d\vec z_\st^2)\ ,\nn\\
\hbox{5-brane}:\!\!\!&&\!\!\! 
ds_{10}^2 = H^{-\ft14}\, (-dt^2 + d\vec x^2 + 
d\vec z_\sw^2) + H^{\ft34}\, (d\vec y^2 +
d\vec z_\st^2)\ ,\label{10ns}\\
{\rm Wave}:\!\!\!&&\!\!\! 
ds_{10}^2 = -H^{-1}\, dt^2 + H\, (dz_\sw + (H^{-1}-1)\, dt)^2
+ (d\vec y^2 +d\vec z_\st^2)\ ,\nn\\
{\rm NUT}:\!\!\!&&\!\!\! 
ds_{10}^2 =-dt^2 + d\vec x^2 + 
d\vec z_\sw^2 + H^{-1}\, \omega^2  
+ H\, d\vec y^2\ .\nn
\eea 

\centerline{R-R single-charge endpoints:}
{\vskip -20pt}
\bea 
\hbox{D0-brane}:\!\!&&\!\! 
ds_{10}^2 = H^{-\ft78}\, (-dt^2 + d\vec x^2 + 
d\vec z_\sw^2) + H^{\ft18}\, (d\vec y^2 +
d\vec z_\st^2)\ ,\nn\\
\hbox{D6-brane}:\!\!&&\!\! 
ds_{10}^2 = H^{-\ft18}\, (-dt^2 + d\vec x^2 + 
d\vec z_\sw^2) + H^{\ft78}\, (d\vec y^2 +
d\vec z_\st^2)\ ,\nn\\
\hbox{D2-brane}:\!\!&&\!\! 
ds_{10}^2 = H^{-\ft58}\, (-dt^2 + d\vec x^2 + 
d\vec z_\sw^2) + H^{\ft38}\, (d\vec y^2 +
d\vec z_\st^2)\ ,\label{10rr}\\
\hbox{D4-brane}:\!\!&&\!\! 
ds_{10}^2 = H^{-\ft38}\, (-dt^2 + d\vec x^2 + 
d\vec z_\sw^2) + H^{\ft58}\, (d\vec y^2 +d\vec z_\st^2)\ ,\nn
\eea 
The $(10-D)$ internal coordinates $z^\a$ are divided into world-volume
coordinates $\vec z_\sw$ and transverse coordinates $\vec z_\st$.  The rules
for how each lower-dimensional single-charge solution oxidises to $D=10$
can be summarised in the following tables, in which the division of the
internal coordinates between world-volume and transverse is given.  Table (2a)
gives the rules for NS-NS oxidations, and Table (2b) gives the rules for R-R
oxidations. 

\bigskip\bigskip

\centerline{
\begin{tabular}{|c|c|c|c|c|c|}\hline
& $ \F31$ & $\F2{1\a}$ & $\F1{1\a\b}$& $\F0{1\a\b\gamma}$ & Endpoint
\\ \hline
Electric $\vec z_\sw=$ &-- &$ z^\a$& $z^\a, z^\b$ & N/A 
& String \\ \hline
Magnetic $\vec z_\st=$ & -- &$z^\a$ &$z^\a, z^\b$ & $z^\a, z^\b, z^\gamma$ 
& 5-brane \\ \hline\hline
&$\cF2\a$ & $\cF1{\a\b}$& $\cF0{\a\b\gamma}$ && \\ \hline
Electric $\vec z_\sw=$ &$z^\a$& $z^\a,z^\b$ & N/A &&
Wave \\ \hline
Magnetic $\vec z_\st=$ &$z^\a$& $z^\a, z^\b$ & $z^a, z^\b, z^\gamma$&&
NUT \\ \hline
\end{tabular}}
\bigskip

\centerline{Table (2a): NS-NS oxidations to $D=10$}

\bigskip\bigskip

\centerline{
\begin{tabular}{|c|c|c|c|c|c|c|}\hline
& $F_4$ & $ \F3{\a}$ & $\F2{\a\b}$ & $F_1^{(\a\b\gamma)}$& 
$F_0^{(\a\b\gamma\delta)}$ & Endpoint
\\ \hline
Electric $\vec z_\sw=$ & -- &$z^\a$ &$z^\a, z^\b$ & $z^\a, z^\b, 
z^\gamma$ & N/A & D2-brane \\ \hline
Magnetic $\vec z_\st=$ & -- &$z^\a$ &$z^\a, z^\beta$ & $z^\a, z^\b, 
z^\gamma$ & $z^\a, z^\b, z^\gamma, z^\delta$& D4-brane \\ \hline\hline
& $\cF21$ & $\cF1{1\a}$ & $\cF0{1\a\beta} $&&&  \\ \hline 
Electric $\vec z_\sw=$ & -- & $z^\a$ & N/A &&& D0-brane\\ \hline
Magnetic $\vec z_\st=$ & -- & $ z^\a$ & $ z^\a, z^\b$ &&& 
D6-brane\\ \hline
\end{tabular}}
\bigskip

\centerline{Table (2b): R-R oxidations to $D=10$}
\bigskip

     The notation in the tables carries over, {\it mutatis mutandis},
from the notation that we described previously for the oxidations to
$D=11$.  In exactly the same way as in for $D=11$, having given the
oxidation rules for single-charge $p$-branes in $D$ dimensions, it is
a straightforward and purely mechanical process to deduce the oxidation
endpoints in $D=10$ when starting from simple multi-charge $p$-branes
in $D$ dimensions.  We shall not present any further examples here,
since no new issues of principle arise.  We may note, however, that
the two examples given in our discussion of oxidations to $D=11$,
namely the dyonic string in $D=6$, and the 3-charge black-hole in
$D=5$, become intersections of a string with a 5-brane, and
intersections of a string, 5-brane and a wave, respectively.

    Again, the classification of all the resulting intersections in
$D=10$ is subsumed by a classification of all possible lower-dimensional
multi-charge $p$-branes.

\section{Classification of $p$-branes in maximal supergravities}

    In this section, we shall address the problem of classifying all
multi-charge $p$-brane solutions in maximal supergravities in $D\ge
3$.  We shall discuss the $D=2$ case in section 5.  This is an
important problem in its own right, since these extremal BPS saturated
solutions are expected to survive as quantum states in compactified
string theory or M-theory.  In addition, as we have seen in the
previous section, their classification also provides a classification
of multiple intersections in M-theory or string theory, where the 
harmonic functions all depend on the coordinates transverse to
the individual world-volumes.

     Our discussion here will divide into two parts.  The first
applies to the $p$-branes in $D$ dimensions that are supported by
4-form, 3-form, 2-form or 1-form field strengths.  Such solutions can
all be viewed as solutions of the standard massless maximal
supergravities that are derived from $D=11$ supergravity by ordinary
Kaluza-Klein dimensional reduction.  The second part of our discussion
will be concerned with $p$-branes in $D$ dimensions that are supported
by ``0-form field strengths.''  These solutions are not seen in the
ordinary massless supergravities, and in fact the 0-form field
strengths are really like cosmological terms in massive
supergravities.  In fact such massive theories, still maximally
supersymmetric, do arise as consistent Kaluza-Klein reductions of
$D=11$ supergravity.  However, they are obtained by making a
generalised reduction of the Scherk-Schwarz type
\cite{ss1,bdgpt,clpst,lpdomain,llp}.  The $p$-brane solutions
supported by the cosmological terms in these massive theories are all
$(D-2)$-branes, and are commonly known as domain walls.  Since some
new features arise in these cases, we shall discuss them separately.

\subsection{$p$-branes in massless supergravities}

          In section 2, we gave a review of the $p$-brane solitons,
supported by certain $n$-rank field strengths, that arise in
$D$-dimensional supergravities.  In the massless maximal
supergravities, we encounter field strengths of degrees $n=4,3,2,1$.
In $D$ dimensions, an $n$-form field strength is dual to a
$(D-n)$-form, and in this paper, such a dualisation will always be
performed if $D-n <n$.  The resulting versions of the supergravities
may be called ``fully dualised.''  Thus in these versions the 4-form
exists for $D\ge 8$; the 3-forms exist for $D\ge6$; 2-forms exist for
$D\ge4$ and 1-forms exist for $D\ge 2$.  

\subsubsection{$p$-branes from 4-form and 3-form field strengths}

        There is only one 4-form field strength, and it gives rise to
electric membrane or magnetic $(D-6)$-brane solutions.  In $D=8$,
there exists a dyonic solution where the 4-form field strength carries
both electric and magnetic charges \cite{ilpt}.  However, in this
solution the contribution from the ${\cal L}_{FFA}$ Wess-Zumino term
to the equations of motion does not vanish, and the solution is
nothing but a perturbative $SL(2,\R)$ transformation of a
singly-charged purely electric or magnetic membrane.  The solution
preserves half the supersymmetry, as in the case of the purely
electric or purely magnetic solutions.  As we discussed in section 2,
such a solution can really be regarded, from the classical point of
view, as a single-charge solution.  We shall return to this example
later, in section 8.

    In $D\ge6$ dimensions there are $(11-D)$ 3-forms in the
fully-dualised supergravities.  However, it is not possible to
construct simple solutions (in the sense defined in section 2) using
more than one 3-form field strength.  This can be seen from the fact
that the dilaton vectors $\vec a_i$ for the 3-form field strengths
$F_3^{(i)}$ satisfy the dot-product relation \cite{lpsol}
\be
\vec a_i \cdot \vec a_j = 2\delta_{ij} -\fft{2(D-6)}{D-2}\ ,\label{3fdot}
\ee
which is not of the form given by (\ref{mdot}).  Of course, there will
exist solutions which are merely U-duality transformations of
singly-charged solutions, and these can involve more than one 3-form
field strength, but we may again view these as being singly-charged,
for the same reasons as we discussed before.  In particular this
implies that extremal solutions supported only by 3-forms will always
preserve half of the supersymmetry, when $D\ge 7$.  In $D=6$, however,
there exist dyonic string solutions, where a single 3-form field
strength carries both electric and magnetic charges.  (In fact we
discussed this solution in section 3.) The dyonic string preserves
$\ft14$ of the supersymmetry, a characteristic of all simple 2-charge
solutions.  There are in all five 3-form field strengths in $D=6$,
giving five possible dyonic string solutions, denoted by the
participating field strengths \cite{lpsol}:
\be
\{*F_3^{(i)}, F_3^{(i)}\}_5\ ,
\ee
where $i=1,\ldots,5$.  We have introduced here a notation that we
shall use throughout the paper, in which a simple $p$-brane solution
is characterised by the list of non-vanishing field strengths that
support it.  In general a field strength labelled with a $*$ signifies
that it carries a magnetic charge if the unstarred field strengths
carry electric charges, and {\it vice versa}. The subscript attached
to the list indicates the multiplicity of such solutions,
corresponding to the different solutions that can be obtained by
making different choices for the internal index values on the
participating field strengths.  Thus in the case of the dyonic string,
the multiplicity of 5 arises because the index $i$ can take five possible
values.

\subsubsection{Multi-charge $p$-branes from 2-form field strengths}

       Simple multi-charge $p$-brane solutions involving multiple
field strengths of degrees 2 or 1 do exist, and their Weyl multiplet
structures were discussed in \cite{lpsweyl}.  First let us discuss the
case of 2-form solutions, in $D\ge 4$. They can be either black holes
or $(D-4)$-branes, and can involve up to four participating field
strengths, in sufficiently low dimensions.  Not surprisingly, the
multiplicities for these $p$-brane solutions grow with decreasing
dimension $D$ (since the range of the internal indices grows).  These
multiplicities, obtained in \cite{lpsweyl}, are given in Table 3
below. Note that the subscripts indicate the fractions of preserved
supersymmetry.   We shall derive these fractions in section 6.

\bigskip\bigskip
\centerline{
\begin{tabular}{|c|c|c|c|c|}\hline Dim.
 &\phantom{for} $N=1$\phantom{for} &\phantom{for}
$N=2$\phantom{for}  &\phantom{for}$N=3$\phantom{for} &
$N=4$\\ \hline\hline
$D=10$   &$1_\fft12$       &      &  &\\ \hline
$D=9$    & $1_\fft12 + 2_\fft12$   & $2_\fft14$    &  &  \\ \hline
$D=8$    & $6_\fft12$   &  $6_\fft14$   &  &  \\  \hline
$D=7$    & $10_\fft12$   & $15_\fft14$    &  & \\  \hline
$D=6$    & $16_\fft12$   &  $40_\fft14$  &  &  \\  \hline
$D=5$    & $27_\fft12$   &  $135_\fft14$   &  $45_\fft18$ & \\  \hline
$D=4$    & $56_\fft12$   & $756_\fft14$  & $2520_\fft18$ & $630_\fft18$\\
\hline
\end{tabular}}
\bigskip
\centerline{Table 3: Multiplicities for $N$-charge supersymmetric 2-form
solutions}
\bigskip

       The $N=1$ solutions can be easily classified, since each field
strength can give rise to one $p$-brane that is either
electrically-charged or magnetically-charged.  All the
singly-charged solutions in a given dimension preserve half the
supersymmetry, and they form an irreducible multiplet under the Weyl
subgroup of the U-duality group \cite{lpsweyl}.  Acting on these
solutions with the full U-duality transformations, we obtain a full
multiplet of solutions that preserve half the supersymmetry.  When
$N\ge 2$ the classification becomes more complicated, since one cannot
obtain simple $N$-charge solutions using an arbitrary set of 2-form
field strengths; only sets whose dilaton vectors satisfy the relations
(\ref{mdot}) will admit such solutions.  It is a straightfoward
matter, given the expressions (\ref{dilatonvec}) and (\ref{gfdot}), to
enumerate the sets of fields which can lead to multi-charge solutions.
We shall now discuss these for each dimension $3\le D\le 9$.

\bigskip
\noindent{$D=9$}
      
      In $D=9$, although there are three 2-form field strengths, their
associated three dilaton vectors $\vec a_{12}$, $\vec b_1$ and $\vec
b_2$ do not all satisfy (\ref{mdot}).  However, two of the three possible
pairs of dilaton vectors, namely $\{\vec a_{12}, \vec b_1\}$ and
$\{\vec a_{12}, \vec b_2\}$, do satisfy (\ref{mdot}).  Thus the
maximum number of 2-form field strengths for simple solutions in $D=9$
is $N_{\rm max} =2$, given by
\be
\{F_2^{(12)}, {\cal F}_2^{(1)}\}\ ,\qquad
\{F_2^{(12)}, {\cal F}_2^{(2)}\}\ .\label{d9n2}
\ee
They form a doublet under $S_2$, the Weyl group of the CJ group
$SL(2,\R)$.  They give rise to 2-charge electric black holes,
or 2-charge magnetic 5-branes. These, and indeed all simple 2-charge
$p$-brane solutions, preserve $\ft14$ of the supersymmetry. Acting with
$SL(2,\R)$, one obtains the full $SL(2,\R)$ multiplet of solutions that
preserve $\ft14$ of the supersymmetry.  This fraction of preserved
supersymmetry distinguishes the multiplet from the one that would be
obtained by acting with $SL(2,\R)$ on a single-charge $p$-brane
solution, which would instead preserve half the supersymmetry (see section 8).

\bigskip
\noindent{$D=8,7,6$}
\bigskip

          From Table 3, in all these dimensions we have $N_{\rm max} =
2$, and the associated solutions have multiplicities $M= 6$, 15 and 40
in $D=8$, 7 and 6.  These are also the dimensions of the irreducible
representations of the associated U Weyl groups \cite{lpsweyl}.  The
pairs of field strengths whose dilaton vectors satisfy (\ref{mdot})
are easily identified, and are given by
\bea
D=8: && \{F_2^{(ij)}, {\cal F}_2^{(i)}\}_6\ ,\label{d8n2}\\
D=7: && \{F_2^{(ij)}, {\cal F}_2^{(i)}\}_{12}\ ,\quad
       \{F_2^{(ij)}, F_2^{(k\ell)}\}_3\ ,\label{d7n2}\\
D=6: && \{F_2^{(ij)}, {\cal F}_2^{(i)}\}_{20}\ ,\quad
       \{F_2^{(ij)}, F_2^{(k\ell)}\}_{15}\ ,\quad
       \{{\cal F}_2^{(i)}, *F_4\}_5\ ,\label{d6n2}
\eea
where the indices $(i,j,k,\ldots)$ are all different, and run from 1
to $11-D$. The subscripts, as usual, denote the multiplicities.  Note
that in $D=7$ and $D=6$ the multiplicities $M=15$ and $M=40$ arise
from more than one kind of structure for the possible pairs of field
strengths.  This phenomenon, which occurs in general in lower
dimensions, is a reflection of the fact that the various field
strengths here are characterised by $SL(11-D,\R)$ indices $i,j\ldots$,
but in the fully-dualised supergravities they assemble into $E_{11-D}$
multiplets \cite{cjlp}.  Thus the multiplets of multi-charge solutions,
although irreducible under the $E_{11-D}$ CJ groups, decompose into
reducible representations under $SL(11-D,\R)$.  

     Recall that when a field strength carries a $*$, this indicates
that it is related by dualisation to the corresponding field in the
truncated Lagrangian (\ref{nlag}).  Thus for example the last
combination in (\ref{d6n2}) corresponds to a truncated Lagrangian with
two 2-form field strengths $F_2^\a$, one chosen from the five $\cF2i$,
and the other being the dualised field $e^{\vec a\cdot \vec\phi}\,
*F_4$.  In this example, the two field strengths $F_2^\a$ either both
carry electric charges or both carry magnetic charges, corresponding
to a 2-charge black-hole or 2-charge membrane respectively.  In terms
of the original fields, the black hole carries an electric charge for
one of the $\cF2i$ fields, and a magnetic charge for the $F_4$ field.
The charge complexions are reversed in the case of the membrane.  Thus
in terms of the original variables, the solutions could be viewed as
being dyonic (of the second kind).

\bigskip
\noindent{$D=5$}
\bigskip

       In five dimensions, we have $N_{\rm max}=3$.  The $N=2$
solutions have multiplicity $M=135$, and form a 135-dimensional
irreducible representation of the $E_6$ Weyl group \cite{lpsweyl}.
The $N=3$ solutions have multiplicity $M=45$, and likewise form a
45-dimensional irreducible representation under the $E_6$ Weyl group.
The sets of allowed participating field strengths for these solutions
are given by
\bea
N=2:&& \{ F_2^{(ij)}, {\cal F}_2^{(i)}\}_{30}\ ,\quad
       \{ F_2^{(ij)}, F_2^{(k\ell)} \}_{45}\ ,\quad
       \{*F_3^{(i)}, {\cal F}_2^{(j)} \}_{30}\ ,\quad
\{*F_3^{(i)}, F_2^{(ij)} \}_{30}\ ,\nn\\
N=3:&& \{ F_2^{(ij)}, F_2^{(k\ell)}, F_2^{(mn)} \}_{15}\ ,\quad
       \{ *F_3^{(i)}, {\cal F}_2^{(j)}, F_2^{(ij)} \}_{30}\ ,
\label{d5n3}
\eea
where the indices $(i,j, \ldots)$ are all different, and
run from 1 to $11-D=6$.

\bigskip
\noindent{$D=4$}
\bigskip

       In $D=4$, the maximal number of participating 2-form field 
strengths is $N_{\rm max}=4$.  The $N=2,3,4$ solutions form
irreducible representations of the $E_7$  Weyl group with dimensions 756,
2520 and 630 \cite{lpsweyl}.  The participating field strengths 
are given by
\bea
N=2:&& \{ F_2^{(ij)}, {\cal F}_2^{(i)}\}_{42+42}\ ,\qquad 
       \{ F_2^{(ij)}, F_2^{(k\ell)} \}_{105+105}\ ,\qquad
       \{ *F_2^{ij}, F_2^{(ik)} \}_{210}\ ,\nonumber\\
    && \{ *F_2^{(ij)}, {\cal F}_2^{(k)} \}_{105+105}\ ,\qquad
       \{ *{\cal F}_2^{(i)}, {\cal F}_2^{(j)} \}_{42}\ ,\label{d4n2}\\
N=3:&& \{F_2^{(ij)}, F_2^{(k\ell)}, F_2^{(mn)} \}_{105+105}\ ,\quad
       \{ F_2^{(ij)}, F_2^{(k\ell)}, *F_2^{(ik)} \}_{420+420}\ ,
\nonumber\\
     &&  \{ F_2^{(ij)}, F_2^{(k\ell)}, *{\cal F}_2^{(m)} \}_{315+315}
\ ,\qquad \{ F_2^{(ij)}, *F_2^{(ik)}, {\cal F}_2^{(j)} \}_{210+210}\ ,
\nonumber\\
  &&\{ F_2^{(ij)}, {\cal F}_2^{(i)}, *{\cal F}_2^{(k)} \}_{210+210}\ ,
\label{d4n3}\\
N=4:&& \{ F_2^{(ij)}, F_2^{(k\ell)}, F_2^{(mn)}, *{\cal F}_2^{p} 
        \}_{105+105}\ ,\quad
       \{F_2^{(ij)}, *F_2^{(ik)}, {\cal F}_2^{(j)}, *{\cal
         F}_2^{(k)} \}_{210}\ ,\nonumber\\
    && \{ F_2^{(ij)}, F_2^{(k\ell)}, *F_2^{(ik)}, *F_2^{(j\ell)}
        \}_{210}\ ,\label{d4n4}
\eea
where the indices $(i,j,\ldots)$ are all different, and 
run from 1 to $11-D=7$.  The pairs of numbers in the multiplicity
subscripts indicate that the 0-brane solutions can be dualised to 
give an equal number of new solutions that are again 0-branes.
Although the 2-form solutions in higher dimensions could also be dualised,
in those cases the solutions dual to 0-branes would be $(D-4)$-branes, and
thus there is no doubling of the multiplicities of $p$-branes with a given
$p$. 

\subsubsection{Comments on 2-form solutions}

\noindent{{\bf 1)}}\ \ \ 
In the previous subsection, we presented the complete results for
simple multi-charge $p$-brane solutions supported by 2-form field
strengths in $D\ge 4$, by listing all their multiple field strength
configurations.  The exact solutions are given by (\ref{dhsol}) and
(\ref{dilsol}), where the dilaton vectors $\vec c_\a$ for each
multiple field configuration can be read off from (\ref{dilatonvec}).
{\it A priori}, we know only that these solutions satisfy the
equations of motion coming from a truncated Lagrangian of the form
(\ref{nlag}).  In fact, they also satisfy the full equations of motion
coming from the complete $D$-dimensional supergravity Lagrangian
(\ref{dgenlag}).  In other words, in these simple solutions, owing to
the specific combinations of multiple field strengths that are
involved, the Kaluza-Klein modifications to field strengths, and the
Wess-Zumino terms ${\cal L}_{FFA}$, make no contribution in the
equations of motion.  It is necessary, and non-trivial, to verify this
point, since our criterion for recognising valid sets of field
strengths for multi-charge solutions was based only on the criterion
that their dilaton vectors should satisfy (\ref{mdot}).

       In order to show that this criterion is in fact necessary and
sufficient, we first note that all our solutions form irreducible
multiplets of the Weyl subgroup of the $E_{11-D}$ CJ group
\cite{lpsweyl}.  In particular, the single-charge solutions form
highest-weight representations of the Weyl group \cite{cjlp}.  In any
simple single-charge solution, all the axions vanish, as do all field
strengths other than the specific 2-form that carries the charge.
This will continue to be true after acting with any element of the
Weyl group, since it simply permutes field strengths, together with
their dilaton vectors \cite{lpsweyl}.  The complete multiplets of
multi-charge solutions that we listed above can be generated from any
given member of the multiplet by acting with Weyl group.  Since the
action of the Weyl group is the same regardless of the number of
charges $N$ in a particular solution, it follows that we only need
verify that one member of the Weyl group multiplet has the simple form
of solution where all non-charge-carrying field strengths vanish, in
order to establish that all members of the multiplet have this
property.  This is a much simpler task than verifying the point for
each member of the multiplet, and indeed one can easily check that it
is true.  It should be emphasised that since the Weyl group preserves
the dot products between dilaton vectors \cite{lpsweyl}, it follows
that the criterion that a set of $N$ field strengths have dilaton
vectors satisfying (\ref{mdot}) is not only necessary, but also
sufficient, as a procedure for generating all simple $N$-charge
solutions.

\bigskip
\noindent{{\bf 2)}}\ \ \ The following provides another argument which
establishes that simple multi-charge 2-form solutions exist if and
only if the dilaton vectors of the participating field strengths
satisfy (\ref{mdot}).  The potentially dangerous terms in the
Lagrangian that could spoil the existence of the simple solution are
either interactions of the form $ \chi\, F_2^\a\cdot F_2^\b$, which we
may call Kaluza-Klein type, or interactions of the form $A\wedge
F_2^\a \wedge F_2^\b$, which we may call Wess-Zumino type.  Here
$\chi$ is an axion, $A$ is a $(D-4)$-form potential, and $F_2^\a$ and
$F_2^\b$ are two of the field strengths that participate in the
solution.  In fact the Wess-Zumino type interactions will always give
contributions to the equations of motion that vanish in the background
of the putative simple multi-charge solution, because the field
strengths $F_2^\a, F_2^\b,\ldots$ involved in the solution either all
carry electric charges or else all carry magnetic charges, given by
(\ref{fans}).\footnote{Recall that, as we discussed in section 2.2,
the field strengths $F_n^\a$ in the truncated Lagrangian (\ref{nlag})
are not necessarily the same as the ones coming by direct reduction
from $D=11$.  In cases where some of the field strengths in a list for
a multi-charge solutions are denoted with a $*$, the corresponding
$F_n^\a$'s are the duals of the directly-reduced fields.  Thus
although the $F_n^\a$ fields themselves all carry electric charges, or
all carry magnetic charges, in terms of the original supergravity
fields the solutions can be dyonic.}  Thus only interactions of the
Kaluza-Klein type would cause trouble, since the equation of motion
for the field $\chi$ in such a term would forbid us from setting it to
zero, spoiling the existence of the simple solution.  One can show
that there is a one-to-one relation between such cubic terms and the
summation rules for the dilaton vectors of the three fields
\cite{cjlp}.  Specifically, it is an easily verified rule that every
term in the $D$-dimensional Lagrangian of the Kaluza-Klein type cubic
form above has the property that the associated dilaton vectors
satisfy $\vec c_\a -\vec c_\b = \pm \vec c_\chi$, where $\vec c_\chi$
is the dilaton vector for the axion $\chi$, and in fact this sum rule
gives the necessary and sufficient condition for the occurrence of
this cubic interaction \cite{cjlp}.  Since dilaton vectors for axions
always satisfy $\vec c_\chi.  \vec c_\chi=4$ (see (\ref{dilatonvec})
and (\ref{gfdot})), it follows that if $\vec c_\a$ and $\vec c_\b$ satisfy
(\ref{mdot}), then for these particular field strengths the worrisome
cubic terms cannot be present in the Lagrangian, and so the existence
of the simple multi-charge solution cannot be spoiled by the
Kaluza-Klein type of interaction terms.  Conversely, if a pair of
dilaton vectors $\vec c_\a$ and $\vec c_\b$ do not satisfy
(\ref{mdot}), then the sum rule {\it is} satisfied (as can easily be
verified from (\ref{dilatonvec}) and (\ref{gfdot})), implying that the
cubic term will occur in the Lagrangian, and so the simple
multi-charge solution involving these field strengths will not exist.

      We have shown that in terms of the field strengths $F^\a$
appearing in the truncated Lagrangian (\ref{nlag}), the requirement
that their dilaton vectors satisfy (\ref{mdot}) implies that the only
cubic interactions that can arise are of the Wess-Zumino type, and
furthermore the contributions that these make in the equations of
motion vanish owing to the purely electric or purely magnetic nature
of the charges carried by the field strengths $F^\a$, as given by
(\ref{fans}).  Of course, we should further verify that all the
interactions of higher than cubic order also give no contributions in
the equations of motion, for the solutions we discussed above.  In
fact, the higher-order interaction terms are also governed by dilaton
vector sum-rules, which satisfy a chain-rule relation \cite{cjlp}.  This
implies that if the solution is immune from all the cubic
interactions, it is immune from all higher-order interactions as well.

       It is worth remarking that applying the same argument to the
case of two 3-form field strengths, whose dilaton vectors always
satisfy (\ref{3fdot}), we find that a cubic interaction $\chi\,
F_3^\a\cdot F_3^\b$ {\it does} exist, which explains why simple
multi-charge solutions involving two or more 3-form field strengths
cannot occur.

\bigskip
\noindent{{\bf 3)}}\ \ \
It is also worth remarking that although we enumerated all of the
$N\ge2$ combinations of field strength configurations allowed by
(\ref{mdot}), in fact the combination rules are already completely
encoded by the allowed 2-charge configurations.  This is because any
$N\ge3$ charge configuration will be allowed by (\ref{mdot}) if and
only if all pairwise sub-combinations of two field strengths are
allowed.  The utility of nevertheless listing the $N\ge3$ combinations
explicitly is that the above rule, although easily stated, is not
necessarily easy to implement by hand in practice.  This point becomes
more acutely apparent in the case of solutions for 1-form or 0-form
field strengths, as we shall see presently.

\bigskip
\noindent{{\bf 4)}}\ \ \         
Note that all the $N=2$ solutions preserve $\ft14$ of the
supersymmetry and all the $N=3,4$ solutions preserve $\ft18$ of the
supersymmetry.  Thus we see that for the cases $N=1,2,3$, each
additional charge breaks one half of the remaining supersymmetry, but
the introduction of the fourth charge does not further break the
supersymmetry \cite{lpsol} (although it does, however, modify the
structure of non-vanishing eigenvalues in the \bog matrix
\cite{lpmult}), if the sign of the new charge is appropriately chosen.
For the other choice of the sign, it will break all the supersymmetry
\cite{lpmult,ko}.  We shall explain this in section 6.  Purely
electric or purely
magnetic 2-form solutions can have a maximum of only $N=3$ charges;
all the $N=4$ solutions are dyonic in the sense that some of the four
participating (original) field strengths carry electric charges and
the others carry magnetic charges.  The Reissner-Nordstr{\o}m black
hole solutions arise when all four of these charges are set equal.
They occur with multiplicity $M=756$.  Note that in the second and
third field configurations given by (\ref{d4n4}), the solution
contains two electric and two magnetic charges.  Reissner-Nordstr{\o}m
black holes of a similar kind arise also in the compactified heterotic
string.  However, the first of the field configurations in
(\ref{d4n4}) is of a different kind, in that three of the charges are
electric and one is magnetic (or {\it vice versa}); solutions of this
type do not occur in the heterotic string.  This can be understood
from the fact that in this case at least two R-R fields are needed,
whereas there are none in the heterotic string.  In $D=5$, there are
in total 135 3-charge black holes, which all give rise to
Reissner-Nordstr{\o}m black holes when the charges are set equal.

\bigskip
\noindent{{\bf 5)}}\ \ \    
All the supersymmetric dyonic solutions in $D=4$ are of the second
kind \cite{lpsol}, in that it is {\it different} field strengths that carry
the electric and the magnetic charges, rather than having one field
strength carrying electric and magnetic charges simultaneously.  In
fact, there does exist a 2-charge solution in $D=4$ in which a single
field strength carries both electric and magnetic charges \cite{gk}.
This is a dyonic black hole of the first kind. However, it breaks all the
supersymmetry.  In fact it can be viewed as a bound state with
positive binding energy \cite{gk}.  When its electric and magnetic
charges are equal, the solution reduces to a $D=4$
Reissner-Nordstr{\o}m black hole.  Thus although the dyonic solution
breaks all the supersymmetry in general, the supersymmetry is fully
restored at the horizon, which is AdS$_2\times S^2$, as well as in the
asymptotically Minkowskian region near infinity.

\subsubsection{Multi-charge $p$-branes from 1-form field strengths}

   We now turn to the case of $p$-brane solutions using 1-form field
strengths, which exist in the fully-dualised supergravities for all
$2\le D\le 9$, although in this section we shall consider only
$D\ge3$.  The discussion for $D=2$ will be given in section 5.
1-form field strengths can support either electric instantons in a
Euclidean-signature space or magnetic $(D-3)$-branes in the usual
Lorentzian-signature spacetime. In the latter case, the transverse
space is two-dimensional.  It turns out that multi-charge solutions
can involve up to $N_{\rm max}=8$ participating field strengths,
although as usual, the maximum number depends upon the dimension
$D$. We find that the multiplicities $M$ of the 1-form solutions are 
given by

\bigskip\bigskip
\centerline{
\begin{tabular}{|c|c|c|c|c|c|c|c|c|c|}\hline Dim.
 &$N=1$&
$N=2$ & $N=3$ & $N=4$ &
$N=5$ & $N=6$ & $N=7$ &
$N=8$ \\ \hline\hline
$D=9$    & $1_\fft12$   &    &  &  & & & & \\ \hline
$D=8$    & $4_\fft12$   & $3_\fft14$    &  &  &&&&\\  \hline
$D=7$    & $10_\fft12$ &  $15_\fft14$  &  & &&&& \\  \hline
$D=6$    & $20_\fft12$  & $70_\fft14$   & $60_\fft18$ & $15_\fft18$ 
&&&& \\  \hline
$D=5$    &  $36_\fft12$  &  $270_\fft14$   &  $540_\fft18$ & 
$135_\fft18$ &&&&  \\  \hline
$D=4$    & $63_\fft12$   & $945_\fft14$  & 
$\begin{array}{c} 3780_\fft18\\ 315_\fft18 \end{array}$
& $\begin{array}{c} 945_\fft18\\ 3780_\fft1{16} \end{array}$ & 
$2835_\fft1{16}$ & $945_\fft1{16}$ & $135_\fft1{16}$ &\\ \hline
$D=3$ & $120_\fft12$ & $3780_\fft14$ & $37800_\fft18$ &
$\begin{array}{c} 9450_\fft1{8}\\ 113400_\fft1{16} \end{array}$& 
$113400_\fft1{16}$ & $56700_\fft1{16}$ 
&$16200_\fft1{16}$ & $2025_\fft1{16}$ \\ \hline
\end{tabular}}
\bigskip
\centerline{Table 4: Multiplicities for supersymmetric 1-form
solutions}
\bigskip

      In the above table, we list the multiplicities $M$ of the
possible field strength configurations.   The dimension of the Weyl
group representation is given by $2^N M$. (The reason for the extra $2^N$
factor, which did not arise in the case of 2-form solutions, is because
of a special feature of 1-form field strengths, related to the fact that
their dilaton vectors, together with the negatives of the dilaton vectors, 
form the roots of the $E_{11-D}$ algebra \cite{lpsweyl}.)  Again, the 
subscripts on the multiplicities indicate the fractions of preserved
supersymmetry.

    The classification of single-charge $p$-branes for 1-form field
strengths is completely straightforward since any one of them can give
rise to such a solution.  As can be seen from the multiplicities
listed in Table 4, the classification of multi-charge solutions
rapidly becomes rather complicated.  This is merely because of the
profusion of combinatoric possibilities, and the underlying structure
is still very simple: any set of $N$ 1-form field strengths whose
dilaton vectors satisfy (\ref{mdot}) will give rise to a simple
$N$-charge solution.  As we discussed in section 3.1.2, the essential
combination rules are in fact already encoded in the results for
2-charge solutions, since the dilaton vectors for a set of $N$ field
strengths will satisfy (\ref{mdot}) if and only if all pairwise
combinations of dilaton vectors satisfy (\ref{mdot}).  Accordingly,
we shall only present the explicit listings of 2-charge combinations
in this section.  The full listings, together with their individual
multiplicities, are relegated to the appendix.  The sums of these
individual multiplicities make up the total multiplicities $M$ given
in Table 4.

     In all the listings, the indices $(i,j,k,\ldots)$ are understood to
be all different, and to run from 1 to $(11-D)$.  

\bigskip
\noindent{$D=8,7$}
\bigskip

          In both $D=8$ and 7 dimensions, the maximum number of 1-form field 
strengths that can satisfy (\ref{mdot}) is $N_{\rm max}=2$, with 
total multiplicities $M=3$ and 15 respectively:
\bea
D=8:&&\{F_1^{(ijk)}, {\cal F}_1^{(ij)}\}_{3}\ ,\\
D=7:&&\{F_1^{(ijk)}, {\cal F}_1^{(ij)}\}_{12}\ ,\qquad
      \{\cF1{ij}, \cF1{k\ell}\}_3\ .\label{d87n21}
\eea

\bigskip
\noindent{$D=6$}
\bigskip

       In this dimension, we have $N_{\rm max} = 4$,
The $N=2$ solutions, numbering 70 in total, are given by
\be
\{ F_1^{(ijk)}, F_1^{i\ell m} \}_{15}\ ,\quad
       \{ F_1^{(ijk)}, {\cal F}_1^{(ij)} \}_{30}\ ,\quad
       \{ F_1^{(ijk)}, {\cal F}_1^{(\ell m)} \}_{10}\ ,\quad
\{ {\cal F}_1^{(ij)}, {\cal F}_1^{(k\ell)} \}_{15}\ .\label{d6n21}
\ee

\bigskip
\noindent{$D=5$}
\bigskip

            As in $D=6$, we have $N_{\rm max} =4$ in
$D=5$.  The $N=2$ solutions, of which there are 270 in total, are given by
\bea
&& \{ F_1^{(ijk)}, F_1^{i\ell m} \}_{90}\ ,\quad
       \{ F_1^{(ijk)}, {\cal F}_1^{(ij)} \}_{60}\ ,\quad
       \{ F_1^{(ijk)}, {\cal F}_1^{(\ell m)} \}_{60}\ ,\nonumber\\
&& \{ {\cal F}_1^{(ij)}, {\cal F}_1^{(k\ell)} \}_{45}\ ,\quad
   \{ *F_4, {\cal F}_1^{(ij)} \}_{15}\ .\label{d5n21}
\eea
Note that in this case there is an additional 1-form field strength $*F_4$,
coming from the dualisation of the 4-form $F_4$.

\bigskip
\noindent{$D=4$}
\bigskip

       In $D=4$, there are a total of 63 1-form field strengths: 35
$F_1^{(ijk)}$, 21 $\cF1{ij}$ and 7 $*\F3{i}$ which come from the
dualisation of the 3-forms.  There can be a up to $N_{\rm max}=7$
1-form field strengths that satisfy (\ref{mdot}).  For $N=2$, there are a 
total of $M=945$ possibilities:
\bea
&&\{ \F1{ijk}, *\F3{i} \}_{105}\ ,\quad
  \{ \F1{ijk}, \F1{i\ell m} \}_{315}\ ,\quad
  \{ \F1{ijk}, \cF1{ij} \}_{105}\ ,\nn\\
&&\{ \F1{ijk}, \cF1{\ell m} \}_{210}\ ,\quad
  \{ \cF1{ij}, *\F3{k} \}_{105}\ ,\quad
  \{ \cF1{ij}, \cF1{k\ell} \}_{105} \ .\label{d4n21}
\eea

\bigskip
\noindent{$D=3$}
\bigskip

      There are a total of 120 1-form field strengths in $D=3$: 56
$F_1^{(ijk)}$ coming from dimensional reduction of the $F_4$ in
$D=11$, 28 ${\cal F}_1^{(ij)}$ coming from the metric, and in addition, 28
$*F_2^{(ij)}$ and 8 $*{\cal F}_2^{(i)}$ coming from dualisation.  The 
maximal number of 1-forms that can satisfy (\ref{mdot}) in $D=3$ is
$N_{\rm max} = 8$.  The $N=2$ solutions, with total multiplicity $M=3780$,
are given by
\bea
&&\{*F_2^{(ij)}, *F_2^{(k\ell)} \}_{210}\ ,\quad
  \{*F_2^{(ij)}, *{\cal F}_2^{(i)}\}_{56}\ ,\quad
  \{F_1^{(ijk)}, *F_2^{(i\ell)} \}_{840}\ ,\quad
  \{F_1^{(ijk)}, F_1^{(i\ell m)} \}_{840}\ \nonumber\\
&&\{ F_1^{(ijk)}, *{\cal F}_2^{(\ell)} \}_{280}\ ,\quad
  \{ F_1^{(ijk)}, {\cal F}_1^{(ij)} \}_{168}\ ,\quad
  \{ F_1^{(ijk)}, {\cal F}_1^{(\ell m)} \}_{560}\ ,\quad
  \{ {\cal F}_1^{(ij)}, *F_2^{(ij)} \}_{48}\ ,\nonumber\\
&&\{{\cal F}_1^{(ij)}, *F_2^{(k\ell)} \}_{420}\ ,\quad
  \{ {\cal F}_1^{(ij)} *{\cal F}_2^{(k)} \}_{168}\ ,\quad
  \{ {\cal F}_1^{(ij)}, {\cal F}_1^{(k\ell)} \}_{210}
\ ,\label{d3n21}
\eea

     All the $N\ge3$ solutions are presented in the appendix.

\subsubsection{Comments on 1-form solutions}

\noindent{{\bf 1)}}\ \ \ 
As in the case of 2-form solutions, so far we have enumerated the
lists of 1-form field strengths whose dilaton vectors satisfy the
condition (\ref{mdot}).  Again, it is necessary now to verify that all
these combinations of field strengths do indeed admit the construction
of simple multi-charge solutions, and furthermore, that these
combinations represent all of the possible simple multi-charge
solutions.  Although the multiplicities of the multi-charge solutions
can be very large (see Table 4), the solutions form irreducible
multiplets under the Weyl groups of the $E_{11-D}$ CJ groups.  The
axions (which are the potentials for the 1-form field strengths) and
the dilatons parametrise homogeneous coset spaces $E_{11-D}/H_{11-D}$,
where $H_{11-D}$ is the maximal compact subgroup of $E_{11-D}$.  In
particular, the dilaton vectors associated with the axions, together
with their negatives, are precisely the root vectors of $E_{11-D}$
\cite{lpsweyl}.  This implies that the axions are equivalent under the
Weyl group, which permutes the root vectors.  Thus verifying that any
one member of a Weyl multiplet is a genuine solution of the full
$D$-dimensional supergravity equations of motion implies that the
entire Weyl multiplet are also genuine solutions.  The task is thus
reduced to a simple one, and we have checked by this means that all
the combinations of field strengths that we list do indeed give rise
to multi-charge solutions.  Since the dot-products between dilaton
vectors are preserved under the Weyl group, it also follows that the
listed combinations represent all the possible simple multi-charge
solutions.  In fact the same argument that we gave in section 4.1.3 can
be applied here, to show that the potentially dangerous interaction
terms in the Lagrangian that might spoil the simple multi-charge
1-form solutions are absent if and only if the participating field
strengths have dilaton vectors that satisfy (\ref{mdot}).  The
argument again involves showing that the dilaton sum rules governing
cubic interactions forbid the occurrence of these dangerous terms for
the sets of field strengths that we are using.  (In the case of 1-form
field strengths there is actually another way to choose a set of field
strengths, whose dilaton vectors do not satisfy (\ref{mdot}), for
which there are again no interaction terms that contribute in the equations
of motion in the $p$-brane solution backgrounds.  This is done by
choosing a set of 1-forms whose dilaton vectors form the simple roots
of the $E_{11-D}$ algebra \cite{lpsln,lmmp}.  However, these $N$-charge
solutions are not expressible in terms of $N$ independent harmonic
functions, and although they can be extremal, the solutions are not
supersymmetric.  In fact they can be viewed as bound states with
negative binding energy \cite{lpsln}.)

\bigskip
\noindent{{\bf 2)}}\ \ \ 
The discussion of the fractions of supersymmetry that are preserved by
the multi-charge 1-form solutions is straightforward for $N\le3$.  In
these cases, just as for 2-form solutions, the fraction of preserved
supersymmetry is $2^{-N}$, so that the addition of each extra charge
halves the remaining supersymmetry.  For 4-charge 1-form solutions, it
turns out that there are now two possibilities.  In some cases, the
introduction of the fourth charge does not break the supersymmetry any
further, and $\ft18$ of the original supersymmetry is preserved.  This
is the same as the situation for 4-charge 2-form solutions that we
discussed in section 3.1.2.  In other cases, when other kinds of
combinations of four field strengths are involved, the 4-charge 1-form
solutions instead preserve $\ft1{16}$ of the original supersymmetry.
For $5\le N\le8$ charges, all the
solutions preserve $\ft1{16}$ of the original supersymmetry.
We shall return to the discussion of supersymmetry in section 6.

     So far, we have completed the classification of simple
multi-charge $p$-branes for all the massless supergravities in
$D\ge3$.  We shall discuss the details of dimensional reduction to
$D=2$ in section 5.  It is interesting to note that in massless
supergravities, the minimum non-vanishing fraction of preserved
supersymmetry for any $p$-brane solution is $\ft1{16}$.

\subsection{Domain-walls in massive supergravities}

      In the previous subsection 3.1, we obtained the simple
multi-charge $p$-brane solutions for $n=4,3,2$ and 1-form field
strengths.  For the solutions of massless maximal supergravities,
these results are complete.  However, we are interested in obtaining
all the BPS solutions in $D=11$, and these do not only come from the
oxidations of $p$-brane solutions in lower-dimensional {\it massless}
supergravities.  Some BPS solutions in $D=11$ come instead from the
oxidation of $p$-brane solutions of the {\it massive} maximal
supergravities that can also be obtained as consistent dimensional
reductions from $D=11$.  The standard toroidal compactifications of
eleven-dimensional supergravity can be generalised, by allowing one or
more axions in $(D+1)$ dimensions to be linearly dependent on the the
compactifying coordinates \cite{bdgpt,clpst,llp}.  The constants of
proportionality become cosmological terms in $D$ dimensions.  The
consistency of the reduction is not spoiled, since the axions that are
involved in the generalised reduction enter the $(D+1)$-dimensional
equations of motion only through their derivatives.  The cosmological
terms can be viewed as 0-form field strengths, labelled using the same
scheme as we have adopted for the higher-degree field strengths.  In
$D$ dimensions, there can be a total of $(11-D)!/(4!(7-D)!)$ of the
form $\F0{ijk\ell}$ and $(11-D)!/(3!(8-D)!)$ of the form $\cF0{ijk}$,
with associated dilaton vectors $\vec a_{ijk\ell}$ and $\vec b_{ijk}$
respectively, defined in (\ref{dilatonvec}).  In addition, there can
be $D$-forms in $D$ dimensions, which can be dualised to give further
cosmological terms.  Note that unlike the field strengths in massless
supergravities, these 0-form field strengths cannot all coexist
simultaneously in one single Lagrangian; there are many different
massive supergravities, each of which contains a subset of the above
list of possible cosmological terms \cite{clpst,llp}.  This is because
a 0-form field strength is really an integration constant in the
Lagrangian, and it either vanishes or it doesn't.  It is not like the
situation with a higher-degree field strength, for which the choice as
to whether or not it will carry a charge remains as yet unsettled in
the Lagrangian.  Thus whilst in the usual massless cases the question
of what possible combinations of field strengths may carry charges
need be decided only at the stage of considering solutions, in the
massive theories the ``charges'' are already present in the
Lagrangian, and the restrictions on possible non-vanishing
combinations are already operative in the construction of the
Lagrangian itself.  From the eleven-dimensional point of view,
however, all the solutions of these different massive supergravities
are equally important in that they are solutions of the $D=11$ theory.

     The $p$-brane solutions supported by 0-form field strengths can
only be magnetic $(D-2)$-branes, since an electric solution would have
to be a $(-2)$-brane, which does not exist. Thus we need consider only
magnetic $(D-2)$-branes, which are also known as domain walls.  These
are more difficult to study than the $p$-branes discussed in the
previous subsection, in that from the lower-dimensional point of view,
the domain-wall solutions can belong to large numbers of different
massive theories.  Furthermore, the CJ groups of the massless
supergravities are broken in the massive theories \cite{clpst}.
However, there is a simple criterion to decide whether a domain wall
solution is possible or not.  First of all, each of the 0-form field
strengths can give a single-``charged'' domain-wall solution.  For
solutions with $N\ge 2$ charges, the selection rule is in fact the
same as in the previous section, namely that the dilaton vectors of
the $N$ 0-form field strengths must satisfy the dot product relation
(\ref{mdot}), with $n=0$.  Thus it simply reduces to the usual
mechanical process of enumerating all the possible combinations of
0-form field configurations that satisfy (\ref{mdot}), using their
associated dilaton vectors as given in (\ref{dilatonvec}).  For
domain-wall solutions, it turns out that we have $N_{\rm max}=8$ when
descend down to $D=3$.  The multiplicities $M$ for each number of
charges $N$ in $3\le D\le 8$ are presented in Table 5.

\bigskip\bigskip
\centerline{
\begin{tabular}{|c|c|c|c|c|c|c|c|c|}\hline Dim.
 &$N=1$&
$N=2$ & $N=3$ & $N=4$ &
$N=5$ & $N=6$ & $N=7$ &
$N=8$ \\ \hline\hline
$D=8$    & $1_\fft12$   &     &    &&&&&\\  \hline
$D=7$    & $5_\fft12$ &  $4_\fft14$  &  & &&&& \\  \hline
$D=6$    & $15_\fft12$  & $31_\fft14$   &    &&&&& \\  \hline
$D=5$    &  $35_\fft12$  &  $211_\fft14$   &  $271_\fft18$ & $54_\fft18$ 
&&&&  \\  \hline
$D=4$    & $71_\fft12$   & $1001_\fft14$  & $3871_\fft18$
& $\begin{array}{c} 777_\fft18\\ 3136_\fft1{16} \end{array}$ 
& $1332_\fft1{16}$ & $316_\fft1{16}$ & $36_\fft1{16}$ &\\ \hline
$D=3$ & $134_\fft12$ & $3836_\fft14$ & $32088_\fft18$ & 
$\begin{array}{c} 6384_\fft18\\ 82632_\fft1{16}
\end{array}$ & $\begin{array}{c} 49232_\fft1{16}\\ 56928_\fft{1}{32}  
\end{array}$&
$\begin{array}{c} 16376_\fft1{16}\\ 48800_\fft1{32} \end{array}$ 
&$\begin{array}{c} 3120_\fft1{16}\\ 14768_\fft1{32} \end{array}$ &
$\begin{array}{c} 240_\fft1{16}\\ 624_\fft1{32} \end{array}$ \\ \hline
\end{tabular}}
\bigskip
\centerline{Table 5: Multiplicities for domain-wall solutions}
\bigskip

\noindent{Note that the occurrence of large prime factors in some of
the multiplicities in the list is consistent with the fact that these
solutions do not in general form multiplets under any group.}  As
usual, the subscripts on the multiplicities indicate the fractions of
preserved supersymmetry.

   As we observed previously in the case of 2-form and 1-form solutions,
the combination rules for sets of $N$ field strengths whose dilaton vectors 
satisfy the condition (\ref{mdot}) are already encoded in the $N=2$ 
combination rules.  Thus we shall present here in this section the
lists of 2-charge solutions for 0-form field strengths, for $3\le D\le 7$.
We shall also give the lists for the maximal numbers of field strengths
in each dimension, since they lead to the maximal numbers of intersections
that can be achieved in $D=10$ or $D=11$.   The multiplicities for
intermediate numbers of charges are given in Table 5.
     
\bigskip
\noindent{$D=7$}
\bigskip
  
     $N=2$ is the maximum number of charges allowed in $D=7$, and there
is a multiplet of four 2-charge combinations, given by  
\be
\{\F0{ijk\ell}, \cF0{ijk}\}_4\ ,\label{d7n20}
\ee

\bigskip
\noindent{$D=6$}
\bigskip

      In this case, we also have $N_{\rm max}=2$, but with 31
solutions:
\be
\{\F0{ijk\ell}, \cF0{ijk}\}_{20}\ ,\quad
           \{\cF0{ijk}, \cF0{i\ell m}\}_{3}\ ,\quad
           \{\cF0{jik}, \cF0{\ell i m}\}_{8}
\ .\label{d6n20}
\ee
Note that one must be careful, in the case of the fields $\cF0{ijk}$, to
take account of the fact that although they can be taken to be antisymmetric
in $jk$, the index $i$ has a distinguished r\^ole, and furthermore they
are defined only for $i<j$ and $i<k$.

\bigskip
\noindent{$D=5$}
\bigskip

   In five dimensions, $N_{\rm max}$ is equal to 4.  The $N=2$ combinations
are given by
 
\bea
&&\{\F0{ijk\ell}, \F0{ijmn}\}_{45}\ ,\quad
  \{\F0{ijk\ell}, \cF0{ijk}\}_{60}\ ,\quad
  \{\F0{ijk\ell}, \cF0{mjn}\}_{40}\ ,\nn\\
&&\{\cF0{ijk}, \cF0{i\ell m}\}_{18}\ ,\quad 
\{\cF0{jik}, \cF0{\ell i m}\}_{48}\ .
\label{d5n20}
\eea
The combinations for $N=N_{\rm max}=4$ are
\be
\{\F0{ijkm}, \F0{i\ell mn}, \cF0{jkm}, \cF0{\ell mn}\}_{48}\ ,\quad
  \{\cF0{ik\ell}, \cF0{imn}, \cF0{jkm}, \cF0{j\ell n}\}_{6}\ .
   \label{d5n40}
\ee

\bigskip
\noindent{$D=4$}
\bigskip

    In four dimensions, the maximum number of charges is $N_{\rm max}=7$.
The allowed combinations for $N=2$ are given by
\bea
&&\{\F0{ijk\ell}, \F0{ijmn}\}_{315}\ ,\quad
\{\F0{ijk\ell}, \cF0{ijk}\}_{140}\ ,\quad
\{\F0{ijk\ell}, \cF0{mjn}\}_{280}\ ,\nn\\
&&\{\cF0{ijk}, *F_4\}_{35}\ ,\quad
\{\cF0{ijk}, \cF0{i\ell m}\}_{63}\ ,\quad
\{\cF0{jik}, \cF0{\ell im}\}_{168}\ .
\label{d4n20}
\eea
The allowed combinations for $N=N_{\rm max}=7$ are 
\bea
&&\{\F0{ijk\ell}, \F0{ijmn}, \F0{ikmp}, \F0{i\ell np}, 
\F0{jknp}, \F0{j\ell mp}, \F0{k\ell mn}\}_{30}\ ,\nn\\
&&\{\cF0{i\ell m}, \cF0{inp}, \cF0{j\ell n}, \cF0{jmp}, 
\cF0{k\ell p}, \cF0{kmn}, *F_4\}_{6}\ .\label{d4n70}
\eea

\bigskip
\noindent{$D=3$}
\bigskip

    In three dimensions, the maximal allowed number of charges is 
$N_{\rm max}=8$.  The possible combinations for $N=2$ are
\bea 
&&\{\F0{ijk\ell}, \F0{ijmn}\}_{1260}\ ,\quad
\{\F0{ijk\ell}, \cF0{ijk}\}_{280}\ ,\quad
\{\F0{ijk\ell}, \cF0{mjn}\}_{1120}\ ,\nn\\
&& \{\cF0{ijk}, *\F3{\ell}\}_{280}\ ,\quad
\{\F0{ijk\ell}, *\F3i\}_{280}\ ,\quad
\{\cF0{ijk}, \cF0{i\ell m}\}_{168}\ ,\label{d3n20}\\
&&\{\cF0{ijk}, \cF0{i\ell m}\}_{448}\ .\nn
\eea
The combinations for $N=N_{\rm max}=8$ are
\bea
&&\{\F0{ijk\ell}, \F0{ijmn}, \F0{ijpq}, \F0{ikmp}, 
\F0{iknq}, \F0{i\ell mq}, \F0{i\ell np}, *\F3i\}_{240}\ ,\nn\\
&&\{\F0{ijkn}, \F0{ikpq}, \F0{i\ell np}, \F0{imnq}, \cF0{jkn}, \cF0{\ell np}, 
\cF0{mnq}, *\F3i\}_{384} \ .\label{d3n80}\\
&&\{\F0{j\ell nq}, \F0{jmpq}, \F0{kmnq}, \F0{k\ell pq}, \cF0{ijk}, 
\cF0{i \ell m}, \cF0{inp}, *\F3q\}_{240} \ .\nn
\eea

\subsubsection{Comments on 0-form solutions}

\noindent{{\bf 1)}}\ \ \ 
As in the cases of solutions for higher-degree forms that we discussed
previously, again here for 0-forms it is still necessary to show that the
configurations that we have listed do indeed give rise to simple multi-charge
solutions of the full equations of motion of the $D$-dimensional 
supergravities.  In other words, again we have to make sure that interaction
terms in the Lagrangian do not spoil the solutions, by preventing us from
setting to zero all the other fields in the theory.  The Weyl group arguments
that we used previously do not help us here, since the standard CJ supergravity
symmetries of the massless theories are broken by the generalised 
Scherk-Schwarz reductions.  We can, however, still use the other argument that
we presented previously, based on the fact that interaction terms in the
Lagrangian occur if and only if the dilaton vectors of the interacting fields 
satisfy appropriate sum rules.  It is straighforward to verify that for
sets of 0-form field strengths whose dilaton vectors satisfy (\ref{mdot}),
cubic interactions of the form $\chi F_0^\a F_0^\beta$ are forbidden, and
hence the simple $N$-charge 0-form solutions of (\ref{nlag}) are indeed
solutions of the full dimensionally-reduced massive supergravity theories.

\bigskip
\noindent{{\bf 2)}}\ \ \ 
We presented the listings of allowed field strength
combinations for 2-charge 0-form solutions, the $N>2$ charge solutions
can be deduced from these by selecting sets of $N$ fields for which all
pairs satisfy the $N=2$ conditions.  We also presented the field combinations
for $N=N_{\rm max}$ in each dimension.  The intermediate-$N$  cases, although
easily generated in principle by a mechanical process, become complicated
when the multiplicities are large.  We have enumerated all these by
computer, and the multiplicities are presented in Table 5.

\bigskip
\noindent{{\bf 3)}}\ \ \ 
As in the case of solutions involving higher-degree field strengths,
the 0-form solutions with $N\le3$ charges all preserve a fraction
$2^{-N}$ of the original supersymmetry.  For $N=4$, some preserve
$\ft18$ whilst others preserve $\ft1{16}$ of the supersymmetry.  For
$5\le N \le 8$ charges, some solutions preserve $\ft1{16}$ whilst
others preserve $\ft{1}{32}$.
In section 6, we shall study the supersymmetry of all the
$p$-branes, and give precise rules that determine the fraction of
preserved supersymmetry for all multi-charge solutions.  With these
rules, all the multi-charge $p$-branes, and their supersymmetry, will
be derivable purely from the knowledge of the 2-charge solutions.

\section{$D=2$ supergravities and their $p$-brane solutions}

    So far in the paper, our discussions have been restricted to 
supergravities in dimensions $D\ge3$.  As can be seen from (\ref{gfdot}), the
Kaluza-Klein reduction scheme that we have been using degenerates when $D=2$.
This is because we cannot any longer choose to work with a metric that is
in the Einstein frame once we descend to $D=2$.   

\subsection{Kaluza-Klein reduction from $D=3$ to $D=2$}

   We shall make the following choice for the Kaluza-Klein reduction of
the three-dimensional metric:
\be
ds_3^2 = e^\varphi\, ds_2^2 + e^{2\varphi}\, (dz_9 +{\cal A}^{(9)}_1)^2\ ,
\label{3metred}
\ee
where $\varphi$ is the new dilatonic scalar, and ${\cal A}^{(9)}_1$ 
is the new
Kaluza-Klein vector potential.  All other fields in the three-dimensional
theory will still be reduced according to $A_n(x,z_9)\rightarrow A_n(x) +
A_{n-1}(x)\wedge dz_9$.  Thus kinetic terms in $D=3$ will reduce to $D=2$
according to the following rules:
\bea
-\ft1{12} e\, F_3^2 &\longrightarrow& -\ft1{4}e\, e^{-2\varphi}\, 
 F_2^2\ ,\nn\\
-\ft14 e\, F_2^2 &\longrightarrow& -\ft14 e\, F_2^2 -
          \ft12 e\,  e^{-\varphi}\, F_1^2 \ ,\label{sss}\\
-\ft12 e\, F_1^2 &\longrightarrow& -\ft12 e\, e^{\varphi}\, F_1^2 -
          \ft12 e\, F_0^2 \ ,\nn\\
-\ft1{2} e\, F_0^2 &\longrightarrow& -\ft1{2} e\, e^{2\varphi} 
\, F_0^2\ .\nn
\eea
The Einstein-Hilbert and dilaton kinetic terms of $D=3$ reduce according
to
\be 
eR -\ft12 e\, (\del\vec\phi)^2 \longrightarrow
e\, e^\varphi\, R +e\, e^{\varphi}\, (\del\varphi)^2-\ft14 e\, 
e^{2\varphi}\, {\cal F}^2  -\ft12
e\, e^{\varphi}\, (\del\vec\phi)^2 \ .
\ee

        Having established the dimensional reduction rules for all the
fields, we can in principle write down all the $D=2$ supergravity
Lagrangians from the ones in $D=3$.  The 2-form field strength in
$D=2$ is not dynamical and can be dualised to a cosmological term.
There can also exist a massless supergravity in $D=2$, which has
$E_9$ global symmetry, whose Lagrangian is given by
\bea
e^{-1}{\cal L} &=&  e^{\varphi}\, R +  e^\varphi\, (\del\varphi)^2 
-\ft12  e^{\varphi}\,(\del\vec\phi)^2  -\ft12 \sum_{i<j<k\le8}
e^{\vec a_{ijk} \cdot \vec\phi+\varphi}\, (\F1{ijk})^2 \nn\\
&&-\ft12 \sum_{i<j\le 8} e^{\vec a_{ij} \cdot \vec\phi-\varphi}
\, (\F1{ij9})^2  -\ft12 \sum_{i<j\le8} 
e^{\vec b_{ij} \cdot \vec \phi+\varphi} \, (\cF1{ij})^2 \label{d2e9lag}\\
&&-\ft12 \sum_{i\le 8} e^{\vec b_{i} \cdot \vec \phi-\varphi}
\, (\cF1{i9})^2
-\ft1{144}\epsilon^{\mu\nu}\,  \del_\mu A_0^{(ijk)}\, 
\del_\nu dA_0^{(\ell mn)}\, A_0^{(pq9)}\, 
\epsilon_{ijk\ell mnpq}\ ,\nn
\eea
where $\vec a_{ijk}$, $\vec b_{ij}$, $\vec a_{ij}$ and $\vec b_i$ are
the dilaton vectors in three dimensions, given by (\ref{dilatonvec})
and (\ref{gfdot}) with $D=3$.  The field strengths $\F1{ij9}$ and
$\cF1{i9}$ are the dimensional reductions of the three-dimensional 2-forms
$\F2{ij}$ and $\cF2i$ respectively.  All the field strengths are
reduced according to the scheme given in (\ref{sss}), and their
Kaluza-Klein modifications are given by the standard formulae obtained
in \cite{lpsol}.

        Of course, there are numerous massive supergravities in $D=2$, where
the theories contain cosmological terms.

\subsection{Instantons in $D=2$}

         There are two types of $p$-branes in $D=2$, namely instanton
solutions using 1-form field strengths and black hole (domain wall)
solutions using cosmological terms.  The instanton solutions can arise
in massless supergravity in $D=2$, whose bosonic Lagrangian is
given by (\ref{d2e9lag}).    As in the higher-dimensional cases that we
discussed earlier, we may consider a truncated Lagrangian of the form
\bea
{\cal L} &=& e\, e^{\varphi}\, R + e\, e^{\varphi}\, (\del\varphi)^2 
-\ft12 e\, e^\varphi\, (\del\vec\phi)^2 \nn\\
&&+\ft12 e\, \sum_\a e^{\vec c_\a\cdot\vec\phi+\varphi}\, (F^\a)^2 
+\ft12 e\, \sum_a e^{\vec d_a\cdot\vec\phi-\varphi}\, (F^a)^2\label{d2trunc}
\ ,
\eea
where $F^\a = d\chi^\a$ and $F^a=d\chi^a$ are 1-form field strengths. The
kinetic terms for the axions $\chi^\a$ and $\chi^a$ have the opposite sign
to the normal ones in a Lorentzian-signature spacetime.  This is because,
in order to obtain instanton solutions, we need to work with a space
of Euclidean signature.  This unusual sign for the kinetic terms can arise
naturally if one obtains the Euclidean-signature theory in $D=2$ by a 
dimensional reduction from $D=11$ in which the original time coordinate 
becomes one of the compactified directions.

   The equations of motion following from the truncated Lagrangian 
(\ref{d2trunc}) are
\bea
&&R_{\mu\nu}= \nabla_\mu\del_\nu \varphi + \ft12 \del_\mu \vec\phi\cdot 
\del_\nu\vec \phi - \ft12 \sum_\a e^{\vec c_\a\cdot\vec\phi} \,
\del_\mu \chi^\a \, \del_\nu\chi^\a \nn\\
&&\qquad\qquad- \ft12 \sum_a e^{\vec d_a\cdot \vec \phi
-2\varphi}\, \Big( \del_\mu\chi^a \, \del_\nu \chi^a - g_{\mu\nu}\, 
(\del \chi^a)^2\Big)\ ,\nn\\
&&\square\varphi + (\del\varphi)^2 =0\ , \\
&&\square\vec\phi + \del^\mu\varphi\, \del_\mu\vec\phi = 
-\ft12 \sum_\a \vec c_\a\, e^{\vec c_\a\cdot\vec\phi}\, (\del\chi^\a)^2
- \ft12 \sum_a \vec d_a\, e^{\vec d_a\cdot\vec\phi-2\varphi}\, 
(\del\chi^a)^2\ ,\nn\\
&&\nabla^\mu(e^{\vec c_\a\cdot\vec\phi +\varphi}\, \del_\mu\chi^\a )=0
\ ,\qquad
\nabla^\mu(e^{\vec d_a\cdot\vec\phi -\varphi}\, \del_\mu\chi^a )=0
\ .\nn
\eea
Note that the equation of motion for the field $\varphi$ has no
sources involving the axionic fields, and so for extremal instanton
solutions we may just set $\varphi=0$.  (This can only be done after
varying the action, however.)  The field strengths $F^\a$ and $F^a$
are on an equal footing after setting $\varphi=0$, and hence the
supersymmetric solutions from the two types of fields will have the
same structure, and their dilaton vectors must satisfy the same
conditions.  Without loss of generality, we can therefore study the
conditions for the existence of multi-charge solutions using the
fields $F^\a$, and then trivially extend the discussion to include the
$F^a$ fields afterwards.  We find that multi-charge instanton
solutions exist if the dilaton vectors $\vec c_\a$ satisfy $\vec
c_\a\cdot \vec c_\b =4\delta_{\a\b}$, which is precisely the usual
requirement (\ref{mdot}) for multi-charge 1-form solutions.  The $D=2$
multi-instantons are given by
\bea
ds_2^2 &=& dr^2 + r^2\, d\theta^2 \ ,\qquad \varphi=0 \ ,\nn\\
\vec\phi &=& \ft12 \epsilon \, \sum_\a \vec c_\a\, \log H_\a\ ,
\eea
where $H_\a = 1 + |Q_\a|\, \log r$, and $\epsilon = +1$ for electric
instantons, and $\epsilon =-1$ for magnetic instantons.  The axions 
are given by $\chi^\a= H_\a^{-1}$ in the electric case, and $\chi^\a =
Q_\a\, \theta$ in the magnetic case.  (As usual, when we present a truncated
Lagrangian, we choose to make the necessary dualisations so that all the
field strengths $F^\a$ carry electric charges, or all of them carry 
magnetic charges.  In terms of the original fields in (\ref{d2e9lag}), 
some charges may be electric and others magnetic.)  Following similar 
arguments to those given earlier in $D\ge3$, we may verify that the 
multi-instanton solutions of the truncated Lagrangian (\ref{d2trunc}) are
also solutions for the full two-dimensional massless Lagrangian 
(\ref{d2e9lag}).

     We have now established the rules that determine the field
strength configurations for multi-charge instanton solutions in $D=2$,
namely that the associated three-dimensional dilaton vectors $\vec
c_\a$ have to satisfy (\ref{mdot}), $\vec c_\a\cdot \vec c_\b
=4\delta_{\a\b}$.  Pairwise configurations of field strengths in $D=3$
that satisfy this dilaton vector dot-product condition are listed in
(\ref{d3n21}).  In $D=2$, there are additional dyonic solutions of the
second kind, involving dualisations of the field strengths, since the
signs of the dilaton vectors do not affect the conditions $\vec c_\a
\cdot \vec c_\b=4\delta_{\a\b}$.  This gives all the pairs of fields
in $D=2$, from which all the higher $N$-charge solutions can then be
obtained.  Since the $\pm$ choices of the signs of the dilaton vectors
$\vec c_\a$ do not affect the dot product conditions $\vec c_\a \cdot
\vec c_\beta= 4\delta_{\a\b}$, it follows that if a given axion carries
electric charge in one multi-charge solution, there is another
multi-charge solution where instead the given axion carries a magnetic
charge, while all the other axions remain unchanged.  Thus the 
multiplicities for an $N$-charge instanton solution
in $D=2$ are $2^N$ times those listed in Table 4 for $D=3$ solutions.
The maximum number of charges in a given solution is $N_{\rm max}=8$.

The $D=2$ multi-charge instanton solutions can oxidise back to give
either instanton or black hole solutions in $D=3$, which were
classified in section 4.1, or to give intersections of instantons and
black holes in $D=3$.  Note that since $c_\a \cdot (-\vec c_\a) = -4$,
it follows that there can be no dyonic solutions of the first kind in
$D=2$, where a single field strength would carry both electric and
magnetic charge.     

     The supersymmetry of the multi-charge instantons in $D=2$ can be 
established using the procedures that we shall discuss in section 6.

\subsection{Black holes in $D=2$}

      Black hole solutions arise in 2-dimensional massive
supergravities.  There are a total of three categories of multi-charge
black hole solutions in $D=2$.  The first comprises those which are
the vertical dimensional reduction of black holes in $D=3$ or the
double dimensional reduction of strings in $D=3$.  All these solutions
have already been completely classified in section 4 for $D=3$.
The second category comprises multi-charge solutions where some
charges are carried by field strengths that were already 0-forms in
$D=3$, while the rest are carried by 0-forms coming from the
dimensional reduction of 1-forms in $D=3$. These solutions will
oxidise back to intersections of black holes and strings in $D=3$.
The third category comprises multi-charge solutions where one charge
is carried by the Kalazu-Klein vector coming from the $D=3$ to $D=2$
reduction (which is dualised to a cosmological term), and the rest are
carried by the 0-forms that were already 0-forms in $D=3$.  These
solutions will oxidise back to the intersections of strings with a
wave in $D=3$.  To see explicitly how these three categories of
solutions arise in $D=2$, we need to consider the relevant
dimensionally-reduced $D=2$ massive Lagrangians.

    The general class of two-dimensional Lagrangians that we shall be
concerned with take the form
\bea
{\cal L} &=& e\, e^\varphi\, R + e\, e^\varphi\, (\del\varphi)^2
-\ft12 e\, e^\varphi\, (\del\vec\phi)^2 \nn\\
&&-\ft12 e\, e^{2\varphi}\,  \sum_\a m_\a^2 \, e^{\vec c_\a\cdot\vec\phi}
-\ft12 e\, \sum_a \tilde m_a^2 \, e^{\vec d_a \cdot\vec \phi}
-\ft12 e\, m_0^2\, e^{-2\varphi}\ ,\label{mclag}
\eea
where the three kinds of cosmological term arise as follows.  Those
with dilaton vectors $\vec c_\a$ correspond to the reductions of
existing 0-forms in $D=3$.  Those with dilaton vectors $\vec d_a$
correspond to the reductions of 1-forms in $D=3$.  Finally, the last
cosmological term in (\ref{mclag}) comes from the dualisation of the
Kaluza-Klein vector ${\cal A}^{(9)}_1$ in (\ref{3metred}).  It should
be understood here that it is not necessarily the case that all the
cosmological terms displayed in (\ref{mclag}) can coexist
simultaneously, for the reasons that we have already discussed in
section 4.  However, any set of cosmological terms which can be used
to construct multi-charge solutions can be present simultaneously in
the Lagrangian. Thus we will suppose that the various mass parameters
(\it i.e.}\ charges) $m_\a$, $\tilde m_a$ and $m_0$ can be turned on
or off at will, to give whichever permitted non-vanishing set we wish to
consider at any time.  

     The equations of motion following from (\ref{mclag}) are
\bea
&&R_{\mu\nu}= \nabla_\mu\del_\nu\varphi +\ft12 \del_\mu\vec\phi \cdot\del_\nu
\vec\phi +\ft14\Big(\sum_\a m_\a^2\, e^{\vec c_\a\cdot\vec\phi+\varphi}
+\sum_a  \tilde m_a^2\, e^{\vec d_a\cdot\vec\phi-\varphi} -3 m_0^2\,
e^{-3\varphi}\Big) \, g_{\mu\nu}\ ,\nn\\
&&\square\varphi+(\del\varphi)^2 = 
-\ft12 \sum_\a m_\a^2\, e^{\vec c_\a\cdot\vec\phi+\varphi}
-\ft12 \sum_a  \tilde m_a^2\, e^{\vec d_a\cdot\vec\phi-\varphi}
-\ft12 m_0^2 \, e^{-3\varphi}\ ,\label{d2coseq} \\
&&\square\vec\phi + \del^\m\varphi\, \del_\mu\vec\phi = 
\ft12 \sum_\a m_\a^2\, \vec c_\a\, e^{\vec c_\a\cdot\vec\phi+\varphi}
+\ft12 \sum_a  \tilde m_a^2\, \vec d_a\, e^{\vec d_a\cdot\vec\phi-\varphi}
\ .\nn
\eea

     Making the metric ansatz $ds_2^2 = -e^{2A}\, dt^2 + e^{2B}\, dy^2$, 
the equations of motion following from (\ref{d2coseq}) are 
\bea
&&A''+{A'}^2 -A'\, B' + A'\, \varphi' =\ft34 S
-\ft14 \sum_\a S_\a +\ft14 \sum_a \wtd S_a \ ,\nn\\
&&A''+{A'}^2 -A'\, B' +\varphi''-B'\, \varphi' +\ft12 \vec\phi'\cdot
\vec\phi'  = \ft34 S -\ft14 \sum_\a S_\a + \ft14 \sum_a \wtd S_a\ 
\ ,\nn\\
&&\varphi'' +{\varphi'}^2 + A'\, \varphi' - B'\, \varphi' =
-\ft12 S -\ft12 \sum_\a S_\a -\ft12\sum_a \wtd S_a \ ,\nn\\
&&\vec\phi'' +\vec\phi'\, (\varphi'+A'-B') =
\ft12\sum_\a \vec c_\a\, S_\a +\ft12\sum_a \vec d_a\, \wtd S_a
\ ,\label{feq}
\eea
where $S_\a = m_\a^2\, e^{\vec c_\a \cdot \vec\phi + 2B + \varphi}$,
$\wtd S_a = \tilde m_a^2\, e^{\vec d_a \cdot \vec\phi + 2B -\varphi}$ and
$S= m_0^2\, e^{2B-3\varphi}$.
It is straightforward to show that these equations admit two different
classes of black-hole solutions.  Firstly, we can find solutions with
$m_0=0$, of the form
\bea
ds_2^2 &=& -(\prod_a \wtd H_a)^{-1/2}\, (\prod_\a H_\a)^{1/2} \, dt^2
+(\prod_a \wtd H_a)^{1/2}\, (\prod_\a H_\a)^{3/2}\, dy^2\ ,\nn\\
\vec\phi &=& -\ft12 \sum_\a \vec c_\a\, \log H_\a -\ft12
\sum_a \vec d_a\, \log \wtd H_a\ ,\label{d2bh1}\\
\varphi &=& =\ft12 \sum_a \log H_\a + \ft12 \sum_a \log \wtd H_a\ ,\nn
\eea
where $H_\a = 1 + m_\a\, |y|$ and $\wtd H_a=1+ \tilde m_a\, |y|$ are the 
independent
harmonic functions for the charges $m_\a$ and $\tilde m_a$, and the 
dilaton vectors
$\vec c_\a$ and $\vec d_a$ satisfy the relations
\be
\vec c_\a\cdot\vec c_\beta = 4\delta_{\a\beta} + 4\ ,\qquad
\vec d_a\cdot \vec d_b = 4\delta_{ab}\ ,\qquad
\vec c_\a\cdot \vec d_a = 2\ .\label{d2dot1}
\ee
These solutions encompass the first category mentioned above (when all charges
$m_\a=0$ or else when all charges $\tilde m_a=0$), and the second category 
(when charges of both the $m_\a$ type and the $\tilde m_a$ type are 
non-vanishing).  Note that the conditions on $\vec c_\a\cdot \vec c_\b$ and
$\vec d_a\cdot \vec d_b$ in (\ref{d2dot1}) are precisely the usual conditions
(\ref{mdot}) in $D=3$, for dilaton vectors for 0-form fields and 1-form
fields respectively.

     A second class of black-hole solutions to the equations (\ref{feq}) 
can be obtained by setting all the $\tilde m_a$ charges to zero, while having
non-vanishing $m_0$ and non-vanishing $m_\a$ charges, with these latter
being associated with dilaton vectors $\vec c_\a$ that satisfy
\be
\vec c_\a\cdot\vec c_\beta = 4\delta_{\a\beta} + 4\ .\label{d2dot2}
\ee
A simple calculation shows that in this case the solutions take the form
\bea
ds_2^2 &=& -(\prod_\a H_\a)^{1/2}\, H^{-3/2}\, dt^2 +
(\prod_\a H_\a)^{3/2}\, H^{-1/2}\, dy^2\ ,\nn\\
\vec\phi &=& -\ft12 \sum_\a \vec c_\a\, \log H_\a\ ,\qquad
\varphi =\ft12 \sum_\a \log H_\a + \ft12 \log H\ ,\label{d2bh2}
\eea
where the harmonic functions $H_\a=1 +m_\a\, |y|$ are the same as in the 
previous solutions, and $H=1+ m_0\, |y|$ is the harmonic function for the 
Kaluza-Klein charge $m_0$.  Solutions of this kind constitute the third
category that we mentioned at the beginning of this subsection.

\subsubsection{Comments on black holes in $D=2$}

\noindent{{\bf 1)}}\ \ \
There are no simple solutions that involve both $m_0$ and $\tilde m_a$
charges.  This implies that there are no intersections between black
holes and waves in $D=3$.  This can be understood from the fact that
vertical dimensional reductions of black holes in $D=3$ are
necessarily of the Scherk-Schwarz type, where the axions that support
the black hole solutions are linearly proportional to the
compactifying coordinate.  In such Scherk-Schwarz reductions, the
Kaluza-Klein vector becomes massive \cite{clpst}, and hence cannot 
participate in supporting simple multi-charge $p$-brane solutions.

\bigskip
\noindent{{\bf 2)}}\ \ \
Having established the necessary requirements for multi-charge black
hole solutions, we may now enumerate all the possible solutions in
$D=2$.  As we mentioned, all $D=2$ black hole solutions can be oxidised back 
to $D=3$, to become intersecting strings and black holes, together with a
wave when $m_0\ne0$.  Since the criterion for the $D=2$ solution is
expressed in (\ref{d2dot1}) and (\ref{d2dot2}), which are dot product
rules for $D=3$ dilaton vectors, it is more convenient to characterise
the $D=2$ solutions in terms of their $D=3$ fields.  For the first
category, the solutions are fully classified in $D=3$, since these are
just solutions of strings and black holes in $D=3$.  Thus the 2-charge
pairs for these solutions are listed in (\ref{d3n21}) and
(\ref{d3n20}).  For the third category, the solutions are also fully
classified, in that all the string solutions in $D=3$ can intersect
with a three-dimensional wave.  In other words, we can take the
dimensional reduction to $D=2$ of any of the multi-charge string
solutions in $D=3$, and add an extra charge, namely that of the new
Kaluza-Klein vector, together with its associated harmonic function.

      It remains for us to classify the second category of solutions.
In terms of the $D=3$ solutions, these are the intersections of
strings and black holes.  Thus in terms of three-dimensional fields,
the combinations of allowed field configurations involve both 1-form
and 0-form field strengths, with dilaton vectors $\vec d_a$ and $\vec
c_\a$ that satisfy (\ref{d2dot1}).  As in the previous case, the $N=2$
charge solutions encode all the combination rules for $N\ge 3$
solutions.  We find that they are given by
\bea
&&\{*\F2{ij}, *\F3k\}_{168}\ ,\quad\!
\{\F1{ijk}, *\F3{\ell}\}_{280}\ ,\quad\!
\{\F1{ijk}, \F0{ij\ell m}\}_{1680}\ ,\quad\!
\{*\cF2i, \F0{jk\ell m}\}_{280}\ ,\nn\\
&&
\{\cF1{ij}, *\F3i\}_{28}\ ,\quad\!
\{\cF1{ij}, \F0{jk\ell m}\}_{560}\ ,\quad\!
\{*\cF2i, \cF0{jk\ell}\}_{280}\ ,\quad\!
\{*\cF2i, *\F3i\}_{8}\ ,\nn\\
&&
\{*\F2{ij}, \F0{ik\ell m}\}_{1120}\ ,\quad
\{*\F2{ij}, \cF0{ijk}\}_{112}\ ,\quad
\{*\F2{ij}, \cF0{k\ell m}\}_{560}\ ,\label{d3int}\\
&&
\{\F1{ijk}, \cF0{ijk}\}_{56}\ ,\quad\!
\{\F1{ijk}, \cF0{\ell jm}\}_{1120}\ ,\quad\!
\{\cF1{ij}, \cF0{ik\ell}\}_{210}\ ,\quad\!
\{\cF1{ij}, \cF0{kj\ell}\}_{350}\ .\nn
\eea
This gives a total of 6812 possible ways for a black hole to intersect
a string in $D=3$.  In computing the multiplicities,
recall that the $i$ index on $\cF0{ijk}$ must be such that $i<j$ and
$i<k$.

     Thus taken together with the 2-charge combination rules
(\ref{d3n21}), (\ref{d3n20}) and (\ref{d3int}), we now have an
enumeration of all the multi-charge black hole solutions in all three
categories in $D=2$.

\bigskip
\noindent{{\bf 3)}}\ \ \ 
The maximal number of field strengths that can participate in simple
multi-charge solutions is $N_{\rm max}=9$.  This can be achieved by
solutions in the third category, taking the Kaluza-Klein field
strength $\cF29$, which is dualised to a cosmological term $m_0$,
together with the any of the 8-field-strength combinations listed in
(\ref{d3n80}).

\section{Supersymmetry of multi-charge $p$-branes}

    We have classified all the simple multi-charge $p$-brane solutions
in all dimensions $D\ge2$.  We have also seen that this implies a
classification of the associated intersections of $p$-branes, waves
and NUTs in any higher dimension, since the set of intersections in a
given dimension are nothing but the oxidations of all the
lower-dimensional $p$-branes. It is important to establish what
fractions of supersymmetry are preserved by the various solutions.
Supersymmetry is fully preserved by the Kaluza-Klein reduction
procedure itself.  This means that the fraction of the original
supersymmetry that is preserved by a particular lower-dimensional
$p$-brane is the same as the fraction that is preserved by its
oxidation to any higher dimension.

    One way, albeit very clumsy, to determine the fraction of
supersymmetry that is preserved by a $D$-dimensional $p$-brane is to
examine the supersymmetry transformation rules of the $D$-dimensional
maximal supergravity, and look for Killing spinors in the background
of the $p$-brane, since these correspond to components of unbroken
supersymmetry.  This method is especially unattractive in low
dimensions, where the multiplicities of the possible non-vanishing
field strengths becomes very large.  Furthermore, it requires that one
know the explicit transformation rules for the maximal supergravity in
question, and these have not in general been obtained for the many
massive supergravities.  An easier method is to oxidise the
lower-dimensional $p$-brane to $D=10$ or $D=11$.  At least in the case
of solutions supported only by the antisymmetric tensors of $D=10$ or
$D=11$, this gives a simpler system of equations for the Killing
spinors.

     Fortunately, there is a much easier procedure for determining the
fraction of supersymmetry that is preserved by any $p$-brane solution.
All that is necessary is to construct the Nester form $N^{AB}$ in
$D=11$, which arises as the anti-commutator of $D=11$ supercharges,
$\{Q_{\epsilon_1}, Q_{\epsilon_2}\}= \int_{\del\Sigma} N^{AB}\,
d\Sigma_{AB}$, and dimensionally reduce it to $D$ dimensions.  Since
it is a purely bosonic object, this is a very simple procedure.  The
\bog matrix ${\cal M}$, defined by $\epsilon_1^\dagger {\cal
M}\epsilon_2 = \int_{\del\Sigma} N^{0r}\, r^{\td d+1}\, d\Omega_{\td
d+1}$ in the asymptotic $r\rightarrow \infty$ limit, is then a
$32\times 32$ Hermitean matrix each of whose zero eigenvalues
corresponds to a component of unbroken supersymmetry.  It is given in
terms of the mass per unit $p$-volume, and the charges, by
\cite{lpsol}
\bea
{\cal M} &=& m\oneone + u\, \Gamma_{012} + u_i\, \Gamma_{01i} +
\ft12 u_{ij}\, \Gamma_{0ij}+ \ft16 u_{ijk}\, \Gamma_{ijk} +
\ft1{24} u_{ijk\ell}\, \Gamma_{ijk\ell} \nn\\
&&+v\,\Gamma_{\hat1\hat2\hat3\hat4\hat5} + v_i\,
\Gamma_{\hat1\hat2\hat3\hat4i}+\ft12 v_{ij}\, \Gamma_{\hat1\hat2\hat3ij}
+ \ft16 v_{ijk}\, \Gamma_{\hat1\hat2ijk} +
\ft1{24} v_{ijk\ell}\, \Gamma_{\hat1 ijk\ell} \label{genbog}\\
&&+p_i \Gamma_{0i}+ \ft12 p_{ij}\, \Gamma_{ij} +\ft16 p_{ijk}\, \Gamma_{ijk}
+ q_i\, \Gamma_{\hat1\hat2\hat3 i}
+\ft12 q_{ij}\, \Gamma_{\hat1\hat2ij}+\ft16 q_{ijk}\, \Gamma_{\hat1ijk}
\ .\nn
\eea
The prefactors of the $\Gamma$ matrices are the various electric and magnetic
charges associated with the various field strengths in $D$ dimensions,
according to the following scheme:
\be
\hbox{
\begin{tabular}{ccccccccc}
 & $F_4$ & $\F3i$ & $\F2{ij}$ & $\F1{ijk}$ & $\F0{ijk\ell}$ &
$\cF2i$ & $\cF1{ij}$ & $\cF0{ijk}$ \\  
Electric & $u$ & $u_i$ & $u_{ij}$ & $u_{ijk}$ & $u_{ijk\ell}$ & 
$p_i$ & $p_{ij}$ & $p_{ijk}$ \\
Magnetic & $v$ & $v_i$ & $v_{ij}$ & $v_{ijk}$ & $v_{ijk\ell}$ & 
$q_i$ & $q_{ij}$ & $q_{ijk}$ \\
\end{tabular}} \label{charges}
\ee
where $u$'s and $p$'s are electric charges, and $v$'s and $q$'s are
magnetic charges.  For a given degree $n$ of antisymmetric tensor
field strength, only the terms with the corresponding charges, as
indicated in (\ref{charges}), will occur in (\ref{genbog}).  The indices 
$0, 1,\ldots$ run over the dimension of the $p$-brane world-volume,
$\hat1,\hat2,\ldots$ run over the transverse space of the $y^m$
coordinates, and $i,j,\ldots$ run over the dimensions that were
compactified in the Kaluza-Klein reduction from 11 to $D$ dimensions.
Note that the electric charges $u_{ijk\ell}$ and $p_{ijk}$ would be
associated with $(-2)$-branes, which presumably have no meaning.  All
the other $\Gamma$-matrix combinations appearing in (\ref{genbog}) are
Hermitean, with the exception of the $\Gamma_{ijk}$ and $\Gamma_{ij}$
combinations, which are anti-Hermitean.  However, these are associated
with instantons, whose existence requires that the ``spacetime'' have
Euclidean signature.  There will then be an extra $i$ factor coming from the
electric charges in such cases, which restores the hermiticity of the
\bog matrix.

     Determining the supersymmetry of any $p$-brane solution in any
dimension $D$ is now reduced to a matter of elementary algebra.  All
that is needed is to substitute the relevant $N$ charges of the
solution, and its mass $m$, into the \bog matrix (\ref{genbog}), and
then to evaluate its 32 eigenvalues.  The number $k$ of zero
eigenvalues implies that a corresponding fraction $k/32$ of the
original supersymmetry is preserved by the solution.  It is very easy
to see that any single-charge solution will give 16 zero eigenvalues,
and hence will preserve $\ft12$ the supersymmetry.  Similarly, any
2-charge solution will preserve $\ft14$, and any 3-charge solution
will preserve $\ft18$. (Of course only sets of charges that correspond
to combinations of field strengths allowed by the dilaton-vector
conditions (\ref{mdot}) are to be considered.)  For $N\ge4$ charges,
as we have indicated in earlier discussion, the fraction of preserved
supersymmetry in general depends on the particular combinations of
field strengths involved.  For example, although all 4-charge 2-form
solutions preserve $\ft18$ of the supersymmetry, in the case of 1-form
or 0-form 4-charge solutions, some preserve $\ft18$ whilst others
preserve $\ft1{16}$.

      It should be noted that since the kinetic terms for field
strengths are quadratic, there are actually $2^N$ different
possibilities for the signs of the charges $Q_\a$ in a simple
$N$-charge $p$-brane solution, where the mass is still given as the
sum of the $N$ positive quantities $|Q_\a|$.  It turns out that when
$N\ge4$ these $2^N$ solutions, although equivalent from a purely
bosonic point of view, can have different properties as far as
supersymmetry is concerned.  (This is because the field strengths
enter linearly in the supersymmetry transformation rules.)  To be
precise, for an $N$-charge $p$-brane solution that can preserve a
fraction $2^{-\tilde N}$ of the supersymmetry, then of the $2^N$
possible sign choices for the charges, $2^{\tilde N}$ will give
solutions that do in fact preserve the fraction $2^{-\tilde N}$ of
supersymmetry, and the remaining $2^N-2^{\tilde N}$ sign choices will
give solutions that preserve no supersymmetry.  In other words, if an
$N$-charge solution preserves a fraction $2^{-N}$ of the supersymmetry
(\ie the successive introduction of each of the $N$ charges breaks a
half of the remaining supersymmetry), then the sign of each of the $N$
charges is immaterial.  If now a new charge can be introduced to give
an $(N+1)$-charge solution that does not further break the
supersymmetry, then the same charge introduced with the opposite sign
will cause {\it all} the supersymmetry to be broken.  In other words,
if a $p$-brane breaks $\ft12$ of the remaining supersymmetry, so will
the anti-$p$-brane.  On the other hand, if a $p$-brane does not break
any further supersymmetry, then the anti-$p$-brane will break it all,
and {\it vice versa}.  We shall present a proof of these statements
below.  Since, as we have noted previously, the smallest non-vanishing
fraction of preserved supersymmetry in any $p$-brane solution is
$\ft1{32}$, it follows that we always have $\tilde N\le5$.
Consequently, for simple $N$-charge $p$-branes where $N$ is large, the
overwhelming majority break {\it all} the supersymmetry, even though
they are extremal and related merely by sign changes of their charges
to solutions that are supersymmetric.  For example, if we consider
8-charge solutions that preserve $\ft1{32}$ of the supersymmetry, then
of the 256 possible choices for the signs of the eight charges, 32
will give supersymmetric solutions, while 224 will give solutions that
break all the supersymmetry.  It is worth remarking that although
these solutions are non-supersymmetric, there is still a no-force
condition between the individual charges.

      To see in detail how the supersymmetry depends on the choice of
charges, we now give a complete analysis based on the \bog matrix.

\subsection{The \bog matrix and supersymmetry}

   To begin, we note that since the mass $m$ per unit $p$-volume for a 
simple $p$-brane with $N$ charges 
$Q_\a$ is given by $m=\sum_\a |Q_\a|$, it follows from (\ref{genbog})
that its \bog matrix is just the sum of the individual \bog matrices
for each of its associated single-charge components: ${\cal M} = 
\sum_\a {\cal M}_\a$, where
\be
{\cal M}_\a = |Q_\a| + Q_\a \, \Gamma_{(\a)}\ ,\label{gamma}
\ee
and we denote by $\Gamma_{(\a)}$ the particular unit-strength $\Gamma$-matrix
product associated with the charge $Q_\a$, as given by (\ref{genbog}).
One can easily show that the individual \bog matrices
commute, $[{\cal M}_\a, {\cal M}_\b]=0$.  

     We may now give an elementary proof of the
previous statements about the fractions of preserved supersymmetry.   
Since the individual ${\cal M}_\a$ matrices commute, it 
follows that they may be simultaneously diagonalised.  Thus the set of
$k$ Killing spinors $\epsilon^a$ ($a=1,\ldots,k$) for an $N$-charge $p$-brane
solution, defined by ${\cal M}
\epsilon^a=0$, can also be chosen to be eigenstates of the individual
${\cal M}_\a$ matrices,
\be
{\cal M}_\a\, \epsilon^a = \lambda_\a^a\, \epsilon^a\ .
\ee
Now the eigenvalues of each ${\cal M}_\a$ must all be non-negative
\cite{klopp}, since otherwise there would be naked singularities in
the solution.  Thus in particular we must have $\lambda_\a^a\ge0$, and
so we have $0={\cal M}\epsilon^a=\sum_\a {\cal M}_\a\, \epsilon^a =
\sum_\a \lambda_\a^a\, \epsilon^a$, implying that $\lambda_\a^a=0$,
and hence the Killing spinors $\epsilon^a$ all satisfy ${\cal
M}_\a\, \epsilon^a=0$.  Each ${\cal M}_\a$ has the form given by
(\ref{gamma}).  Since ${\rm tr}\, \Gamma_{(\a)}=0$ and
$(\Gamma_{(\a)})^2=1$, it follows that each ${\cal M}_\a$ has sixteen
zero eigenvalues and sixteen non-zero eigenvalues $2|Q_\a|$.  In
an $N$-charge solution, there will be $N$ individual \bog matrices
${\cal M}_\a$, with their associated $\Gamma_{(\a)}$ matrices.  We
must now distinguish between two cases.  If the $\Gamma_{(\a)}$
matrices are all independent, in the sense that none of them can be
written in terms of products of any of the rest, then it follows that
the number of zero eigenvalues in ${\cal M}$ is $2^{5-N}$.  This can
be seen from the fact that in the diagonalised basis, any such set of
$N$ independent $\Gamma_{(\a)}$ matrices can be chosen from the set
\bea
&&s_1=\sigma\times\oneone\times\oneone\times\oneone\times\oneone \ ,\qquad
s_2=\oneone\times\sigma\times\oneone\times\oneone\times\oneone \ ,\qquad
s_3=\oneone\times\oneone\times\sigma\times\oneone\times\oneone \ ,\nn\\
&&s_4=\oneone\times\oneone\times\oneone\times\sigma\times\oneone \ ,\qquad
s_5=\oneone\times\oneone\times\oneone\times\oneone\times\sigma \ ,\label{sis}
\eea
where $\sigma$ is the Pauli matrix $\sigma_3$, and $\oneone$ denotes the
$2\times2$ unit matrix.  The eigenvalues $\mu$ of ${\cal M}$ are therefore 
given by
\be
\mu= \sum_\a |Q_\a| \pm Q_1\pm Q_2\cdots \pm Q_N\ ,
\ee
where the sign choices are all independent.  In particular, a 
fraction $2^{-N}$ of the 32 eigenvalues are zero.   Note that it is 
manifest, for example from (\ref{sis}), that the maximum possible number of 
independent $\Gamma_{(\a)}$ matrices is 5, leading to a fraction $\ft1{32}$
of preserved supersymmetry.

    If not all the $\Gamma_{(\a)}$ matrices are independent, in the sense that
some can be expressed as products of others, then let us assume that $\tilde N$
of them {\it are} independent.  It then follows that the solution either
preserves a fraction $2^{-\tilde N}$ of the supersymmetry, or it preserves
none at all.  Which of these occurs depends upon the signs of the charges.  
To see this, consider an $N$-charge
solution that preserves a fraction $k/32$ of the supersymmetry, with Killing 
spinors $\epsilon^a$.  If we now introduce an $(N+1)$'th charge $Q_{N+1}$, 
with its individual \bog matrix ${\cal M}_{N+1} = |Q_{N+1}| + Q_{N+1}\,
\Gamma_{(N+1)}$, where
$\Gamma_{(N+1)}$ is expressible as a product of some of the previous
$\Gamma_{(\a)}$ matrices, $\Gamma_{(N+1)} = \prod_{\b \in
\{\a\}}\Gamma_{(\b)}$, then ${\cal M}_{N+1}$ can be expressed as
\be
{\cal M}_{N+1} = |Q_{N+1}| + Q_{N+1}\, \prod_{\b\in\{\a\}}\fft1{Q_\b} \,
({\cal M}_\b -|Q_\b|)\ .
\ee
Thus for one sign choice for $Q_{N+1}$, the matrix ${\cal M}_{N+1}$ is
expressed as polynomials in the ${\cal M}_\b$ with no term of zero'th order
in the ${\cal M}_\b$.  For this sign choice, the original Killing spinors 
$\epsilon^a$ of the $N$-charge solution will also satisfy ${\cal M}_{N+1}
\epsilon^a=0$, and hence they will all continue to be Killing spinors in
the $(N+1)$-charge solution.  In this case, there is no further breaking of
supersymmetry when the $(N+1)$'th charge is introduced.  On the other hand,
if the $Q_{N+1}$ charge is chosen with the opposite sign, the previous
Killing spinors $\epsilon^a$ will satisfy ${\cal M}_{N+1}\epsilon^a=2
|Q_{N+1}|\, \epsilon^a$, and thus all the supersymmetry will be broken
when the $(N+1)$'th charge is introduced.  (It is worth remarking
that in simple mult-charge solutions, if a gamma matrix $\Gamma_{(\a)}$
is not indepedent, it is always a product of three other gamma
matrices associated with the charges in this solution.  This explains
why $N$-charge solutions with $N\le 3$ always preserve $2^{-N}$ of the
supersymmetry, and it is only when $N\ge 4$ that the complications
set in.)

    Iterating the above argument, we see that an $N$-charge solution for 
which $\tilde N$ of the $\Gamma_{(\a)}$ matrices are independent will 
preserve a fraction $2^{-\tilde N}$ of the supersymmetry for $2^{\tilde N}$
out of the total of $2^N$ sign choices for the charges, and it will preserve
no supersymmetry for the remaining sign choices.

   Having understood that a given $N$-charge extremal $p$-brane may
have versions that break all the supersymmetry, as well as versions that
preserve a fraction $2^{-\tilde N}$ of the supersymmetry, we shall
in general assume that the sign choices for the charges are made so
that the supersymmetric versions are obtained, unless we have specific
reasons for wanting to discuss the non-supersymmetric versions.

\subsection{Comments on supersymmetry}

\bigskip
\noindent{{\bf 1)}}\ \ \ 
In the previous section, we showed
that the fraction of preserved supersymmetry is given by $2^{-\tilde N}$, 
where $\tilde N$ is the number of independent $\Gamma_{(\a)}$ matrices
associated with the $N$ charges.  This provides us with a simple way
to determine the fraction of preserved supersymmetry without needing to
perform any explicit computation of the eigenvalues of the \bog matrix.
The $\Gamma$-matrix products in the \bog matrix are governed by
the internal coordinate indices of the charges of the $p$-brane
solution.  All one needs to do is to identify the maximal subset of 
$\tilde N$ independent $\Gamma_{(\a)}$ matrices, implying that a fraction
$2^{-\tilde N}$ of the supersymmetry is preserved.

     To illustrate this, let us consider 4-charge 1-form solutions in
$D=3$, for which the allowed field strength combinations are listed in
(\ref{d5n41}).  We shall consider two specific examples, using the
fields $\{\F1{123},*\F2{34},*\cF24,\F1{12}\}$, or instead
$\{\F1{123},*\F2{34},*\cF24, \F1{56}\}$.  (Note that the star indicate
that the 2-forms are dualised to 1-forms in $D=3$.) We can use these
to give 4-charge black hole solutions, where the unstarred field
strengths carry magnetic charges, and the starred ones carry electric
charges.  From (\ref{genbog}), the \bog matrix for the first example
is
\be 
{\cal M} =  \sum_\a |Q_\a| + Q_1\, \Gamma_{\hat1\hat2 \td 1\td 2 \td 3} + 
Q_2\, \Gamma_{0\td 3 \td 4} + Q_3\, \Gamma_{0\td 4} + 
Q_4\, \Gamma_{\hat1\hat2\td 1\td 2}\ , 
\ee 
where we denote explicit internal index values by $\td 1, \td
2,\ldots$.  It is
easy to verify that any of the $\Gamma_{(\a)}$ matrices can be expressed as a
product of the other three, implying that this 4-charge black hole
preserves $2^{-3}=\ft18$ of the original supersymmetry.  (Of course, in line
with our earlier discussion, eight of the possible sign choices for the
four charges will give solutions with this $\ft18$ supersymmetry, while the 
other eight choices will give solutions with no supersymmetry.)

     The \bog matrix for the second example is given by 
\be 
{\cal M} =  \sum_\a |Q_\a| + Q_1\, \Gamma_{\hat1\hat2 \td 1\td 2 \td 3} + 
Q_2\, \Gamma_{0\td 3 \td 4} + Q_3\, \Gamma_{0\td 4} + 
Q_4\, \Gamma_{\hat1\hat2\td 5\td 6}\ , 
\ee 
and in this case we see that no product of two or three of the
$\Gamma$ matrices can give the fourth.  In this case, therefore, the
solution preserves $2^{-4}=\ft1{16}$ of the supersymmetry, and the
sign choices for the charges have no effect on the supersymmetry.
Note that the only difference between these two examples is that the
fourth field strength has different internal indices. This illustrates
the fact that one can read off the supersymmetry of a $p$-brane just
by inspecting the internal indices on the field strengths that support
the solution.

\bigskip
\noindent{{\bf 2)}}\ \ \
Some observations related to the above were presented in
\cite{bdejs,roo} for the case of intersections purely involving
M-branes or D-branes, but no waves or NUTs.  These involved looking at
the contributions of the antisymmetric tensor terms in the Killing
spinor equations in $D=11$ or $D=10$, and recognising certain
projection-operator combinations involving the associated $\Gamma$
matrices that lead to halvings of the numbers of Killing spinors.  Our
proofs in this section provide a complete analysis of the supersymmetry
for all simple multi-charge $p$-branes, including the cases where
there are waves and NUTs in the higher-dimensional intersections, and
for all choices of the signs of the charges.

\bigskip
\noindent{{\bf 3)}}\ \ \ 
The supersymmetry of any simple multi-charge $p$-brane can also be found
by means of the following rule, which we shall refer to as
``casting out charges,'' and is formulated in the following five steps:
\bigskip

\noindent{1)}\  {\it Using the rules given in Tables (1a) and (1b), list the
world-volume and the transverse space directions $z^i$ of the compactification
coordinates for each of the $N$ single-charge solutions obtained by setting
in turn all but one of the $N$ charges to zero.}

\noindent{2)} \ {\it If any charge is such that its removal contracts the
total list of
world-volume $z^i$ directions or transverse-space $z^i$ directions, then
delete this charge, and accumulate a factor of $\ft12$.}

\noindent{3)}\ {\it If the removal of no single charge can achieve the
above contraction, then delete an arbitrarily chosen charge that is
associated with a $z$ coordinate that appears only twice,\footnote{Note
that this can always be done.} and accumulate a factor of 1.}

\noindent{4)}\  {\it Repeat the above steps on the remaining $N-1$ charges,
until eventually all have been removed.}

\noindent{5)}\  {\it The product of the accumulated $\ft12$ and 1 factors
gives the  fraction of preserved supersymmetry for the original $N$-charge
solution.}
\bigskip

    We obtained the fractions of preserved supersymmetry using the
above rules for all the $p$-branes, and we verified that they are
consistent with the explicit computations of the eigenvalues of the
\bog matrix.  The results for the preserved supersymmetry are
summarised in Tables 3, 4, and 5.  The advantage of the casting out
charges rule is that the determination of the supersymmetry
of a simple multi-charge solution can be done by inspection
of the configuration of participating field strengths, rather than by
computing the eigenvalues of a $32\times 32$ matrix.

     We may illustrate this ``casting out charges'' rule with some
examples.  First, consider the example of the dyonic string given by
(\ref{dyonic}).  From Table (1a), we see that the compactification
coordinates $z^i$ divide between the world-volume and the transverse
space as follows:
\be
\hbox{
\begin{tabular}{ccc}
 & World-volume & Transverse Space \\
$\F31$ & 1 & 2, 3, 4, 5  \\
$*\F31$ & 2,3,4,5 & 1 \\
\end{tabular}} \label{cocdyon}
\ee
We see that either of the charges satisfies rule 2, and so we cast out one
of them, accumulating a factor of $\ft12$.  Casting out the remaining one
gives another $\ft12$, implying that the fraction of supersymmetry preserved
by the dyonic string is $\ft12 \times\ft12 = \ft14$.

     For a more complicated example, consider the two 4-charge black
holes solutions in $D=3$ discussed in comment (1) above. For the first
case, we see from Tables (1a) and (1b) that we have
\be \hbox{
\begin{tabular}{ccc}
 & World-volume & Transverse Space \\
$\F1{123}$ &4,5,6,7,8   & 1,2,3  \\
$*\F2{34}$ &3,4     & 1,2,5,6,7,8 \\
$*\cF24$   &4       & 1,2,3,5,6,7,8 \\
$\cF1{12}$  &3,4,5,6,7,8 & 1,2\\
\end{tabular}} \label{cocd51}
\ee
There is no single charge that can be deleted so as to contract the
total list of world-volume or transverse-space directions.  Thus we
apply rule 3, and arbitrarily delete a charge, say number four,
accumulating a factor of 1.  Deleting the first charge the removes 5
(and 6) from the list of world-volume directions, and so we accumulate
a factor of $\ft12$ by rule 2.  Deleting the second charge now removes
3 from the list of world-volume directions, and so we accumulate
another $\ft12$ factor.  Deletion of the last two charges accumulates
two more factors of $\ft12$, leading to the conclusion that this
4-charge black hole preserves $1\times \ft12 \times\ft12\times \ft12 =
\ft18$ of the supersymmetry, in agreement with our previous derivation
using the \bog matrix.

    Taking the other example for a 4-charge black hole, we have
\be
\hbox{
\begin{tabular}{ccc}
 & World-volume & Transverse Space \\
$\F1{123}$ &4,5,6,7,8   & 1,2,3  \\
$*\F2{34}$ &3,4     & 1,2,5,6,7,8 \\
$*\cF24$   &4       & 1,2,3,5,6,7,8 \\
$\cF1{56}$  &1,2,3,4,7,8 & 5,6\\
\end{tabular}} \label{cocd52}
\ee
Here, deleting the first charge removes 6 from the total list of
world-volume directions, giving a $\ft12$ factor.  Deleting the third
charge then removes 3 from the transverse space, giving another factor
of $\ft12$.  Deleting the remaining two charges gives two more factors
of $\ft12$.  Therefore, this 4-charge black hole solution preserves a
fraction $\ft12\times \ft12 \times \ft12\times \ft12 = \ft1{16}$ of
the supersymmetry, again agreeing with the result we previously
derived using the \bog matrix.

       Both the 4-charge $D=3$ black hole solutions (\ref{cocd51}) and
(\ref{cocd52}) can be interpreted as intersections of a membrane, a
5-brane, a wave and a NUT in $D=11$.  The different fractions of
supersymmetry that the two solutions preserve is related to the
different ways in which these eleven-dimensional objects intersect
each other.  Although in both examples, the full list of
transverse-space directions is the same, the list of world-volume
directions in (\ref{cocd52}) contains $z^5$ and $z^6$, which are not
contained in the world-volume list for (\ref{cocd51}).  This shows
that more supersymmetry is broken by intersecting M-branes, waves or
NUTs when they occupy more directions in either the world-volume or
the transverse space.  In other words, the intersecting objects tend to
preserve less supersymmetry if they are spread over more directions in
either the world-volume or the transverse space, and conversely they
preserve more supersymmetry if they are confined to fewer directions.

   Let us consider another pair of examples, namely two configurations
of quadruple intersections of 5-branes in $D=11$.  They can both be
reduced to $D=4$, where they become 4-charge magnetic strings, with
field strengths given by $\{\F1{123}, \F1{145}, \F1{246}, \F1{167}\}$,
and $\{\F1{123}, \F1{145}, \F1{246}, \F1{356}\}$ respectively.
Clearly the second solution describes quadruple intersections of
5-branes that fit into fewer transverse directions, in that the total
list of transverse-space coordinates is smaller ($z^7$ is not in the
list of transverse-space directions of any of the four 5-branes).  In
the first solution, the total list of transverse-space coordinates is
larger, since it includes $z^7$ as well.  And indeed, the first
solution preserves only $\ft1{16}$ of the supersymmetry, whilst the
second preserves $\ft18$.

         Thus the ``casting out charges'' rule exploits the relation
between the spacetime geometrical structures of intersecting objects
and the supersymmetries they preserve.

\section{Harmonic intersections in M-theory or type II strings}

      In the previous sections, we have obtained a complete
classification of all the simple $N$-charge extremal $p$-branes in all
dimensions $2\le D\le 11$, and given procedures for determining the
fractions of the original $D=11$ supersymmetry that each of them
preserves.  We have also shown how they can be oxidised back to $D=11$
(or $D=10$), by retracing the steps of the Kaluza-Klein dimensional
reductions to the $D$-dimensional maximal supergravities.  The
oxidation rules are very simple, and are summarised in (\ref{11s}) and
Tables (1a) and (1b) (for $D=11$), and in (\ref{10ns}), (\ref{10rr})
and Tables (2a) and (2b) (for $D=10$).  In $D=11$ or $D=10$, the
lower-dimensional $N$-charge $p$-branes become intersections of
$p$-branes, waves and NUTs (\ie monopoles).  Thus the classification
of the lower-dimensional multi-charge $p$-branes also provides a
classification of the intersections in M-theory or string theory where
the harmonic functions depend only on the coordinates transverse to
all the individual world-volumes.  The classification of more general
intersections (\ie including those that do not dimensionally reduce to 
$p$-branes) has been studied in \cite{bdejs,roo,bdejs2}.

     Just as the rules for $N$-charge $p$-branes in lower dimensions
are encoded in the rules for 2-charge $p$-branes, so the rules for the
intersections of $N$ objects are encoded in the basic rules for the
intersections of all possible pairs of objects.  In $D=11$, there are
four basic objects, namely the membrane, 5-brane, wave and NUT.  (The
NUTs subdivide into three sub-categories, NUT$_i$ given by equations
(\ref{d3b1}--\ref{d3b3}).)  The solutions for all allowed pairwise
intersections are characterised by the overlap of the spatial
world-volume directions of the two basic objects.  For example, the
intersection of a membrane and a 5-brane is given in (\ref{25int}),
and we see that they have the one common spatial world-volume
coordinate $x$.  This particular example came from the oxidation of a
dyonic string in $D=6$; one can easily verify that all other 2-charge
$p$-branes that oxidise to an intersection of a membrane with a
5-brane exhibit the same feature, of an overlap of one spatial
world-volume coordinate.  Thus, knowing that the intersection of a
membrane and a 5-brane always shares one common spatial world-volume
coordinate, the solution is uniquely determined, up to the relabelling
of coordinates.  We may summarise all the required information for
constructing arbitrary multiple intersections of all the basic objects
in a table listing the world-volume overlaps of all allowed pairs.
This is given in Table 6.

\bigskip\bigskip
\centerline{
\begin{tabular}{|c||c|c|c|c|c|c|}\hline
 & Membrane & 5-brane & Wave & NUT$_1$ & NUT$_2$ & NUT$_3$ \\ \hline\hline
Membrane& 0 & 1 & 1 & 2 & 2 & 2   \\ \hline
5-brane &   & 3 & 1 & 5 & 5 & 5   \\ \hline 
Wave    &   &   & --& 1 & 1 & 1   \\ \hline
NUT$_1$ &   &   &   &-- &-- &--   \\ \hline
NUT$_2$ &   &   &   &   & 4 & 4   \\ \hline   
NUT$_3$ &   &   &   &   &   & 4   \\ \hline   
\end{tabular}}
\bigskip

\centerline{Table 6. Spatial world-volume overlaps of harmonic
intersections in $D=11$}
\bigskip

     It should be emphasised that the rules in Table 6 are for 
pairwise intersections that can dimensionally reduce to $p$-branes.  Such
pairwise intersection rules in the type IIA theory can be
obtained by dimensional reduction of the ones for $D=11$.  This was
done in \cite{bdejs,roo,bdejs2}.  From these intersection rules, one
can derive all the possible intersections of $N$ basic objects.  Their
supersymmetry can be determined by the ``casting out charges'' rule
described in section 6.2.

\subsection{Comments on intersections}

\noindent{{\bf 1)}}\ \ \ 
The maximum possible number of intersecting $M$-branes is $N=8$,
achieved as the intersection of seven 5-branes and one membrane.  The
solutions can be dimensionally reduced to $D=3$, where they become
8-charge string solutions in maximal massive supergravities, as
discussed in section 4.2.  For intersections of M-branes to be reduced
instead to solutions of {\it massless} supergravities in lower
dimensions, the maximum number of intersections is $N=7$, namely four
5-branes and three membranes, or seven 5-branes.  The solutions can be
reduced to 7-charge black holes in $D=3$.

      The maximum possible number of intersecting objects is $N=9$
\cite{bdejs2}.  For example, the above eight intersecting M-branes can
further intersect with a wave.  These solutions reduce to
two-dimensional 9-charge black holes, as we discussed in section 5.
There can also be intersections of one membrane, four 5-branes,
one wave and three NUTs.

\bigskip
\noindent{{\bf 2)}}\ \ \ 
In order to oxidise the solutions we obtained in the previous sections
to the type IIA theory, we need to split the internal index $i=(1,
\a)$ and distinguish between the NS-NS and R-R fields, as explained in
section 3.2.  The perturbative solutions in lower dimensions are those
which are supported only by NS-NS field strengths carrying electric
charges.  In Minkowskian-signature spacetime, we note that all the
$N\ge 3$ charge solutions are non-perturbative.  The only perturbative
$N=2$ charge solutions are those described by intersections between
the NS-NS string and a wave.
  
       In particular, this implies that two NS-NS strings cannot
intersect in Minkowskian-signature spacetime.  This can be seen from
the fact that the spatial world-volume overlap of two intersecting
membranes, as listed in Table 6, is zero, and hence the only way to
obtain intersections of two NS-NS strings is by compactifying the
intersecting membranes on the time coordinate, giving a $D=10$
Euclidean-signature space.  In fact intersections in string theory are
typically a non-perturbative phenomenon.

\bigskip
\noindent{{\bf 3)}}\ \ \
The type IIA and type IIB theories are related by T-duality, and so
all the intersections in the type IIA theory can be mapped into
intersections in the type IIB theory by this duality transformation.
If we wish to oxidise the lower-dimensional solutions to the $D=10$
type IIB theory directly, we need to split the internal index as
$i=(1,2,\a)$, since M-theory compactified on a two-torus is T-dual to
the type IIB theory compactified on a circle, and the fields of the
two theories can then be identified by T-duality transformations.

         For example, there are five dyonic string solutions in $D=6$,
using the field strengths $\{\F3i, *\F3i\}$.  All five of them become
intersections of a membrane and a 5-brane in $D=11$.  Oxidising
instead to the $D=10$ type IIA theory, the $i=1$ solution gives an
intersection of a string and a 5-brane; while for $i=2,3,4,5$, they
are intersections of a membrane and a 4-brane.  Oxidising back to
$D=10$ type IIB instead, the $i=1$ solution gives an intersection of
an NS-NS string and an NS-NS 5-brane; for $i=2$, it is the
intersection of a R-R string and a R-R 5-brane; and for $i=3,4,5$,
they are nothing but intersections of two self-dual 3-branes.

         This illustrates that once all the lower-dimensional simple
multi-charge $p$-brane solutions are classified, all the associated
intersections in higher dimensions are also completely understood.
The oxidations of the lower-dimensional solutions follow a few simple
rules, given in section 3.

\section{U multiplets and non-harmonic intersections}

       In the previous sections, we obtained all the simple
multi-charge solutions in maximal supergravity theories.  These
solutions can be viewed as harmonic intersections of $p$-branes, waves
and NUTs, in M-theory or type II strings.  In lower dimensional
supergravities, there are in general global symmetries that can be
used to generate multiplets from any given solution.  In particular,
in massless maximal supergravities, there are $E_{11-D}$ CJ global
symmetry groups.  Acting on the simple solutions we obtained in the
previous sections, we obtain full multiplets of solutions (see
\cite{clps} for a discussion of the analogous spectrum-generating
symmetries at the quantum level).  These can
also be oxidised back to $D=10$ or $D=11$, where they acquire
interpretations as intersections of the basic objects.  However, in
this case, the number of harmonic functions involved in the
intersections is less than the number of basic objects that are
involved.  We shall refer to these as ``non-harmonic intersections.''

     To find simple examples, we shall study $p$-brane multiplets of
an $SL(2,\R)$ global symmetry.   In $D=9$, the two Kaluza-Klein vectors
${\cal A}^{(1)}_1$ and ${\cal A}^{(2)}_1$ form a doublet under the
$SL(2,\R)$ factor of the $GL(2,\R)$ global symmetry group.  The relevant 
bosonic Lagrangian is given by
\bea
e^{-1}{\cal L} &=&R -\ft12 (\del \phi)^2 + \ft12(\del\varphi)^2
-\ft12(\del\chi)^2 e^{-2\phi}\nn\\
&&-\ft14 e^{-\phi +3\varphi/\sqrt7}(\cF21)^2 
-\ft14 e^{\phi +3\varphi/\sqrt7}(\cF22)^2\ ,\label{d92flag}
\eea
where $\chi=-{\cal A}_0^{(12)}$, $\cF21=d\td{\cal A}_1^{(1)} -
\chi d\td{\cal A}_1^{(2)}\equiv
d{\cal A}_1^{(1)} - d\chi\wedge {\cal A}_1^{(2)}$ 
and $\cF22=d\td{\cal A}_1^{(2)} \equiv d{\cal A}_1^{(2)}$.   The dilatonic
scalar fields $\phi$ and $\varphi$ are related to the usual $\phi_1$ and 
$\phi_2$ fields appearing in (\ref{dgenlag}) as follows \cite{lpsweyl}:
\be
\pmatrix{\phi\cr \varphi} = 
\pmatrix{\ft34 & -\ft{\sqrt7}{4} \cr
         -\ft{\sqrt7}{4} & -\ft34}
\pmatrix{\phi_1\cr\phi_2}\ .
\ee
The Lagrangian (\ref{d92flag}) is invariant under the $SL(2,\R)$ 
transformations
\bea
\tau \longrightarrow \tau'=\fft{a\tau + b}{c \tau +d}\ ,\quad
\pmatrix{ \td{\cal A}_1^{(1)}\cr \td{\cal A}_1^{(2)}}
\longrightarrow
\pmatrix{ {{\cal A}_1^{(1)}}'\cr {{\cal A}_1^{(2)}}'}
=\pmatrix{a & b\cr c & d} 
\pmatrix{ \td{\cal A}_1^{(1)}\cr \td{\cal A}_1^{(2)}}\ ,\label{sl2r}
\eea
where $\tau=\chi + ie^{\phi}$ and $ad-bc=1$.   Starting from the simple
single-charge black hole solution supported by the field strength $\cF21$,
\bea
&& ds_9^2 =-H^{-\ft67}\, dt^2 + H^{\ft17}\, d\vec y^2 \ ,\nn\\
&&e^{\phi} = H^{-\ft12}\ ,\quad
e^{\varphi}= H^{\ft3{2\sqrt7}}\ ,\quad
\chi =0\ ,\quad {\cal A}_1^{(1)} = H^{-1}\,dt\ ,
\eea
we make an $SL(2,\R)$ transformation using (\ref{sl2r}) to obtain a new 
solution, where the metric and the scalar $\varphi$ are unchanged, but 
the other fields become
\bea
&&e^{\phi} = \fft{H^{\ft12}}{c^2 + d^2 H}\ ,\qquad
\chi = -\fft{ac+bd H}{c^2 + d^2 H}\ ,\nn\\
&& {\cal A}_1^{(1)} = \fft{d}{c^2 +d^2 H}\, dt\ ,\qquad
{\cal A}_1^{(2)} = cH^{-1}\, dt\ .
\eea
It is straightforward to oxidise this solution back to $D=10$ and $D=11$,
leading to the metrics
\bea
ds_{10}^2\!\!\!&=&\!\!\! -(c^2+d^2 H)^{\ft18} H^{-1}\, dt^2 +
(c^2 +d^2H)^{-\ft78} H(dz_2 + {\cal A}^{(2)}_1)^2 +
(c^2 + d^2 H)^{\ft18}\, d\vec y^2\ ,\nn\\
ds_{11}^2\!\!\! &=&\!\!\! -H^{-1} dt^2 + (c^2 + d^2H)^{-1} H\, (dz_2 +
{\cal A}_1^{(2)})^2 \nn\\
&&+ (c^2 + d^2H) (dz_1 + {\cal A}_1^{(1)} - \chi dz_2)^2
+ d\vec y^2\ ,\label{wavewave}
\eea
The $SL(2,\R)$ transformation interpolates between two basic objects, namely 
a wave and a D0-brane in $D=10$, or between two waves in $D=11$.  Thus
the above
two metrics describe non-harmonic intersections of a wave and a D0-brane 
in $D=10$, and two waves in $D=11$.  Of course since they are simply related
to the oxidations of a simple single-charge black-hole by an $SL(2,\R)$ 
transformation, the transformed $D=9$ solution and its oxidations all 
preserve the same fraction $\ft12$ of the supersymmetry as does the simple 
$D=9$ solution itself.    In the above example, we have considered
the electric black hole solutions.  If instead, we consider magnetic
5-brane solutions in $D=9$, then they will oxidise to become non-harmonic
intersections of two NUTs in $D=11$.

    For another example, consider the string solutions in $D=9$, which
again form a doublet under $SL(2,\R)$.  The relevant part of the $D=9$
Lagrangian is
\bea
e^{-1}{\cal L} &=&R -\ft12 (\del \phi)^2 + \ft12(\del\varphi)^2
-\ft12(\del\chi)^2 e^{-2\phi}\nn\\
&&-\ft1{12} e^{\phi -\varphi/\sqrt7}(\F31)^2 
-\ft1{12} e^{-\phi +\varphi/\sqrt7}(\F32)^2\ ,\label{d93flag}
\eea
where $\F31=dA_2^{(1)}$ and $\F32= dA_2^{(2)} + \chi\, dA_2^{(1)}$.  This is
invariant under $SL(2,\R)$ transformations where $\phi$ and $\chi$ transform
as in (\ref{sl2r}), and the 2-form potentials transform in the 
contragedient fashion
\be
\pmatrix{ A_2^{(1)}\cr  A_2^{(2)}}
\longrightarrow
\pmatrix{ {A_2^{(1)}}'\cr {A_2^{(2)}}'}
=\pmatrix{d & -c\cr -b & a} 
\pmatrix{ A_2^{(1)}\cr A_2^{(2)}}\ ,\label{sl2r3}
\ee

     Starting from a simple single-charge string solution using the 
field strength $\F31$, and applying the $SL(2,\R)$ transformation, 
we obtain the more general string solution
\bea
&&ds_9^2 = H^{-\ft57}\, dx^\mu \, dx_\mu + H^{\ft27}\, d\vec y^2\ ,
\qquad e^\varphi= H^{-\ft1{2\sqrt7}}\ ,\nn\\
&&e^{\phi} = \fft{H^{\ft12}}{d^2 + c^2 H}\ ,\qquad
\chi = -\fft{bd+ ac H}{d^2 + c^2 H}\ ,\nn\\
&& A_2^{(1)} = (d^2x\wedge dH^{-1})\, d \ ,\qquad
A_2^{(2)} = -(d^2x\wedge dH^{-1})\, b\ .
\eea
It is straightforward to oxidise this to $D=10$ and $D=11$, where we obtain
the metrics
\bea
ds_{10}^2 &=& H^{-\ft34}\, (d^2 + c^2 H)^{\ft18}\, dx^\mu \, dx_\mu
+ H^{\ft14}\, (d^2 + c^2 H)^{\ft18}\, d\vec y^2 \nn\\
&&+ H^{\ft14}\, (d^2 + c^2 H)^{-\ft78}\, dz_2^2\ ,\nn\\
ds_{11}^2 &=& H^{-\ft23}\, dx^\m\, dx_\mu + H^{\ft13}\, d\vec y^2 
+ H^{\ft13}\, (d^2 + c^2 H)^{-1}\, dz_2^2\nn\\
&&+ H^{-\ft23}\, (d^2 + c^2 H) \, (dz_1 -\chi \, dz_2)^2\ .
\eea
In $D=10$, this solution interpolates between a string and a membrane.  In
$D=11$, it interpolates between two membranes.  In all cases, the solution
preserves $\ft12$ of the supersymmetry.  Again, if we instead oxidise
the magnetic 4-brane solutions in $D=9$, they will become non-harmonic
intersections of 5-branes in $D=11$.

        Finally, let us consider a dyonic membrane in $D=8$, where
the relevant part of the Lagrangian is given by
\be
e^{-1}{\cal L} =R -\ft12 (\del \phi)^2 + \ft12(\del\varphi)^2
-\ft12(\del\chi)^2 e^{-2\phi} - \ft1{48}e^{\phi} F_4^2 + \ft1{48}
\chi F_4\cdot *F_4\ ,
\ee
with $\chi=A_0^{(123)}$ and $\phi= \vec a\cdot \vec \phi$.  The dyonic
solution was obtained in \cite{ilpt}, and is given by
\bea
&&ds^2_8= H^{-\ft12}\, dx^\mu\,dx_\mu +
        H^{\ft12}\, d\vec y^2\ ,\nn\\
&&e^{\phi} = \fft{H^{\ft12}}{d^2 + c^2 H}\ ,\qquad
\chi = \fft{bd+ ac H}{d^2 + c^2 H}\ ,\\
&& F_4 = (d^3x\wedge dH^{-1})\,d - H^{\ft12}\,  *(d^3x\wedge
dH^{-1})\,c\ ,\nn
\eea
where $a$, $b$, $c$ and $d$ are the parameters of the $SL(2,\R)$ factor in
the $SL(3,\R)\times SL(2,\R)$ global symmetry group, satisfying $ad-bc=1$.
Performing the oxdations to $D=10$ and $D=11$, we obtain the metrics
\bea
ds_{10}^2&=& H^{-\ft58} (d^2 + c^2 H)^{\ft14}\, dx^\mu\,dx_\mu +
             H^{\ft38} (d^2 + c^2H)^{\ft14} d\vec y^2\nn\\
&& + H^{\ft38} (d^2 + c^2H)^{-\ft34} (dz_2^2 + dz_3^2)\ ,\\
ds_{11}^2&=&H^{-\ft23} (d^2 + c^2 H)^{\ft13}\, dx^\mu\,dx_\mu +
             H^{\ft13} (d^2 + c^2H)^{\ft13} d\vec y^2\nn\\
&& + H^{\ft13} (d^2 + c^2H)^{-\ft23} (dz_1^2+dz_2^2 + dz_3^2)\ .
\eea
In $D=10$, this solution interpolates between a membrane and a 4-brane.  In
$D=11$, it interpolates between a membrane and 5-brane.  In all cases, the 
solution preserves $\ft12$ of the supersymmetry.  

\subsection{Further comments}

\noindent{{\bf 1)}}\ \ \ 
All the 2-charge solutions (simple or non-simple) in lower dimensions
can be characterised by the dilaton vectors of the two field strengths
that are involved.  Defining 
\be
\Delta = \ft14 (\vec c_a + \vec c_\b)^2 +
\fft{2(n-1)(D-n-1)}{D-2}\ , 
\ee
where $\vec c_\a$ and $\vec c_\b$ are the dilaton vectors and $n$ is
the degree of the field strengths, {\it all} the 2-charge solutions
can be categorised into three types, namely $\Delta =1, 2$ and 3.  The
$\Delta =2$ configurations give rise to simple 2-charge solutions, as
discussed earlier, where the solutions involve two independent
harmonic functions.  All the $\Delta=3$ configurations give rise to
solutions that are $SL(2,\R)$ rotations of simple single-charge
solutions, and hence the solutions involve only one harmonic function.
(Note that $\Delta=3$ solutions arise only in supergravities with
global symmetries that contain $SL(2,\R)$ subgroups.) The $\Delta=1$
type solutions were discussed in \cite{lpsln}, where the equations of
motion were cast into the form of 1-dimensional $SL(3,\R)$ Toda
equations; these solutions involve no harmonic functions.  The masses
(at the self-dual point), the charges, and the eigenvalues of \bog
matrix for the above three types of 2-charge solutions are given by
\bea
\Delta=1:&& m=(Q_1^{2/3} + Q_2^{2/3})^{3/2}\ ,\qquad                
\mu = m\pm \sqrt{Q_1^2 + Q_2^2} \ ,\nn \\
\Delta=2:&& m=|Q_1| + |Q_2|\ ,\qquad\quad\quad\ \mu = m \pm Q_1 \pm Q_2\ ,\\
\Delta=3:&& m=\sqrt{Q_1^2 + Q_2^2}\ ,\qquad\quad\ \ \,\,\,
\mu = m \pm \sqrt{Q_1^2 + Q_2^2}\ .\nn
\eea
Thus the $\Delta=3$ solutions preserve $\ft12$ of the supersymmetry,
and can be viewed as bound states with positive binding energy.
The $\Delta=2$ solutions preserve $\ft14$ of the supersymmetry,
and can be viewed as bound states of zero binding energy.  The eigenvalues
of \bog matrix for $\Delta=1$ solutions are all positive definite, and
hence all the supersymmetry is broken.  They can be viewed as bound
states with negative binding energy \cite{gk,lpsln}.

         Having obtained all structures for $p$-brane solutions for
all possible pairs of field strengths, it is straightforward to
generalise to multi-field strength solutions for all possible sets of
field-strength configurations.  For example, in an $N$-field-strength
solution, if there are $\tilde N$ field strengths that are pairwise of
the $\Delta =2$ type, and the rest are of the $\Delta=3$ type, then
this $N$-charge solution is a U-duality transformation of a simple
$\tilde N$-charge solution.

\bigskip
\noindent{{\bf 2)}}\ \ \ 
Note that as we listed in Table 6, there are no harmonic intersections
of two waves.  The example of the $SL(2,\R)$ multiplets of black hole
solutions in $D=9$ suggests that there can, however, be non-harmonic
intersections of waves.  Similarly, there can be no world-volume
spatial overlap in harmonic intersecting membranes, however, as we saw
earlier, for non-harmonic intersections the world-volume spatial
overlap can be 1.  These suggest that the intersections that are
harmonically impossible can be supplemented by non-harmonic
intersections.  The classification of all possible pair-wise
intersections in $D=11$ is of course subsumed by the classification of
all possible pairwise 2-charge solutions described in comment 2.  This
leads to two more types of possible intersections in $D=11$, namely
non-harmonic but supersymmetric intersections ($\Delta=3$), 
presented in table 7,
and non-supersymmetric intersections ($\Delta=1$) presented in Table 8.

\bigskip\bigskip
\centerline{
\begin{tabular}{|c||c|c|c|c|c|c|}\hline
 & Membrane & 5-brane & Wave & NUT$_1$ & NUT$_2$ & NUT$_3$ \\ \hline\hline
Membrane& 1 & 2 & 0 & 1 & 1 & 1   \\ \hline
5-brane &   & 4 & 0 & 4 & 4 & 4   \\ \hline 
Wave    &   &   & 0& -- & 0 & 0   \\ \hline
NUT$_1$ &   &   &   &5 &-- &--   \\ \hline
NUT$_2$ &   &   &   &   & 5 & 5   \\ \hline   
NUT$_3$ &   &   &   &   &   & 5   \\ \hline   
\end{tabular}}
\bigskip

\centerline{Table 7. Spatial world-volume overlaps of non-harmonic
intersections in $D=11$}
\bigskip

\bigskip\bigskip
\centerline{
\begin{tabular}{|c||c|c|c|c|c|c|}\hline
 & Membrane & 5-brane & Wave & NUT$_1$ & NUT$_2$ & NUT$_3$ \\ \hline\hline
Membrane& -- & 0 & -- & -- & 1 & 1   \\ \hline
5-brane &   & -- & -- & -- & 4 & 4   \\ \hline 
Wave    &   &   & --& 0 & 0 & 0   \\ \hline
NUT$_1$ &   &   &   &-- &-- &--   \\ \hline
NUT$_2$ &   &   &   &   & 5 & 5   \\ \hline   
NUT$_3$ &   &   &   &   &   & 5   \\ \hline   
\end{tabular}}
\bigskip

\centerline{Table 8. Spatial world-volume overlaps of non-supersymmetric
intersections in $D=11$}
\bigskip

    Tables 6, 7 and 8 give to all possible pairwise intersections in
$D=11$ that can dimensionally reduce to $p$-branes.

\section{Conclusion}

     In this paper, we have performed a classification of all simple
$N$-charge $p$-brane solutions in the massless and massive
supergravities that arise from the toroidal dimensional reduction of
$D=11$ supergravity.  These solutions are extremal and expressed in
terms of $N$ independent harmonic functions. (Note that they all admit
non-extremal generalisations where the mass becomes an additional
parameter, independent of the charges \cite{dlp}.) We have also
discussed in detail the procedures for determining the fractions of
supersymmetry that are preserved by each such extremal solution.
Included in this discussion was the question of how the supersymmetry
fractions are affected by the possible sign choices for the charges.

     A by-product of the classification of lower-dimensional
$p$-branes is a classification of certain kinds of intersections in
$D=10$ or $D=11$, namely those where the harmonic functions all depend
on a common set of transverse coordinates.  In fact there are distinct
advantages in classifying them in terms of the lower-dimensional
$p$-branes, since it then becomes much easier to study the multiplets
of solutions that are related by U-duality transformations.  There are
also other kinds of intersections \cite{kr1,bbj,gkt} that do not
dimensionally reduce to $p$-branes, although they can still give rise
to some lower-dimensional solutions.  Their significance in string
theory is less clear. However the lower-dimensional solutions are
supersymmetric, and it would be interesting to give a classification
of them, if they do indeed play a r\^ole in the spectrum of the
string.

     We also discussed the structure of the multiplets that are
generated by acting with U-duality transformations on the simple
multi-charge solutions.  In particular, any supersymmetric solution
involving two charges is either a simple 2-charge solution, which
preserves $\ft14$ of the supersymmetry, or it is an $SL(2,\R)$
rotation of a simple single-charge solution, which preserves $\ft12$
of the supersymmetry.  (The $SL(2,\R)$ is in general a subgroup of the
global symmetry group of the supergravity theory.)  A third kind of
2-charge solution, which preserves no supersymmetry, can also arise in
certain special cases \cite{gk,lpsln,lmmp}.  Any pair of field
strengths of equal degree will give rise to solutions of one of these
three types.

\section*{Acknowlegement}

        H.L. and C.N.P. are grateful for hospitality at SISSA, Trieste,
T.A.T. is grateful for hospitality at the ICTP, Trieste and H.L. and
K.W.X. are grateful for hospitality at TAMU, College Station.

\appendix
\section{$(N\ge3)$-charge 1-form solutions} 

       In this appendix, we give all the 1-form field strength
configurations for $(N\ge3)$ charge solutions for $D\ge 2$.  Together
with the single-charge and 2-charge solutions listed in section 4.1,
we have all the simple multi-charge solutions using 1-form field
strengths in $D\ge 2$.  The $(N\ge3)$-charge solutions arise in $D\le
6$.  Note that all the $N=3,4'$ solutions preserve $\ft18$ of the
supersymmetry and all the $N=4,5,6,7,8$ solutions preserve $\ft1{16}$.

\bigskip
\noindent{$D=6$}
\bigskip

       In this dimension, we have $N_{\rm max} = 4'$, The field strength 
configurations are
given by
\bea
N=3:&& \{ F_1^{(ijk)}, F_1^{(i\ell m)}, {\cal F}_1^{(jk)} \}_{30}
\ ,\quad \{F_1^{(ijk)}, {\cal F}_1^{(ij)}, {\cal F}_1^{(\ell m)}
\}_{30}\ ,\label{d6n31}\\
N=4':&& \{ F_1^{(ijk)}, F_1^{(i\ell m)}, {\cal F}_1^{jk},
{\cal F}_1^{(\ell m)} \}_{15}\ .\label{d6n41}
\eea

\bigskip
\noindent{$D=5$}
\bigskip

            As in the case of $D=6$, we have $N_{\rm max} =4'$ in
$D=5$:
\bea
N=3:\!\!\!\!&&\!\!\!\! \{ F_1^{(ijk)}, F_1^{(i\ell m)}, 
F_1^{(j\ell n)} \}_{120}\ ,\quad
\{ F_1^{(ijk)}, F_1^{(i\ell m)}, {\cal F}_1^{(jk)} \}_{180}
\ ,\quad \{F_1^{(ijk)}, {\cal F}_1^{(ij)}, {\cal F}_1^{(\ell m)}
\}_{180}\ ,\nonumber\\
\!\!\!\!&&\!\!\!\! \{ {\cal F}_1^{(ij)}, {\cal F}_1^{(k\ell)}, 
{\cal F}_1^{(mn)} \}_{15}
\ ,\quad \{ *F_4, {\cal F}_1^{(ij)}, {\cal F}_1^{(\ell m)} \}_{45}
\ ,\label{d5n31}\\
N=4':\!\!\!\!&&\!\!\!\! \{ F_1^{(ijk)}, F_1^{(i\ell m)}, F_1^{(j\ell n)}, 
F_1^{(kmn)} \}_{30}\ ,\quad
\{ F_1^{(ijk)}, F_1^{(i\ell m)}, {\cal F}_1^{jk},
{\cal F}_1^{(\ell m)} \}_{90}\ .\nonumber\\
&&\{*F_4, {\cal F}_1^{(ij)}, {\cal F}_1^{(k\ell)}, {\cal F}_1^{(mn)} \}_{30}
\label{d5n41}
\eea

\bigskip
\noindent{$D=4$}
\bigskip

        For $N=3$, we have $M=4095$, given by
\bea
&&\{\F1{ijk}, \F1{i\ell m}, *\F3{i}\}_{315}\ ,\quad
  \{\F1{ijk}, \F1{i\ell m}, \F1{inp}\}_{105}\ ,\quad
  \{\F1{ijk}, \F1{i\ell m}, \F1{j\ell n}\}_{840}\ ,\nn\\
&&\{\F1{ijk}, \F1{i\ell m}, \cF1{jk}\}_{630}\ ,\quad
  \{\F1{ijk}, \F1{i\ell m}, \cF1{np}\}_{315}\ ,\quad
  \{\F1{ijk}, \cF1{ij}, *\F3{k}\}_{105}\ ,\nn\\
&&\{\F1{ijk}, \cF1{\ell m}, *\F3k\}_{630}\ ,\quad
  \{\F1{ijk}, \cF1{ij}, \cF1{\ell m}\}_{630}\ ,\quad
  \{\F1{ijk}, \cF1{\ell m}, \cF1{pq}\}_{105}\ ,\nn\\
&&\{\cF1{ij}, \cF1{k\ell}, *\F3m\}_{315}\ ,\quad
  \{\cF1{ij}, \cF1{k\ell}, \cF1{mn}\}_{105}\ .\label{d4n31}
\eea

      For $N=4'$, we have $M=945$, given by
\bea
&&\{\F1{ijk}, \F1{i\ell m}, \F1{j\ell n}, \F1{kmn}\}_{210}\ ,\quad
  \{\F1{ijk}, \F1{i\ell m}, \cF1{np}, *\F3i\}_{315}\ ,\nn\\
&&\{\F1{ijk}, \F1{i\ell m}, \cF1{jk}, \cF1{\ell m}\}_{315}\ ,\quad
  \{\cF1{ij}, \cF1{k\ell}, \cF1{mn}, *\F3p\}_{105}\ ,\label{d4n4'1}
\eea

        For $N=4$ solutions,we have $M=3780$, given by
\bea
&&\{\F1{ijk}, \F1{i\ell m}, \F1{inp}, *\F3{i}\}_{105}\ ,\quad
  \{\F1{ijk}, \F1{i\ell m}, \F1{inp}, \F1{j\ell n}_{840}\ ,\nn\\
&&\{\F1{ijk}, \F1{i\ell m}, \F1{inp}, \cF1{jk}\}_{315}\ ,\quad
  \{\F1{ijk}, \F1{i\ell m}, \cF1{jk}, *\F3i\}_{630}\ ,\nn\\
&&\{\F1{ijk}, \F1{i\ell m}, \cF1{jk}, \cF1{np}\}_{630}\ ,\quad
  \{\F1{ijk}, \cF1{ij}, \cF1{\ell m}, *\F3k\}_{630}\ ,\nn\\
&&\{\F1{ijk}, \cF1{\ell m}, \cF1{np}, *\F3i\}_{315}\ ,\quad
  \{\F1{ijk}, \cF1{ij}, \cF1{\ell m}, \cF1{np}\}_{315}\ .
\label{d4n41}
\eea

    The detail for the $N=5$ solutions is given by
\bea
&&N=5\ ,\qquad M=2835\ ,\nn\\
&&\{\F1{ijk}, \F1{i\ell m}, \F1{inp}, \F1{j\ell n}, \F1{jmp}\}_{630}
  \ ,\quad
  \{\F1{ijk}, \F1{i\ell m}, \F1{inp}, \cF1{jk}, *\F1{i} \}_{315}
  \ ,\nn\\
&&\{\F1{ijk}, \F1{i\ell m}, \F1{inp}, \cF1{jk}, \cF1{\ell m}\}_{315}
  \ ,\quad
  \{\F1{ijk}, \F1{i\ell m}, \cF1{jk}, \cF1{\ell m}, *\F1{i}\}_{315}
  \ ,\nn\\
&&\{\F1{ijk}, \F1{i\ell m}, \cF1{jk}, \cF1{np}, *\F1{i}\}_{630}
  \ ,\quad
  \{\F1{ijk}, \F1{i\ell m}, \cF1{jk}, \cF1{\ell m}, \cF1{np}\}_{315}
  \ ,\nn\\
&&\{\F1{ijk}, \cF1{ij}, \cF1{\ell m}, \cF1{np}, *\F3{k}\}_{315}
  \ ,\label{d4n51}
\eea

        There are 945 cases of $N=6$ solutions, given by
\bea
&&N=6\ ,\qquad M=945\ ,\nn\\
&&\{\F1{ijk}, \F1{i\ell m}, \F1{inp}, \F1{j\ell n}, \F1{jmp},
       \F1{k\ell p}\}_{210}\ ,\nn\\
&&\{\F1{ijk}, \F1{i\ell m}, \F1{inp}, \cF1{jk}, \cF1{jk}, 
        \cF1{\ell m}, *\F3{i}\}_{315}\ ,\nn\\
&&\{\F1{ijk}, \F1{i\ell m}, \F1{inp}, \cF1{jk}, \cF1{\ell m},
         \cF1{np} \}_{105}\ ,\nn\\
&&\{\F1{ijk}, \F1{i\ell m}, \cF1{jk}, \cF1{\ell m},\cF1{np}, 
         *\F1{i}\}_{315} \ ,\label{d4n61}
\eea

          Finally in $D=4$, there are a total of 135 of $N=7$
solutions, given by
\bea
&&N=7\ ,\qquad M=135\ ,\nn\\
&&\{\F1{ijk}, \F1{i\ell m}, \F1{inp}, \F1{j\ell n}, \F1{jmp},
       \F1{k\ell p}, \F1{kmn} \}_{30}\ ,\nn\\
&&\{\F1{ijk}, \F1{i\ell m}, \F1{inp}, \cF1{jk}, \cF1{\ell m},
         \cF1{np}, *\F3{i} \}_{105}\ ,\label{d4n71}
\eea

\bigskip
\noindent{$D=3$}
\bigskip

      The $N=3$ solutions have a multiplicity 37800, with 23 different
field strength configurations, given by
\bea
&&N=3\ ,\qquad M=37800\ ,\nonumber\\
&&\{ *F_2^{(ij)}, *F_2^{(k\ell)}, *F_2^{(mn)} \}_{420}\ ,\quad
  \{F_1^{(ijk)}, *F_2^{(i\ell)}, *F_2^{(jm)} \}_{3360}\ ,\quad
  \{F_1^{(ijk)}, *F_2^{(i\ell)}, *{\cal F}_1^{(\ell)} \}_{840}
  \ ,\nonumber\\
&&\{F_1^{(ijk)}, F_1^{(i\ell m)}, *F_2^{(in)} \}_{2520}\ ,\quad
  \{F_1^{(ijk)}, F_1^{(i\ell m)}, *F_2^{(i\ell)} \}_{3360}\ ,\quad
  \{F_1^{(ijk)}, F_1^{(i\ell m)}, F_1^{(inp)} \}_{840}\ ,\nonumber\\
&&\{F_1^{(ijk)}, F_1^{(i\ell m)}, F_1^{(j\ell n)} \}_{3360}\ ,\quad
  \{F_1^{(ijk)}, F_1^{(i\ell m)}, *{\cal F}_2^{(n)} \}_{2520}\ ,\quad
  \{F_1^{(ijk)}, F_1^{(i\ell m)}, {\cal F}_1^{(jk)} \}_{1680}
  \ ,\nonumber\\
&&\{F_1^{(ijk)}, F_1^{(i\ell m)}, {\cal F}_1^{(np)} \}_{2520}\ ,\quad
  \{F_1^{(ijk)}, {\cal F}_1^{(ij)}, *F_2^{(k\ell)} \}_{840}\ ,\quad
  \{F_1^{(ijk)}, {\cal F}_1^{(mn)}, *F_2^{(k\ell)} \}_{5040}
  \ ,\nonumber\\
&&\{F_1^{(ijk)}, {\cal F}_1^{(ij)}, *{\cal F}_2^{(\ell)} \}_{840}
  \ ,\quad
  \{F_1^{(ijk)}, {\cal F}_1^{(mn)}, *{\cal F}_2^{(\ell)} \}_{1680}
  \ ,\quad
  \{F_1^{(ijk)}, {\cal F}_1^{(ij)}, {\cal F}_1^{(\ell m)} \}_{1680}
  \ ,\nonumber\\
&&\{F_1^{(ijk)}, {\cal F}_1^{(\ell m)}, {\cal F}_1^{(np)} \}_{840}
  \ ,\quad
  \{{\cal F}_1^{(ij)}, *F_2^{(ij)}, *F_2^{(\ell m)} \}_{420}\ ,\quad
  \{{\cal F}_1^{(ij)}, *F_2^{(k\ell)}, *F_2^{(mn)} \}_{1260}\ ,
  \nonumber\\
&&\{ {\cal F}_1^{(ij)}, *F_2^{(k\ell)}, *{\cal F}_2^{(k)} \}_{840}\ ,
   \quad
  \{ {\cal F}_1^{(ij)}, {\cal F}_1^{(k\ell)}, *F_2^{(ij)} \}_{420}
   \ ,\quad
  \{{\cal F}_1^{(ij)}, {\cal F}_1^{(k\ell)}, *F_2^{(mn)} \}_{1260}
   \ ,\nonumber\\
&&\{{\cal F}_1^{(ij)}, {\cal F}_1^{(k\ell)}, *{\cal F}_2^{(m)}
   \}_{420}\ ,\quad
  \{{\cal F}_1^{(ij)}, {\cal F}_1^{(k\ell)}, {\cal F}_1^{(mn)}
  \}_{420}\ ,\label{d3n31}
\eea

        The $N=4'$ results are the following:
\bea
&&N=4'\ ,\qquad M=9450\ :\nn\\
&&\{F_1^{(ijk)}, F_1^{(i\ell m)}, *F_2^{(j\ell)}, *F_2^{(km)}\}_{1680}
  \ ,\quad
  \{F_1^{(ijk)}, F_1^{(i\ell m)}, F_1^{(j\ell 6)}, F_1^{(kmn)}\}_{840}
  \ ,\nn\\
&&\{F_1^{(ijk)}, F_1^{(i\ell m)},F_1^{(inp)}, 
            *{\cal F}_2^{(q)}\}_{840}\ ,\quad
  \{F_1^{(ijk)}, F_1^{(i\ell m)}, {\cal F}_1^{(np)},
            *F_2^{(iq)} \}_{2520}\ ,\nn\\
&&\{F_1^{(ijk)}, F_1^{(i\ell m)}, {\cal F}_1^{(jk)}, {\cal F}_1^{(\ell
                    m)}\}_{840}\ ,\quad
  \{F_1^{(ijk)}, {\cal F}_1^{(ij)}, *F_2^{(k\ell)},
           *{\cal F}_2^{(\ell)}\}_{840}\ ,\nn\\
&&\{F_1^{(ijk)}, {\cal F}_1^{(\ell m)}, {\cal F}_1^{(np)},
             *{\cal F}_2^{(q)} \}_{840}\ ,\quad 
  \{ \cF1{ij}, *\F2{k\ell}, *\F2{mn}, *\F2{pq} \}_{420}\ ,\nn\\
&&\{\cF1{ij}, \cF1{k\ell}, *\F2{ij}, *\F2{k\ell} \}_{210}\ ,\quad
  \{\cF1{ij}, \cF1{k\ell}, \cF1{mn}, *\F2{pq} \}_{420}\ .
  \label{d3n4'1}
\eea 

      The $N=4$ results are the following:
\bea
&&N=4\ ,\qquad M=113400\ :\nn\\
&&\{*\F2{ij}, *\F2{k\ell}, *\F2{mn}, *\F2{pq} \}_{105}\ ,\quad
  \{\F1{ijk}, *\F2{i\ell}, *\F2{jm}, *\F2{kn} \}_{3360}\ ,\nn\\
&&\{\F1{ijk}, \F1{i\ell m}, *\F2{in}, *\F2{j\ell}\}_{10080}\ ,\quad
  \{\F1{ijk}, \F1{i\ell m}, *\F2{in}, *\cF2{n}\}_{2520}\ ,\nn\\
&&\{\F1{ijk}, \F1{i\ell m}, \F1{inp},*\F2{iq}\}_{840}\ ,\quad
  \{\F1{ijk}, \F1{i\ell m}, \F1{j\ell n}, *\F2{in}\}_{10080}\ ,\nn\\
&&\{\F1{ijk}, \F1{i\ell m}, \F1{inp}, \F1{j\ell n}\}_{6720}\ ,\quad
  \{\F1{ijk}, \F1{i\ell m}, \F1{j\ell n},*\cF2{p}\}_{6720}\ ,\nn\\
&&\{\F1{ijk}, \F1{i\ell m}, \F1{inp}, \cF1{jk}\}_{2520}\ ,\quad
  \{\F1{ijk}, \F1{i\ell m}, \F1{j\ell n},\cF1{pq}\}_{3360}\ ,\nn\\
&&\{\F1{ijk}, \F1{i\ell m}, \cF1{jk}, *\F2{in}\}_{5040}\ ,\quad
  \{\F1{ijk}, \F1{i\ell m}, \cF1{np}, *\F2{j\ell}\}_{10080}\ ,\nn\\
&&\{\F1{ijk}, \F1{i\ell m}, \cF1{jk}, *\cF2{n}\}_{5040}\ ,\quad
  \{\F1{ijk}, \F1{i\ell m}, \cF1{np}, *\cF2{q}\}_{2520}\ ,\nn\\
&&\{\F1{ijk}, \F1{i\ell m}, \cF1{jk}, \cF1{np}\}_{5040}\ ,\quad
  \{\F1{ijk}, \cF1{\ell m}, *\F2{in}, *\F2{jp}\}_{10080}\ ,\nn\\
&&\{\F1{ijk}, \cF1{\ell m}, *\F2{in}, *\cF2{n}\}_{5040}\ ,\quad
  \{\F1{ijk}, \cF1{ij}, \cF1{\ell m}, *\F2{kn}\}_{5040}\ ,\nn\\
&&\{\F1{ijk}, \cF1{\ell m}, \cF1{np}, *\F2{iq}\}_{2520}\ ,\quad
  \{\F1{ijk}. \cF1{ij}, \cF1{\ell m}, *\cF2{n}\}_{5040}\ ,\nn\\
&&\{\F1{ijk}, \cF1{ij}, \cF1{\ell m}, \cF1{np}\}_{2520}\ ,\quad
  \{\cF1{ij}, *\F2{ij}, *\F2{k \ell}, *\F2{mn}\}_{1260}\ ,\nn\\
&&\{\cF1{ij}, \cF1{k\ell}, *\F2{ij}, *\F2{mn}\}_{2520}\ ,\quad
  \{\cF1{ij}, \cF1{k\ell}, *\F2{mn}, *\F2{pq}\}_{630}\ ,\nn\\
&&\{\cF1{ij}, \cF1{k\ell}, *\F2{mn}, *\cF2{m}\}_{2520}\ ,\quad
  \{\cF1{ij}, \cF1{k\ell}, \cF1{mn}, *\F2{ij}\}_{1260}\ ,\nn\\
&&\{\cF1{ij}, \cF1{k\ell}, \cF1{mn}, *\cF2{p}\}_{840}\ ,\quad
  \{\cF1{ij}, \cF1{k\ell}, \cF1{mn}, \cF1{pq}\}_{105}\ ,
\label{d3n41}
\eea

      The $N=5$ results are the following:
\bea
&&N=5\ ,\qquad M=113400\ :\nn\\
&&\{\F1{ijk}, \F1{i\ell m}, *\F2{in}, *\F2{j\ell}, *\F2{km}\}_{5040}\ ,\quad
  \{\F1{ijk}, \F1{i\ell m}, \F1{j\ell n}, *\F2{in}, *\F2{jm}\}_{10080}\ ,\nn\\
&&\{\F1{ijk}, \F1{i\ell m}, \F1{inp},*\F2{iq}, *\cF2{q}\}_{840}\ ,\quad
  \{\F1{ijk}, \F1{i\ell m}, \F1{j\ell n}, \F1{kmn},*\F2{in}\}_{2520}\ ,\nn\\
&&\{\F1{ijk}, \F1{i\ell m}, \F1{inp}, \F1{j\ell n}, \F1{jmp}\}_{5040}\ ,\quad
  \{\F1{ijk}, \F1{i\ell m}, \F1{inp}, \F1{j\ell n},*\cF2{q}\}_{6720}\ ,\nn\\
&&\{\F1{ijk}, \F1{i\ell m}, \F1{j\ell n}, \F1{kmn}, \cF2{p}\}_{1680}\ ,\quad
  \{\F1{ijk}, \F1{i\ell m}, \F1{j\ell n}, \F1{kmn}, \cF1{pq}\}_{840}\ ,\nn\\
&&\{\F1{ijk}, \F1{i\ell m}, \F1{inp}, \cF1{jk}, *\F2{iq}\}_{2520}\ ,\quad
  \{\F1{ijk}, \F1{i\ell m}, \F1{j\ell n}, \cF1{pq}, *\F2{in}\}_{10080}\ ,\nn\\
&&\{\F1{ijk}, \F1{i\ell m}, \F1{inp}, \cF1{jk}, *\cF2{q}\}_{2520}\ ,\quad
  \{\F1{ijk}, \F1{i\ell m}, \F1{inp}, \cF1{jk}, \cF1{\ell m}\}_{2520}\ ,\nn\\
&&\{\F1{ijk}, \F1{i\ell m}, \cF1{np}, *\F2{iq}, *\F2{j\ell}\}_{10080}\ ,\quad
  \{\F1{ijk}, \F1{i\ell m}, \cF1{np}, *\F2{j\ell}, *\F2{km}\}_{5040}\ ,\nn\\
&&\{\F1{ijk}, \F1{i\ell m}, \cF1{jk}, *\F2{ip}, *\cF2{p}\}_{5040}\ ,\quad
  \{\F1{ijk}, \F1{i\ell m}, \cF1{np}, *\F2{iq}, *\cF2{q}\}_{2520}\ ,\nn\\
&&\{\F1{ijk}, \F1{i\ell m}, \cF1{jk}, \cF1{\ell m}, *\F2{in}\}_{2520}\ ,\quad
  \{\F1{ijk}. \F1{i\ell m}, \cF1{jk}, \cF1{np}, *\F2{iq}\}_{5040}\ ,\nn\\
&&\{\F1{ijk}, \F1{i\ell m}, \cF1{jk}, \cF1{\ell m}, *\cF2{n}\}_{2520}\ ,\quad
  \{\F1{ijk}, \F1{i\ell m}, \cF1{jk}, \cF1{np}, *\cF2{q}\}_{5040}\ ,\nn\\
&&\{\F1{ijk}, \F1{i\ell m}, \cF1{jk}, \cF1{\ell m}, \cF1{np}\}_{2520}\ ,\quad
  \{\F1{ijk}, \cF1{\ell m}, *\F2{in}, *\F2{jp}, *\F2{kq}\}_{3360}\ ,\nn\\
&&\{\F1{ijk}, \cF1{ij}, \cF1{\ell m}, *\F2{mn}, *\cF2{n}\}_{5040}\ ,\quad
  \{\F1{ijk}, \cF1{\ell m}, \cF1{np}, *\F2{iq}, *\cF2{q}\}_{2520}\ ,\nn\\
&&\{\F1{ijk}, \cF1{ij}, \cF1{\ell m}, \cF1{np}, *\F2{kq}\}_{2520}\ ,\quad
  \{\cF1{ij}, \cF1{ij}, *\F2{k\ell}, *\F2{mn}, *\F2{pq}\}_{420}\ ,\nn\\
&&\{\cF1{ij}, \cF1{k\ell}, *\F2{ij}, *\F2{k\ell}, *\F2{mn}\}_{1260}\ ,\quad
  \{\cF1{ij}, \cF1{k\ell}, *\F2{ij}, *\F2{mn}, *\F2{pq}\}_{1260}\ ,\nn\\
&&\{\cF1{ij}, \cF1{k\ell}, \cF1{mn}, *\F2{ij}, *\F2{k\ell}\}_{1260}\ ,\quad
  \{\cF1{ij}, \cF1{k\ell}, \cF1{mn}, *\F2{ij}, *\F2{pq}\}_{1260}\ ,\nn\\
&&\{\cF1{ij}, \cF1{k\ell}, \cF1{mn}, *\F2{pq}, *\cF2{p}\}_{840}\ ,\quad
  \{\cF1{ij}, \cF1{k\ell}, \cF1{mn}, \cF1{pq}, *\F2{ij}\}_{420}\ ,
\label{d3n51}
\eea
 
       The $N=6$ results are the following:
\bea
&&N=6\ ,\qquad M=56700\ :\nn\\
&&\{\F1{ijk}, \F1{i\ell m}, \F1{j\ell n}, *\F2{in}, *\F2{jm}, 
*\F2{k\ell}\}_{3360}\ ,\nn\\
&&  \{\F1{ijk}, \F1{i\ell m}, \F1{j\ell n}, \F1{kmn}, *\F2{in}, 
*\F2{jm}\}_{2520}\ ,\nn\\
&&\{\F1{ijk}, \F1{i\ell m}, \F1{inp}, \F1{j\ell n}, \F1{jmp}, 
\F1{k\ell p}\}_{1680}\ ,\nn\\
&&  \{\F1{ijk}, \F1{i\ell m}, \F1{inp}, \F1{j\ell n}, \F1{jmp}, 
*\cF2{q}\}_{5040}\ ,\nn\\
&&\{\F1{ijk}, \F1{i\ell m}, \F1{j\ell n}, \F1{kmn}, \cF1{pq}, 
*\F2{in}\}_{2520}\ ,\nn\\
&&  \{\F1{ijk}, \F1{i\ell m}, \F1{j\ell n}, \cF1{pq}, *\F2{in}, 
*\F2{jm}\}_{10080}\ ,\nn\\
&&\{\F1{ijk}, \F1{i\ell m}, \F1{inp}, \cF1{jk}, *\F2{iq}, 
  \cF2{q}\}_{2520}\ ,\nn\\
&&  \{\F1{ijk}, \F1{i\ell m}, \F1{j\ell n}, \F1{kmn}, 
  \cF1{pq}\}_{840}\ ,\nn\\
&&\{\F1{ijk}, \F1{i\ell m}, \F1{inp}, \cF1{jk}, \cF1{\ell m}, 
   *\cF2{q}\}_{2520}\ ,\nn\\
&&  \{\F1{ijk}, \F1{i\ell m}, \F1{inp}, \cF1{jk}, \cF1{\ell m}, 
\cF1{np}\}_{840}\ ,\nn\\
&&\{\F1{ijk}, \F1{i\ell m}, \F1{1np}, \cF1{jk}, *\cF2{q}\}_{2520}\ ,\nn\\
&&  \{\F1{ijk}, \F1{i\ell m}, \cF1{np}, *\F2{iq}, *\F2{j\ell}, 
*\F2{km}\}_{5040}\ ,\nn\\
&&\{\F1{ijk}, \F1{i\ell m}, \cF1{jk}, \cF1{\ell m}, *\F2{in}, 
   *\cF2{n}\}_{2520}\ ,\nn\\
&&  \{\F1{ijk}, \F1{i\ell m}, \cF1{jk}, \cF1{np}, *\F2{iq}, 
   *\cF2{q}\}_{5040}\ ,\nn\\
&&\{\F1{ijk}, \F1{i\ell m}, \cF1{jk}, \cF1{\ell m}, \cF1{np}, 
         *\F2{iq}\}_{2520}\ ,\nn\\
&&  \{\F1{ijk}, \F1{i\ell m}, \cF1{jk}, \cF1{\ell m}, 
           \cF1{np}, *\cF2{q}\}_{2520}\ ,\nn\\
&&\{\F1{ijk}, \cF1{ij}, \cF1{\ell m}, \cF1{np}, *\F2{kq}, 
            *\cF2{q}\}_{2520}\ ,\nn\\
&&  \{\cF1{ij}. \cF1{k\ell}, *\F2{ij}, *\F2{k\ell}, *\F2{mn}, 
         *\F2{pq}\}_{630}\ ,\nn\\
&&\{\cF1{ij}, \cF1{k\ell}, \cF1{mn}, *\F2{ij}, *\F2{k\ell}, 
           *\F2{mn}\}_{1680}\ ,\nn\\
&&  \{\cF1{ij}, \cF1{k\ell}, \cF1{mn}, \cF1{pq}, *\F2{ij}, 
         *\F2{k\ell}\}_{630}\ ,\label{d3n61}
\eea

          The $N=7$ results are the following:
\bea
&&N=7\ ,\qquad M=16200\ :\nn\\
&&\{\F1{ijk}, \F1{i\ell m}, \F1{j\ell n}, \F1{kmn}, *\F2{in}, 
           *\F2{jm}, *\F2{k\ell}\}_{840}\ ,\nn\\
&&\{\F1{ijk}, \F1{i\ell m}, \F1{inp}, \F1{j\ell n}, 
             \F1{jmp}, \F1{k\ell p}, \F1{kmn}\}_{240}\ ,\nn\\
&&\{\F1{ijk}, \F1{i\ell m}, \F1{inp}, \F1{j\ell n}, 
              \F1{jmp}, \F1{k\ell p}, *\cF2{q}\}_{1680}\ ,\nn\\
&&\{\F1{ijk}, \F1{i\ell m}, \F1{j\ell n}, \F1{kmn}, 
                  \cF1{pq}, *\F2{in}, *\F2{jm}\}_{2520}\ ,\nn\\
&&\{\F1{ijk}, \F1{i\ell m}, \F1{j\ell n}, \cF1{pq}, 
                   *\F2{in}, *\F2{jm},*\F2{k\ell}\}_{3360}\ ,\nn\\
&&\{\F1{ijk}, \F1{i\ell m}, \F1{inp}, \cF1{jk}, \cF1{\ell m},  
                             *\F2{iq}, *\cF2{q}\}_{2520}\ ,\nn\\
&&\{\F1{ijk}, \F1{i\ell m}, \F1{inp}, \cF1{jk}, 
                     \cF1{\ell m}, \cF1{np}, *\F2{iq}\}_{840}\ ,\nn\\
&&\{\F1{ijk}, \F1{i\ell m}, \F1{inp}, \cF1{jk}, 
                     \cF1{\ell m}, \cF1{np}, *\cF2{q}\}_{840}\ ,\nn\\
&&\{\F1{ijk}, \F1{i\ell m}, \cF1{jk}, \cF1{\ell m}, 
                         \cF1{np}, *\F2{iq}, *\cF2{q}\}_{2520}\ ,\nn\\
&&\{\cF1{ij}. \cF1{k\ell}, \cF1{mn}, *\F2{ij}, *\F2{k\ell}, 
                          *\F2{mn}, *\F2{pq}\}_{420}\ ,\nn\\
&&\{\cF1{ij}, \cF1{k\ell}, \cF1{mn}, \cF1{pq}, *\F2{ij},  
                     *\F2{k\ell}, *\F2{mn}\}_{420}\ ,\label{d3n71}
\eea

          The $N=8$ results are the following:
\bea
&&N=8\ ,\qquad M=2025\ :\nn\\
&&\{\F1{ijk}, \F1{i\ell m}, \F1{inp}, \F1{j\ell n}, \F1{jmp}, 
                     \F1{k\ell p}, \F1{kmn}, *\cF2{q}\}_{240}\ ,\nn\\
&&\{\F1{ijk}, \F1{i\ell m}, \F1{j\ell n}, \F1{kmn}, 
                  \cF1{pq}, *\F2{in}, *\F2{jm}, *\F2{k\ell}\}_{840}\ ,\nn\\
&&\{\F1{ijk}, \F1{i\ell m}, \F1{inp}, \cF1{jk}, 
             \cF1{\ell m}, \cF1{np}, *\F2{iq}, *\cF2{8}\}_{840}\ ,\nn\\
&&\{\cF1{ij}, \cF1{k\ell}, \cF1{mn}, \cF1{pq}, *\F2{ij}, 
               *\F2{k\ell}, *\F2{mn},\ *\F2{pq}\}_{105}\ ,\label{d3n81}
\eea
All of the solutions preserve $\ft1{16}$ of the supersymmetry.

\bigskip
$D=2$
\bigskip

      As we discussed in section 2, all the possible field
configurations for $D=2$ multi-charge instanton solutions are given by
the 1-form field configurations of $D=3$, except that in the case of
$D=2$, each field strength in the list can either be dualised or
un-dualised.


\begin{thebibliography}{99}

\bm{ht} C.M. Hull and P.K. Townsend, {\sl Unity of superstring
dualities}, Nucl. Phys. {\bf B438} (1995) 109-137: hep-th/9410167

\bm{cj1} E. Cremmer and J. Julia, {\sl The $N=8$ supergravity
theory-1-the Lagrangian}, Phys. Lett. {\bf B80} (1978) 48;
{\sl The SO(8) supergravity}, Nucl. Phys. {\bf B156} (1979) 141.

\bm{cj2} B. Julia, {\sl Group disintegrations}; E. Cremmer, 
{\sl Supergravities in 5 dimensions},
in {\sl Superspace and Supergravity}, Eds. S.W. Hawking and M. 
Rocek (Cambridge Univ. Press, 1981) 331; 267.

\bm{pt} G. Papadopoulos and P.K. Townsend,
{\sl Intersecting $M$-branes},
Phys. Lett. {\bf B380} (1996) 273: hep-th/9603087.

\bm{tsey1}A.A. Tseytlin,
{\sl Harmonic superpositions of $M$-branes},
 Nucl. Phys. {\bf B475} (1996) 149: hep-th/9604035.

\bm{cjs} E. Cremmer, B. Julia and J. Scherk, {\sl Supergravity theory
in eleven dimensions}, Phys. Lett. {\bf B76} (1978) 409.

\bm{lpsol} H. L\"u and C.N. Pope, {\sl $p$-brane solitons in maximal
supergravities}, Nucl. Phys. {\bf B465} (1996) 127: hep-th/9512012

\bm{clpst} P.M. Cowdall, H. L\"u, C.N. Pope, K.S. Stelle and P.K. Townsend,
{\sl Domain walls in massive supergravities}, Nucl. Phys. {\bf B486} 
(1997) 49: hep-th/9608173.

\bm{lpmult} H. L\"u and C.N. Pope, {\sl Multi-scalar $p$-brane solitons,}
Int. J. Mod. Phys. {\bf A12} (1997) 437: hep-th/9512153.

\bm{lpss1} H. L\"u, C.N. Pope, K.S. Stelle and E. Sezgin, {\sl Stainless
super $p$-branes,} Nucl. Phys. {\bf B456} (1995) 669: hep-th/9508042.

\bm{dfkr} M.J. Duff, S. Ferrara, R.R. Khuri and J. Rahmfeld, {\sl
Supersymmetry and dual string solitons}, Phys. Lett. {\bf B356}
(1995) 479: hep-th/9506057.

\bm{r} J. Rahmfeld, {\sl Extremal black holes as bound
states}, Phys. Lett. {\bf B372} (1996) 198: hep-th/9512089.

\bm{kklp} N. Khviengia, Z. Khviengia, H. L\"u and C.N. Pope, {\sl
Intersecting M-branes and bound states}, Phys. Lett. {\bf B388}
(1996) 21: hep-th/9605077.

\bm{drbound} M.J. Duff and J. Rahmfeld, {\sl Bound states
of black holes and other $p$-branes}, Nucl. Phys. {\bf B481}
(1996) 332: hep-th/9605085.

\bibitem{dghr}A. Dabholkar, G.W. Gibbons, J.A. Harvey and
F. Ruiz Ruiz, {\sl Superstrings and solitons}, 
Nucl. Phys. {\bf B340} (1990) 33.

\bm{str} A. Strominger, {\sl Heterotic solitons}, Nucl. Phys.
{\bf B343} (1990) 167.

\bm{dust} M.J. Duff and K.S. Stelle, {\sl Multi-membrane solutions
of $D=11$ supergravity}, Phys. Lett. {\bf B253} (1991) 113.

\bm{dl1} M.J. Duff and J.X. Lu, {\sl Elementary fivebrane solutions of
$D=10$ supergravity}, Nucl. Phys. {\bf B354} (1991) 141;
{\sl Strings from fivebranes}, Phys. Rev. Lett. {\bf 66} (1991) 1402;
{\sl The selfdual type IIB superthreebrane}, Phys. Lett. {\bf B273}
(1991) 409.

\bm{chs} C.G. Callan, J.A. Harvey and A. Strominger, {\sl World sheet
approach to heterotic instantons and solitons}, Nucl. Phys. 
{\bf B359} (1991) 611; {\sl Worldbrane actions for string solitons},
Nucl. Phys. {\bf B367} (1991) 60.

\bm{guv} R. G\"uven, {\sl Black $p$-brane solutions of $D=11$
supergravity theory}, Phys. Lett. {\bf B276} (1992) 49.

\bm{k1} R.R. Khuri, {\sl A comment on the stability of string
monopoles}, Phys. Lett. {\bf B307} (1993) 302: hep-th/9301035;
{\sl Classical dynamics of macroscopic strings}, Nucl. Phys. 
{\bf B403} (1993) 335: hep-th/9212029.

\bm{cy} M. Cvetic and D. Youm, {\sl Static four-dimensional abelian
black holes in Kaluza-Klein theory}, Phys. Rev. {\bf D52}, (1995)
2144: hep-th/9502099; {\sl Kaluza-Klein black holes within heterotic
string theory on a torus}, Phys. Rev. {\bf D52} (1995) 2574:
hep-th/9502119; {\sl All the four-dimensional static, spherically
symmetric solutions of abelian Kaluza-Klein theory},
Phys. Rev. Lett. {\bf 75} (1995) 4165: hep-th/9503082.

\bm{dklreview} M.J. Duff, R.R. Khuri and J.X. Lu, {\sl String
solitons}, Phys. Rep. {\bf 259} (1995) 213: hep-th/9412184.

\bm{lpsweyl} H. L\"u, C.N. Pope and K.S. Stelle, {\sl Weyl group invariance
and $p$-brane multiplets}, Nucl. Phys. {\bf B476} (1996) 89: hep-th/9602140.

\bm{aafft} L. Andrianopoli, R. D'Auria, S. Ferrara, P. Fr\'e and
M. Trigiante, {\sl $E_{7(7)}$ duality, BPS black-hole evolution and
fixed scalars,} hep-th/9707087.

\bm{kr1} R.R Khuri, {\sl A comment on string solitons}, Phys. Rev. 
{\bf D48} (1993) 2947: hep-th/9305143.

\bm{kt} I.R. Klebanov and A.A. Tseytlin, {\sl Intersecting M-branes as
four dimensional black holes}, Nucl. Phys. {\bf B475} (1996) 179:
hep-th/9604166.

\bm{bbj} K. Behrndt, E. Bergshoeff and B. Janssen, {\sl Intersecting
D-branes in ten and six dimensions}, Phys. Rev. {\bf D55} (1997) 3785:
hep-th/9604168.

\bm{gkt} J.P. Gauntlett, D.A. Kastor and J. Traschen, {\sl Overlapping
branes in M-theory}, Nucl. Phys. {\bf B478} (1996) 544: hep-th/9604179.

\bm{lpsvert} H. L\"u, C.N. Pope and K.S. Stelle, {\sl Vertical versus
diagonal dimensional reduction for $p$-branes}, Nucl. Phys. {\bf B481}
(1996) 313: hep-th/9605082.

\bm{bdejs} E. Bergshoeff, M. de Roo, E. Eyras, B. Janssen, J.P. van
der Schaar, {\sl Multiple intersections of D-branes and M-branes},
hep-th/9612095.

\bm{roo} M. de Roo, {\sl Intersecting branes and supersymmetry,} 
hep-th/9703124.  

\bm{bdejs2}  E. Bergshoeff, M. de Roo, E. Eyras, B. Janssen, J.P. van
der Schaar, {\sl Intersections involving monopoles and waves in eleven
dimensions,} hep-th/9704120.

\bm{pol1} J. Polchinski, {\sl Dirichlet branes and Ramond-Ramond charges,}
Phys. Rev. Lett. {\bf 75} (1995) 4724: hep-th/9510017.

\bm{ss1} J. Scherk and J.H. Schwarz, {\sl How to get masses from extra 
dimensions}, Nucl. Phys. {\bf B153} (1979) 61.

\bm{bdgpt} E. Bergshoeff, M. de Roo, M.B. Green, G. Papadopoulos and
P.K. Townsend, {\sl Duality of type II 7-branes and 8-branes}, Nucl. Phys. 
{\bf B470} (1996) 113: hep-th/9601150.

\bm{lpdomain} H. L\"u and C.N. Pope, {\sl Domain walls from M-branes},
Mod. Phys. Lett. {\bf A12} (1997) 1087: hep-th/9611079.

\bm{llp} I.V. Lavrinenko, H. L\"u and C.N. Pope, {\sl From topology to
generalised dimensional reduction}, Nucl. Phys. {\bf B492} (1997) 278:  
hep-th/9611134.

\bm{ilpt} J.M. Izquierdo, N.D. Lambert, G. Papadopoulos and P.K.
Townsend, {\sl Dyonic membranes}, Nucl. Phys. {\bf B460} (1996) 560:
hep-th/9508177.

\bm{cjlp} E. Cremmer, B. Julia, H. L\"u and C.N. Pope, work in progress.

\bm{gk} G.W. Gibbons and R.E. Kallosh, {\sl Topology, entropy and
Witten index of dilaton black holes}, Phys. Rev. {\bf D51} (1995)
2839: hep-th/9407118.

\bm{ko} R.R. Khuri and T. Ortin, {\sl A non-supersymmetric dyonic extreme
Reissner-Nordstr{\o}m black hole}, Phys. Lett. {\bf B373} (1996) 56:
hep-th/9512178.

\bm{lpsln} H. L\"u and C.N. Pope, {\sl $SL(N+1, R)$ Toda solitons in 
supergravities}, Int. J. Mod. Phys. {\bf A12} (1997) 2061: hep-th/9607027.

\bm{lmmp} H. L\"u, J. Maharana, S. Mukherji and C.N. Pope, {\sl Cosmological
solutions, $p$-branes and the Wheeler-DeWitt equation},  hep-th/9707182.

\bm{klopp} R. Kallosh, A. Linde, T. Ortin, A. Peet and A. Van Proeyen,
{\sl Supersymmetry as a cosmic censor}, Phys. Rev. D46 (1992) 5278:
hep-th/9205027. 

\bm{clps} E. Cremmer, H. L\"u, C.N. Pope and K.S. Stelle, {\sl 
Spectrum-generating symmetries for BPS solitons}, hep-th/9707207.

\bm{dlp} M.J. Duff, H. L\"u and C.N. Pope, {\sl The black branes of
M-theory}, Phys. Lett. {\bf B382} (1996) 73: hep-th/9604052.

\end{thebibliography}
\end{document}